\documentclass[preprint,notoc]{JHEP3}
\usepackage{epsfig}
\def\eslt{E_T^{\rm miss}}
\def\to{\rightarrow}

\def\bi{\begin{itemize}}
\def\ei{\end{itemize}}
\def\te{\tilde e}

\def\tu{\tilde u}

\def\tb{\tilde b}
\def\tf{\tilde f}

\def\tQ{\tilde Q}

\def\tst{\tilde t}
\def\ttau{\tilde \tau}
\def\tmu{\tilde \mu}
\def\tg{\tilde g}
\def\tnu{\tilde\nu}
\def\tell{\tilde\ell}
\def\tq{\tilde q}

\def\tw{\widetilde W}
\def\tz{\widetilde Z}
\def\alt{\stackrel{<}{\sim}}
\def\agt{\stackrel{>}{\sim}}
\def\be{\begin{equation}}  
\def\ee{\end{equation}}  

\title{Collider and Dark Matter Phenomenology\\ 
of Models with Mirage Unification}

\author{Howard Baer$^{a,b}$, Eun-Kyung Park$^a$, Xerxes Tata$^c$ 
and Ting T. Wang$^{a,b}$\\
$^a$Department of Physics, Florida State University, Tallahassee, 
FL 32306, USA\\
$^b$Department of Physics, University of Wisconsin, Madison, WI,
53706, USA\\
$^c$Department of Physics and Astronomy, University of Hawaii,
Honolulu, HI 96822, USA\\
E-mail: \email{baer@hep.fsu.edu},\email{epark@hep.fsu.edu},
\email{tata@phys.hawaii.edu},\email{tingwang@hep.fsu.edu} }

\preprint{\vbox{\hbox{FSU-HEP-070228, UH-511-1100-07}}} 

\abstract{
We examine supersymmetric models with mixed modulus-anomaly mediated
SUSY breaking (MM-AMSB) soft terms which get comparable 
contributions to SUSY breaking from moduli-mediation and anomaly-mediation. 
The apparent (mirage) unification
of soft SUSY breaking terms at $Q=\mu_{\rm mir}$ not associated with any
physical threshold is the hallmark of this scenario. 
The MM-AMSB structure of soft terms arises in
models of string compactification with fluxes, where the addition of
an anti-brane leads to an uplifting potential and a de Sitter universe,
as first constructed by Kachru {\it et al.}.
The phenomenology mainly depends on the relative strength of moduli- and
anomaly-mediated SUSY breaking contributions, and on
the Higgs and matter field modular weights, 
which are determined by the location of these fields in the extra dimensions.
We delineate 
the allowed parameter space for a low and high value of $\tan\beta$,
for a wide range of modular weight choices. We calculate the neutralino 
relic density and display the WMAP-allowed regions. We 
show the reach of the CERN LHC and of the International Linear Collider.
We discuss aspects of MM-AMSB models for Tevatron, LHC and ILC
searches, muon $g-2$ and $b\to s\gamma $ branching fraction. 
We also calculate direct and indirect dark matter detection rates,
and show that almost all WMAP-allowed models should be accessible to a 
ton-scale noble gas detector. Finally, we comment on the potential of 
colliders to measure the mirage unification scale
and modular weights in the difficult case where $\mu_{\rm mir}\gg 
M_{\rm GUT}$. 
}


\begin{document}

\section{Introduction}
\label{sec:intro}

Superstring theory allows for a consistent merging of gravitational
physics with quantum mechanics, while also containing within it
the possibility of describing the gauge interactions of the Standard Model. 
This impressive theoretical framework does not, however, allow the extraction
of predictions of physical phenomena at experimentally accessible
energy scales.
A major obstruction to obtaining a predictive superstring theory is the
lack of understanding of how the degeneracy associated with the many
flat directions in the space of scalar fields (the moduli) is lifted to
yield the true ground state, the problem being that many couplings
necessary for the extraction of ``observables quantities'' are
determined by the ground state values of these moduli.

The discovery of a new class of compactifications, where the extra
spatial dimensions are curled up to small sizes with fluxes of
additional fields trapped along these extra dimensions has been
exploited by Kachru {\it et al.} (KKLT)\cite{kklt} to construct a
concrete model with a stable, calculable ground state with a positive
cosmological constant and broken supersymmetry.  This toy model is based
on type-IIB superstrings including compactification of the extra
dimensions to a Calabi-Yau orientifold, with fluxes along these extra
directions.  While the background fluxes serve to stabilize the dilaton
and the moduli that determine the shape of the compact manifold, it is
necessary to invoke a non-perturbative mechanism such as gaugino
condensation on a $D7$ brane to stabilize the size of the compact
manifold.  Finally, a non-supersymmetric anti-brane ($\overline{D3}$) --
included in order to break supersymmmetry -- also yields a de Sitter
universe as required by observations.  The KKLT construction, which
yields an example of a low energy theory that has no unwanted light
moduli, broken supersymmetry and a positive cosmological constant,
may be viewed as
a starting point for the program of discovering a string ground
state that may lead to the (supersymmetric) Standard Model at low
energies, and which is consistent with various constraints from
cosmology.

These considerations have 
recently motivated several authors to analyze 
the structure of the soft SUSY breaking (SSB) terms in models based on 
a generalization of the KKLT set-up\cite{choi}. 
The key observation is that because of the mass hierarchy
\begin{equation}
m_{\rm moduli}\gg m_{3/2}\gg m_{\rm SUSY}
\label{eq:hierarchy}
\end{equation}
that develops in these models,
the SSB terms receive comparable contributions via both
modulus (gravity) and anomaly mediation of SUSY breaking\cite{amsb}, 
with their relative size parametrized by one new parameter $\alpha$. 
The hierarchy (\ref{eq:hierarchy}) that leads to mixed modulus-anomaly
mediated SUSY breaking seemingly allows the moduli to decay early enough
not to disrupt Big Bang nucleosynthesis. It has, however, been pointed
out that these decays of the moduli would inevitably produce an
unacceptably large number of gravitinos (or other sparticles)
\cite{endo} which would subsequently decay to the LSP, unless the
gravitino is itself heavier than $\sim 100$~TeV, or the density of
particles is reduced by some mechanism such as a period of thermal
inflation\cite{lyth}. Here, we will assume that such a mechanism is
operative and that the observed dark matter, in our case the lightest
neutralino, is produced thermally upon subsequent reheating.  Upon
integrating out the heavy dilaton field and the shape moduli, we are
left with an effective broken supergravity theory of the observable
sector fields denoted by $\hat{Q}$ and the size modulus field $\hat{T}$.
The K\"ahler potential depends on the location of matter and Higgs
superfields in the extra dimensions via their modular weights $n_i = 0 \
(1)$ for matter fields located on $D7$ ($D3$) branes, or $n_i=1/2$ for
chiral multiplets on brane intersections, while the gauge kinetic
function $f_a={\hat{T}}^{l_a}$, where $a$ labels the gauge group, is
determined by the corresponding location of the gauge supermultiplets,
since the power $l_a= 1 \ (0)$ for gauge fields on $D7$ ($D3$) branes
\cite{choi3}.\footnote{More specifically, these modular weights for
chiral superfields are obtained for examples with toroidal
compactifications, possibly with singularities \cite{ibanez}. This is
not generic as Calabi-Yau compactifications allow for more general choices:
for instance $n_i= \frac{2}{3}$ may also be allowed (see
Refs. \cite{conlon,choi4}).}

Within the MM-AMSB model,
the SSB gaugino mass parameters, trilinear SSB
parameters and sfermion mass parameters, all renormalized just below the
unification scale (taken to be $Q=M_{\rm GUT}$), are given (in
the convention used in the event generator Isajet \cite{isajet}) by,
\begin{eqnarray}
M_a&=& M_s\left( l_a \alpha +b_a g_a^2\right),\label{eq:M}\\
A_{ijk}&=& M_s \left( -a_{ijk}\alpha +\gamma_i +\gamma_j +\gamma_k\right),
\label{eq:A}\\
m_i^2 &=& M_s^2\left( c_i\alpha^2 +4\alpha \xi_i -
\dot{\gamma}_i\right) ,\label{eq:m2}
\end{eqnarray}
where $M_s\equiv\frac{m_{3/2}}{16\pi^2}$, $b_a$ are the gauge $\beta$
function coefficients for gauge group $a$ and $g_a$ are the
corresponding gauge couplings. The coefficients that appear in
(\ref{eq:M})--(\ref{eq:m2}) are given by $c_i =1-n_i$,
$a_{ijk}=3-n_i-n_j-n_k$ and $\xi_i=\sum_{j,k}a_{ijk}{y_{ijk}^2 \over 4}
- \sum_a l_a g_a^2 C_2^a(f_i).$ Finally, $y_{ijk}$ are the
superpotential Yukawa couplings, $C_2^a$ is the quadratic Casimir for
the a$^{th}$ gauge group corresponding to the representation to which
the sfermion $\tf_i$ belongs, $\gamma_i$ is the anomalous dimension and
$\dot{\gamma}_i =8\pi^2\frac{\partial\gamma_i}{\partial \log Q}$.
Expressions for the last two quantities involving the anomalous
dimensions can be found in the Appendix of Ref. \cite{flm}, whose
notation we adopt here.\footnote{We note that $\alpha$ defined in
Ref. \cite{choi3} differs from the definition in Ref. \cite{flm} that we
use here by $\alpha_{\rm Ref.\cite{choi3}}=
\frac{16\pi^2}{\ln(M_P/m_{3\over 2})}{1\over \alpha_{\rm our}}$. The
original KKLT construction yields $\alpha_{\rm Ref.\cite{choi3}}\simeq
1$, corresponding to $\alpha \sim 4.6$ for $m_{3/2}\sim 1$~TeV, in the
notation used in this paper.}

The MM-AMSB model is completely specified by the parameter set,
\begin{equation}
\ m_{3/2},\alpha ,\ \tan\beta ,\ sign(\mu ),\ n_i,\ l_a.
\label{eq:ps}
\end{equation}
The mass scale for the MSSM SSB parameters is dictated by $M_s \equiv
\frac{m_{3/2}}{16\pi^2}$. The phenomenological parameter
$\alpha$, which could be of either sign, determines the relative
contributions of anomaly mediation and gravity mediation to the soft
terms, and as mentioned above $|\alpha| \sim {\cal O}(1)$ is the
hallmark of this scenario. Non-observation of large flavor changing
neutral currents implies common modular weights of particles with the
same gauge quantum numbers: within this framework, this suggests a
common location for these fields in the extra dimensions. 
Grand Unification implies matter particles
within the same GUT multiplet have common modular weights, and that the
$l_a$ are universal. We will assume that all $l_a=l$ and, for
simplicity, a common modular weight for all matter particles, but allow
a different (common) one for the two Higgs doublets of the MSSM.

The universality of the $l_a$ leads to the
phenomenon of {\it mirage unification}\cite{choi3,flm} of gaugino
masses. 
In mirage unification, the splitting of the gaugino masses at
$Q=M_{\rm GUT}$ is proportional to $\beta_a(g_a)/g_a$,
where $\beta_a$ are the beta-functions of their associated
gauge groups. As the gaugino masses run from $M_{\rm GUT}$ to lower energy 
scales, the RG running exactly cancels the GUT scale mass splitting, leading
(for $\alpha >0$) to unified gaugino masses at some {\it intermediate} 
energy scale 
$Q=\mu_{\rm mir} \not= M_{\rm GUT}$, the scale of unification
of gauge couplings. Indeed, the observation of gaugino unification at
the mirage unification scale,
\be
\mu_{\rm mir}=M_{\rm GUT}e^{-8\pi^2/(l \alpha)} ,
\label{eq:mumir}
\ee
is the smoking gun of the MM-AMSB scenario. If $\alpha < 0$, 
$\mu_{\rm   mir}> M_{\rm GUT}$, though one would have to continue extrapolation
{\em still using MSSM RGEs} beyond $M_{\rm GUT}$ to discover this! 
We will assume hereafter that $l\not=0$, 
since $l=0$ would be distinguished by a gaugino mass pattern as in
the AMSB framework. 
While $\mu_{\rm mir}$ determines $l\alpha$, the (unified) value of the 
the gaugino masses extrapolated to $Q=\mu_{\rm mir}$ is $M_a(\mu_{\rm
  mir})= M_s \times (l\alpha)$, and so gives the value of $M_s$ (and hence
$m_{3/2}$). We note here that the soft SUSY breaking scalar masses
of the first two generations also unify at $Q=\mu_{\rm mir}$, which allows for
experimental corroboration of mirage unification of gaugino masses.
In fact, for cases with $n_H+2 n_m =2$, third generation and Higgs SSB
terms also unify at $\mu_{\rm mir}$ \cite{choi3}.

Phenomenologically attractive features of the MM-AMSB scenario, also
referred to as the {\it mirage mediation} model,
are that it provides natural
solutions to 1)~the negative slepton mass squared
problem inherent to AMSB models, 2)~the SUSY flavor
problem, plausibly assuming common location for matter fermions
in the extra dimensions, and the concomitant universality of their
modular weights,
and 3) the SUSY $CP$ problem, in that (up to ${\cal O}({1\over
{4\pi^2}})$ corrections) there are no physical $CP$ violating phases in $\mu$,
gaugino masses and $A$-terms: moreover, there are mechanisms that also yield a
real value for $b\equiv B\mu$\cite{choi3,choi4}.

For these reasons, a number of authors have begun exploring the
associated collider and dark matter phenomenology of these models.
Indeed, in Ref. \cite{choi3} the phenomena of mirage unification was noted,
while mass spectra were computed in Ref's \cite{choi3,flm,endo2,kklt1}.
In Ref. \cite{kn}, it was emphasized that bringing the mirage unification
scale down close to the weak scale ameliorates fine-tuning problems in 
supersymmetric theories. In Ref. \cite{kklt1}, mass spectra were computed
in the MM-AMSB scenario for two choices of modular weights 
$(n_H,\ n_m)=(0,\ 0)$ and $(1,\ {1\over 2})$. Regions of parameter
space where the neutralino is the lightest supersymmetric particle, and
where its 
predicted relic abundance (assuming it is a thermal relic in standard
Big Bang cosmology) is in agreement 
with WMAP measurements\cite{wmap} were mapped out. Allowing that 
the observed DM may consist of more than one component, 
$\Omega_{\tz_1}h^2\alt 0.13$. 
Collider reaches in MM-AMSB parameter space were also presented.
In Ref. \cite{nojiri}, aspects of  LHC
detection for MM-AMSB spectra where sparticle masses are roughly 
degenerate were examined. 
In Ref. \cite{kklt2}, it was pointed out that measurement of 
soft SUSY breaking terms at the LHC and ILC could measure the matter field
and in some cases the Higgs field modular weights, by taking ratios of 
scalar to gaugino masses at $\mu_{\rm mir}$.
In Ref. \cite{choi_dm}, the neutralino 
relic abundance and direct and indirect dark matter detection rates were
presented for four modular weight choices.

In this paper, we present allowed parameter space regions for a wide
range of Higgs and matter field modular weights at low and high
$\tan\beta$ values. We also compute the neutralino relic abundance, and
note where it is consistent with WMAP measurements. We present the
approximate reach of the CERN LHC with 100~fb$^{-1}$ of integrated
luminosity, and the reach of a $\sqrt{s}=0.5$ and 1 TeV international
linear $e^+e^-$ collider (ILC).  We organize the parameter space
discussion and delineate the WMAP allowed regions in Sec.~2.  In
Secs.~3-6, we present various features of MM-AMSB models from a scan
over a wide range of modular weight choices, confining ourselves to the
regions consistent with WMAP measurements.  In Sec.~3, we present plots
of $(g-2)_\mu$ and the branching fraction for $b\to s\gamma $
decays. The value of $(g-2)_\mu$ is found to rule out most of the
MM-AMSB parameter space at large negative values of $\alpha$ and $\mu
>0$.  In Sec.~4, we comment on features of WMAP allowed MM-AMSB models
relevant for SUSY searches at the Fermilab Tevatron, CERN LHC and the
ILC.  In Sec.~5, we discuss projections for signals via various direct
and indirect searches for dark matter. We find that almost all
WMAP-allowed parameter space should be accessible to ton-size direct
dark matter search experiments. Signals from indirect searches
(especially those from high energy gamma rays) are (very) sensitive to
the dark matter halo profile in our Galaxy. An optimistic choice for
this implies that GLAST would be sensitive to essentially the entire
parameter space, but a different (yet currently viable) choice suggests
that many models may be below the level of sensitivity for a wide range
of $m_{\tz_1}$.  In Sec.~6, we extend our earlier discussion
\cite{kklt2}, and consider the possibility of determining mirage
unification and matter modular weights in the case where $\alpha <0$ where
$\mu_{\rm mir}>M_{\rm GUT}$.  In Sec.~7, we present our conclusions.

\section{WMAP-Allowed Parameter Space of MM-AMSB models}
\label{sec:model}

To facilitate calculations within the MM-AMSB framework, we have
incorporated it as model line number 9 in the Isajet 7.75 event
generator\cite{isajet}.  For a given set of parameters~(\ref{eq:ps}),
Isajet runs the measured gauge and Yukawa couplings from the weak scale
to the GUT scale, where the GUT scale is determined by where gauge
couplings $g_1=g_2$. The boundary conditions (\ref{eq:M})-(\ref{eq:m2})
are imposed at $Q=M_{\rm GUT}$, and the values of weak scale SSB
parameters are obtained by numerically solving the complete set of
26 coupled 2-loop renormalization group equations. Sparticle mixing
matrices are determined by freezing the parameters at a scale
$Q=\sqrt{m_{\tst_L}m_{\tst_R}}$, while non-mixing soft parameters (that
determine masses) are frozen out at a scale equal to their
value\cite{kraml}. Next, the RG-improved one-loop effective potential is
minimized at an optimized scale (which accounts for leading two-loop
terms), allowing the magnitude of $\mu$ to be determined, and complete
one-loop radiative corrections to sparticle and Higgs masses are
obtained. The Yukawa couplings are updated due to threshold effects. The
SSB parameters evolved back to $Q=M_{\rm GUT}$, of course, do not match their
input boundary values because of the differences in the upward and
downward evolution. 
The procedure is iterated until a stable solution is
obtained. The neutralino relic
density $\Omega_{\tz_1}h^2$, $BF(b\to s\gamma )$, $(g-2)_\mu$,
$BF(B_s\to \mu^+\mu^- )$ and the direct dark matter detection cross
section $\sigma (\tz_1 p )$ are then obtained using the Isatools package
\cite{bbb}.  We interface to
DarkSUSY\cite{darksusy} to obtain indirect dark matter detection rates.

Throughout our analysis, we take all the $l_a=1$, and assume that all
matter fields have a common modular weight $n_m$, but allow the Higgs
fields to have a different modular weight $n_H$ not necessarily equal to
$n_m$. We organize our discussion of the models by specifying the values
of ($n_H$, $n_m$) where each of these is allowed to take the values, 0,
1/2 and 1. The phenomenology of models with intermediate values (1/3 and
2/3) of modular weights that may be allowed by more general
compactifications is presumably bracketed by the nine cases that we
study. For any choice of $\tan\beta$ and the modular weights, the
$\alpha -m_{3/2}$ plane provides a convenient panorama for an overview
of the phenomenology. We begin by delineating the regions of this plane
that are allowed by theoretical considerations and by the constraint on
$\Omega_{\tz_1}h^2$ from WMAP, starting with the examination of the
three $n_m$ cases with $n_H=0$.

\subsection{${\bf n_H=0}$ cases}

Our first results are presented in Fig. \ref{fig:psnH0}. Here we show
the allowed parameter space in the $\alpha\ vs.\ m_{3/2}$ plane for
$n_H=0$ and {\it a}) $n_m=0,\ \tan\beta =10$, {\it b}) $n_m=0,\
\tan\beta =30$, {\it c}) $n_m=1/2,\ \tan\beta =10$, {\it d}) $n_m=1/2,\
\tan\beta =30$ {\it e}), $n_m=1,\ \tan\beta =10$ and {\it f}) $n_m=1,\
\tan\beta =30$.  We take $\mu >0$ and $m_t=175$ GeV throughout, and
require points to be compatible with approximate sparticle and Higgs
mass constraints from LEP2 searches: $m_{\tw_1}>103$ GeV,
$m_{\ttau_1}>95$ GeV, $m_{\tz_1}+m_{\tz_2}>120$ GeV and $m_h>110$ GeV
\cite{lep2}.\footnote{While the LEP2 SM Higgs mass constraint
$m_h>114.4$ GeV translates to a corresponding constraint on $h$ as long
as $m_A$ is large, we require a somewhat lower bound owing to an
expected 3-4 GeV uncertainty on the theory calculation of $m_h$; for a
discussion of how this is affected if $m_A \alt 150-200$~GeV, see
Ref.\cite{sasha}.}  The white unshaded regions do not lead to an
acceptable sparticle mass spectrum due to a wrong pattern of
electroweak symmetry breaking, signalled by tachyonic sfermion 
masses, a negative value of $\mu^2$
or one of the Higgs squared masses. The turquoise-shaded region leads to a
top-squark LSP, while the magenta-shaded region leads to a stau LSP:
these regions would likely give rise to stable colored or charged relics
from the Big Bang, and hence are also excluded. The blue-dotted region
leads to an acceptable sparticle mass spectrum with a neutralino LSP,
but in this case the calculated relic density $\Omega_{\tz_1}h^2>0.5$,
in violation of WMAP limits. The green-shaded region has a lower relic
density, $0.13<\Omega_{\tz_1}h^2<0.5$ but is also excluded. The
red-shaded regions have $\Omega_{\tz_1}h^2<0.13$ in accord with WMAP
and hence are allowed. Frames {\it a}) and {\it b}) are repeated, but
updated, from Ref. \cite{kklt1}. As noted in Ref. \cite{kklt1} (and also
in Ref. \cite{choi_dm} and the updated Ref. \cite{flm}), in frame {\it
a}) the allowed red-shaded region for $\alpha >0$ occurs because the
$\tst_1$ is quite light, with $m_{\tst_1}\sim m_{\tz_1}$. This is due to
the large value of the $A_t$ parameter occurring in Eq. \ref{eq:A}.  A
large $A_t$ feeds into the running of the soft SUSY breaking terms
$m_{\tst_L}^2$ and $m_{\tst_R}^2$ via large values of
$X_t=m_{\tst_L}^2+m_{\tst_R}^2+m_{H_u}^2+A_t^2$ in the corresponding RGEs, which
then accentuates the impact of the large top-quark Yukawa coupling in driving
these to small values. The resultant light top squark $\tst_1$
enhances neutralino annihilation in the early universe via top-squark
co-annihilation.  There is also a red-shaded region around $\alpha\sim
-2$ where the weak scale gaugino masses $M_1\sim -M_2$, so that
bino-wino co-annihilation (BWCA) acts to reduce the neutralino relic
density\cite{bwca} to the required level.  

In frame {\it b}), for $\tan\beta =30$,
once again we see a top-squark-coannihilation region adjacent to the
turquoise-shaded stop-LSP region. But in this case, at large $\alpha$ and low
$m_{3/2}\sim 5$ TeV, there is also a region of stau-coannihilation.
Note in frame {\it b}) that the WMAP-allowed BWCA region at 
$\alpha\sim -2$ has been eliminated. 
On these plots we also denote the approximate reach of the CERN LHC with 
100 fb$^{-1}$ of integrated luminosity. The LHC reach has been evaluated in
Ref. \cite{lhcreach} and found to extend to $m_{\tg}\sim 3.1$ TeV in the case
where $m_{\tq}\simeq m_{\tg}$ 
(as is the case for MM-AMSB models, see Sec. 4).\footnote{While the LHC reach 
was calculated in Ref. \cite{lhcreach} for the case of the mSUGRA model, 
it was found that the reach mainly depends on the squark and gluino masses, 
and not on their particular decay modes, 
so long as there is ample energy release in the squark and 
gluino cascade decays. For this reason, we expect that the reach results
of Ref. \cite{lhcreach} will also apply approximately to the case of 
MM-AMSB models, the exception being where the energy released in the
primary decays is suppressed for kinematic reasons \cite{nojiri}.}
The 100 fb$^{-1}$ LHC reach extends up to $m_{3/2}\sim 60$ TeV,
essentially covering all the region with $m_{\tg}\sim m_{\tq}\alt 3.1$ TeV.
The reach of a $\sqrt{s}=0.5$ and 1 TeV ILC is also indicated. The ILC
reach is determined mainly by the kinematic limit for 
$\tw_1^+\tw_1^-$ or $\ttau_1^+\ttau_1^-$
pair production processes\cite{bbkt}, 
and is somewhat smaller than the reach of the CERN LHC.
\FIGURE[htb]{
\epsfig{file=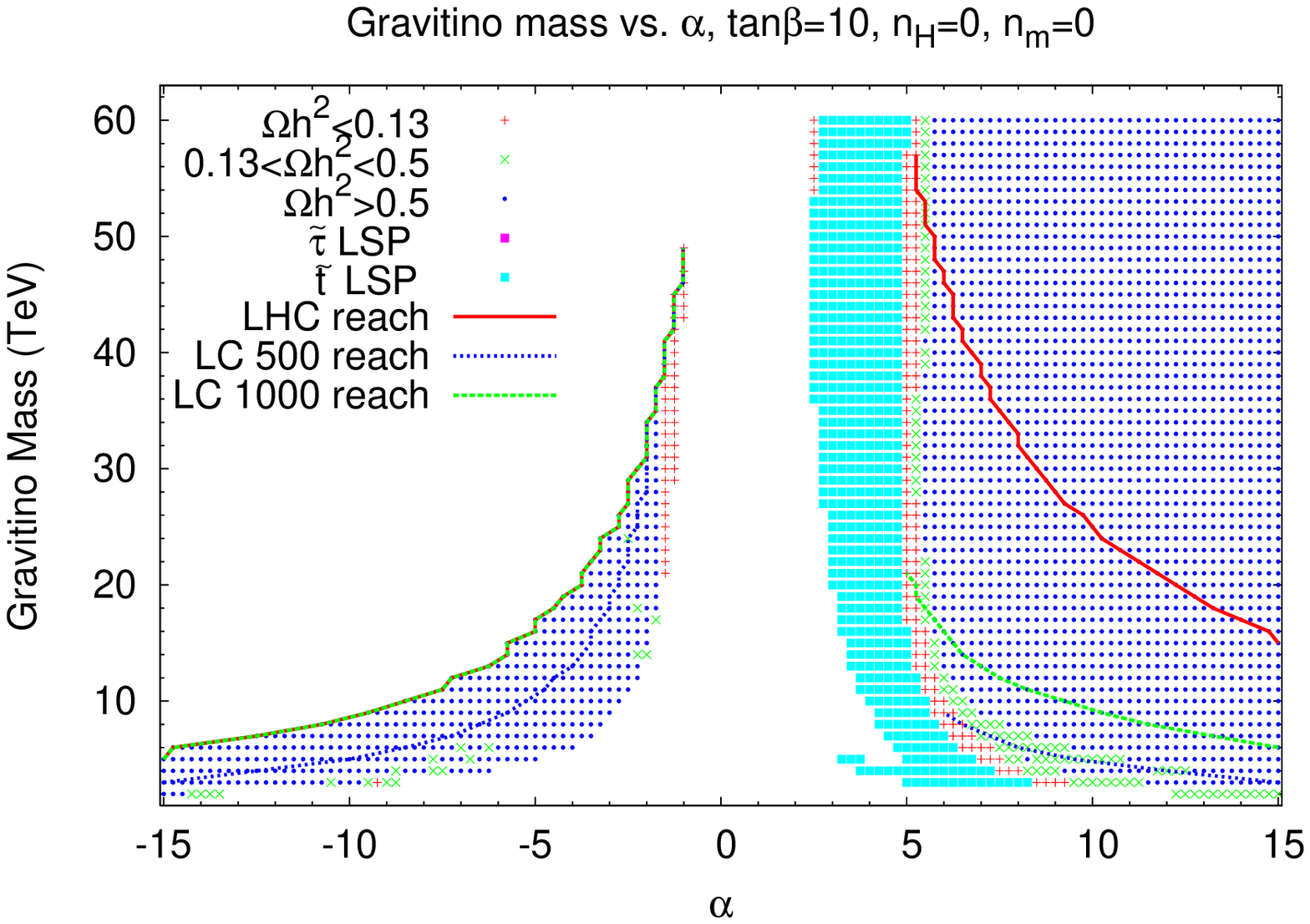,height=5.cm, angle=0}
\epsfig{file=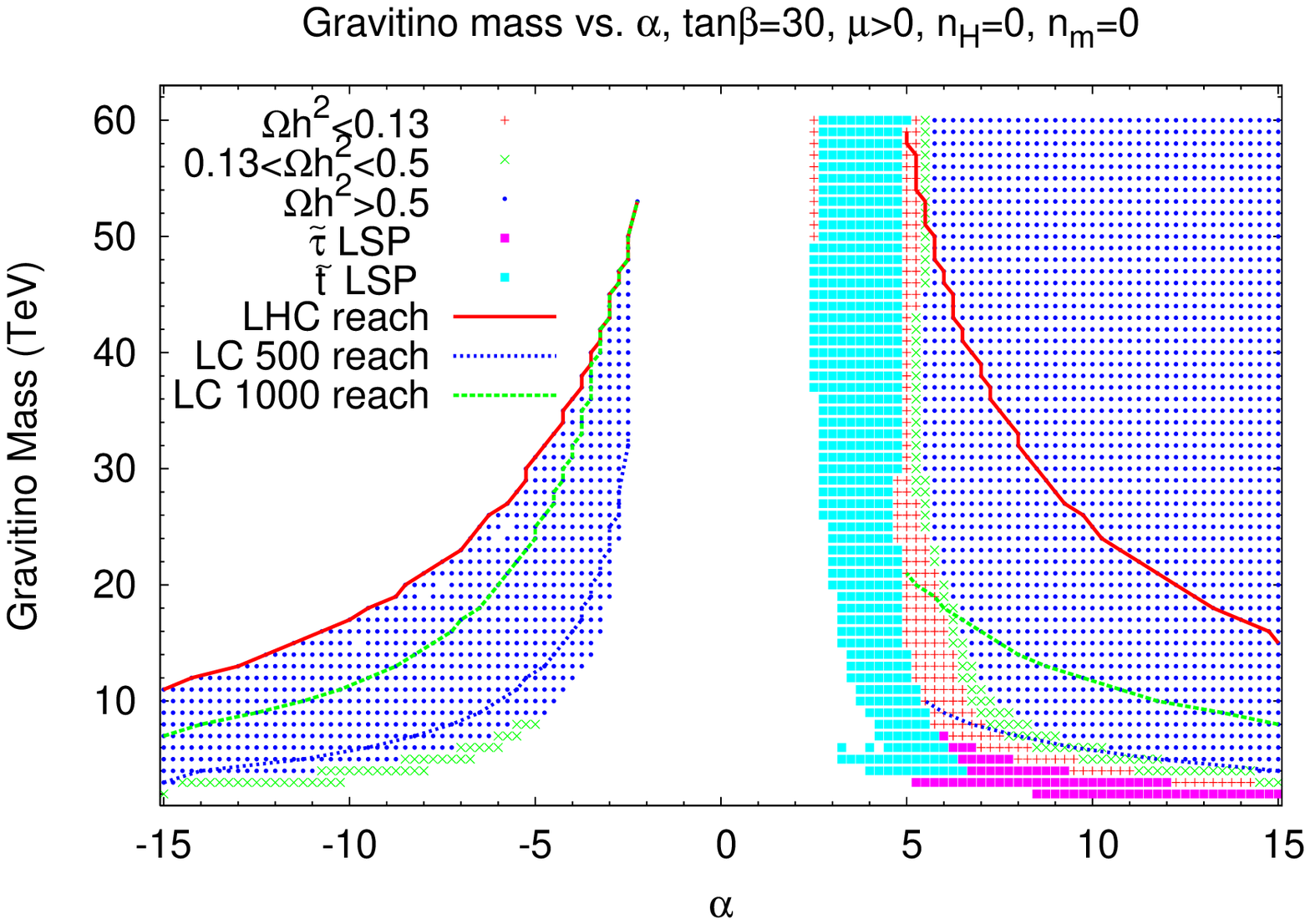,height=5.cm,angle=0} \\
\epsfig{file=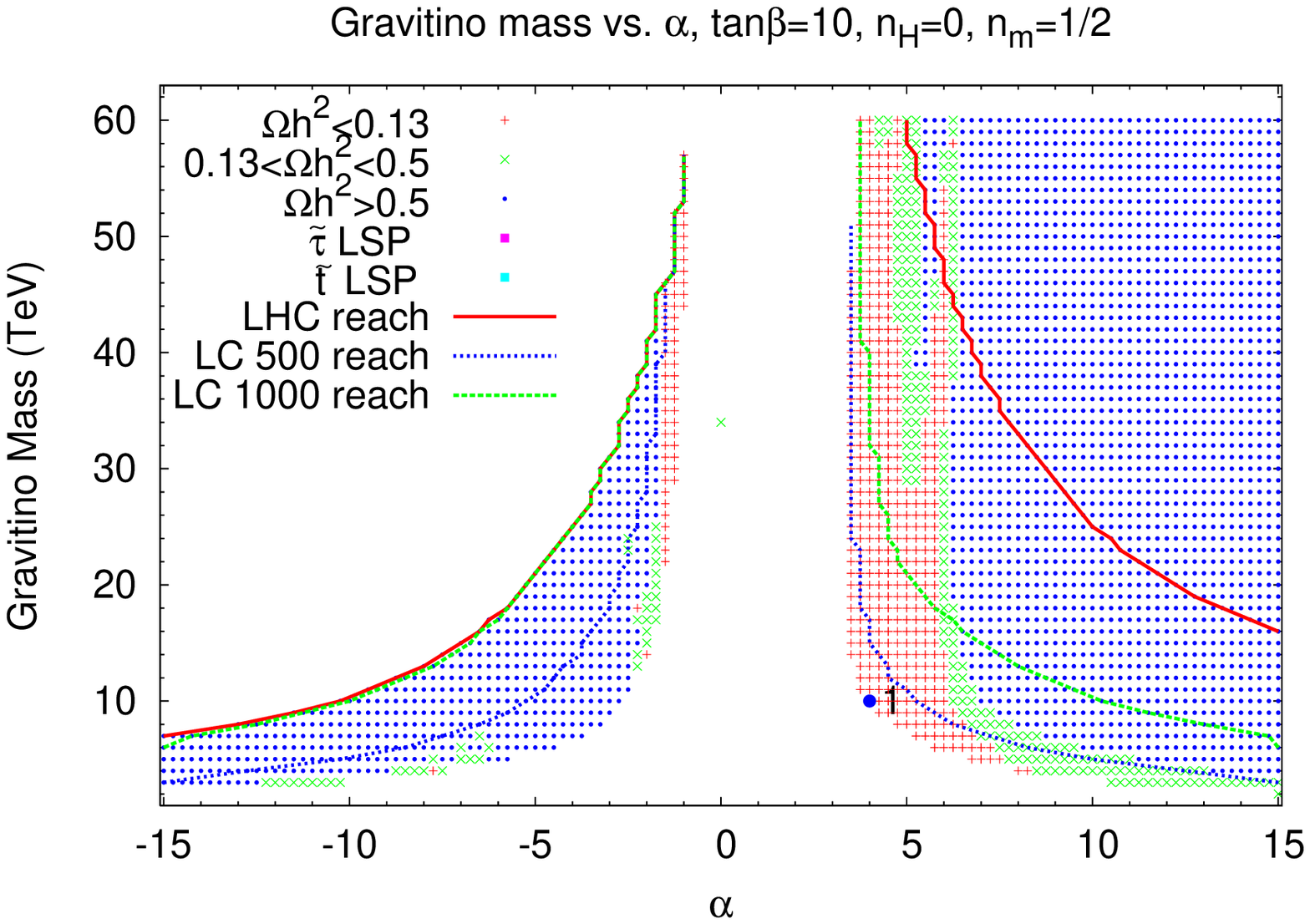,height=5.cm, angle=0}
\epsfig{file=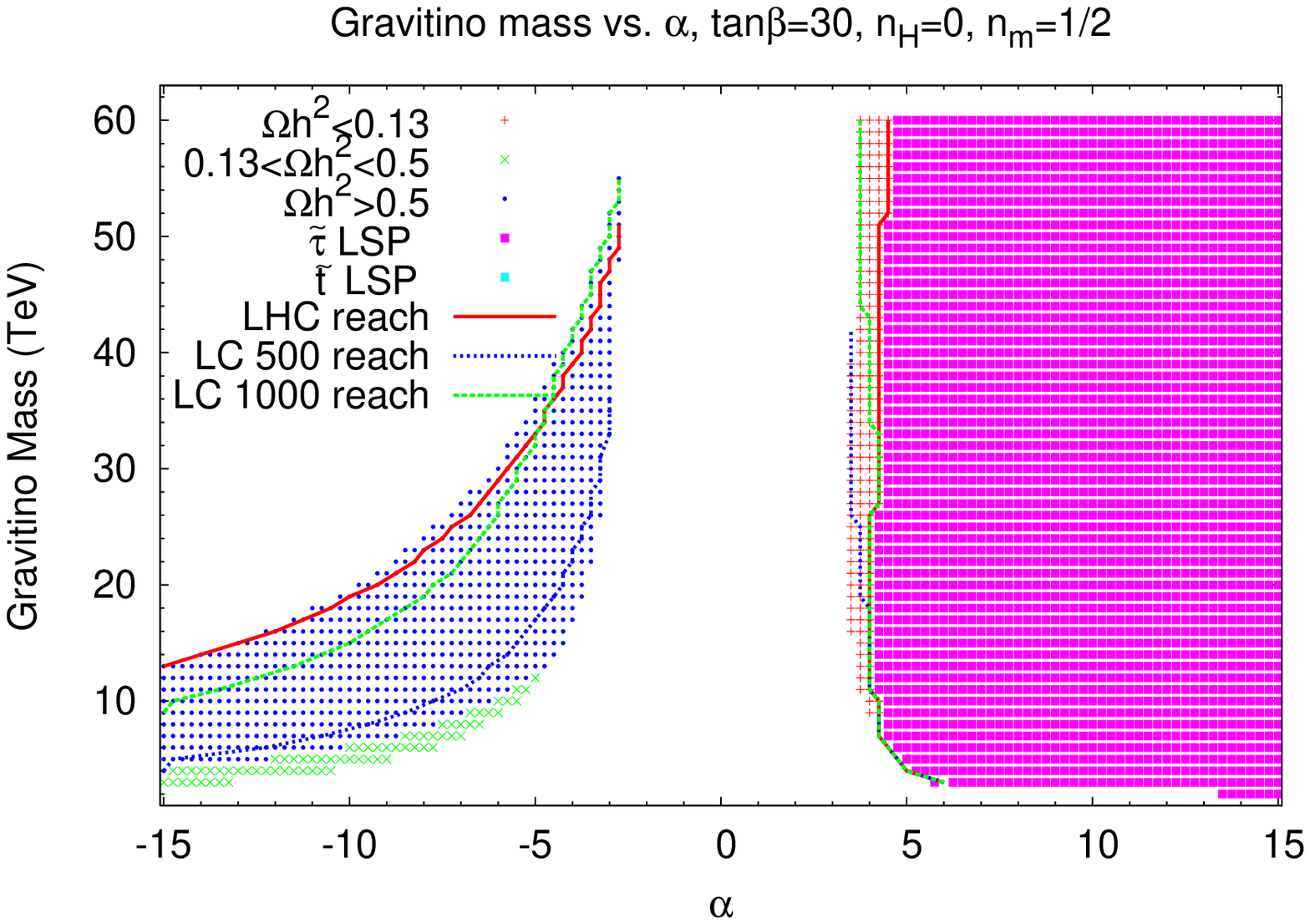,height=5.cm,angle=0} \\
\epsfig{file=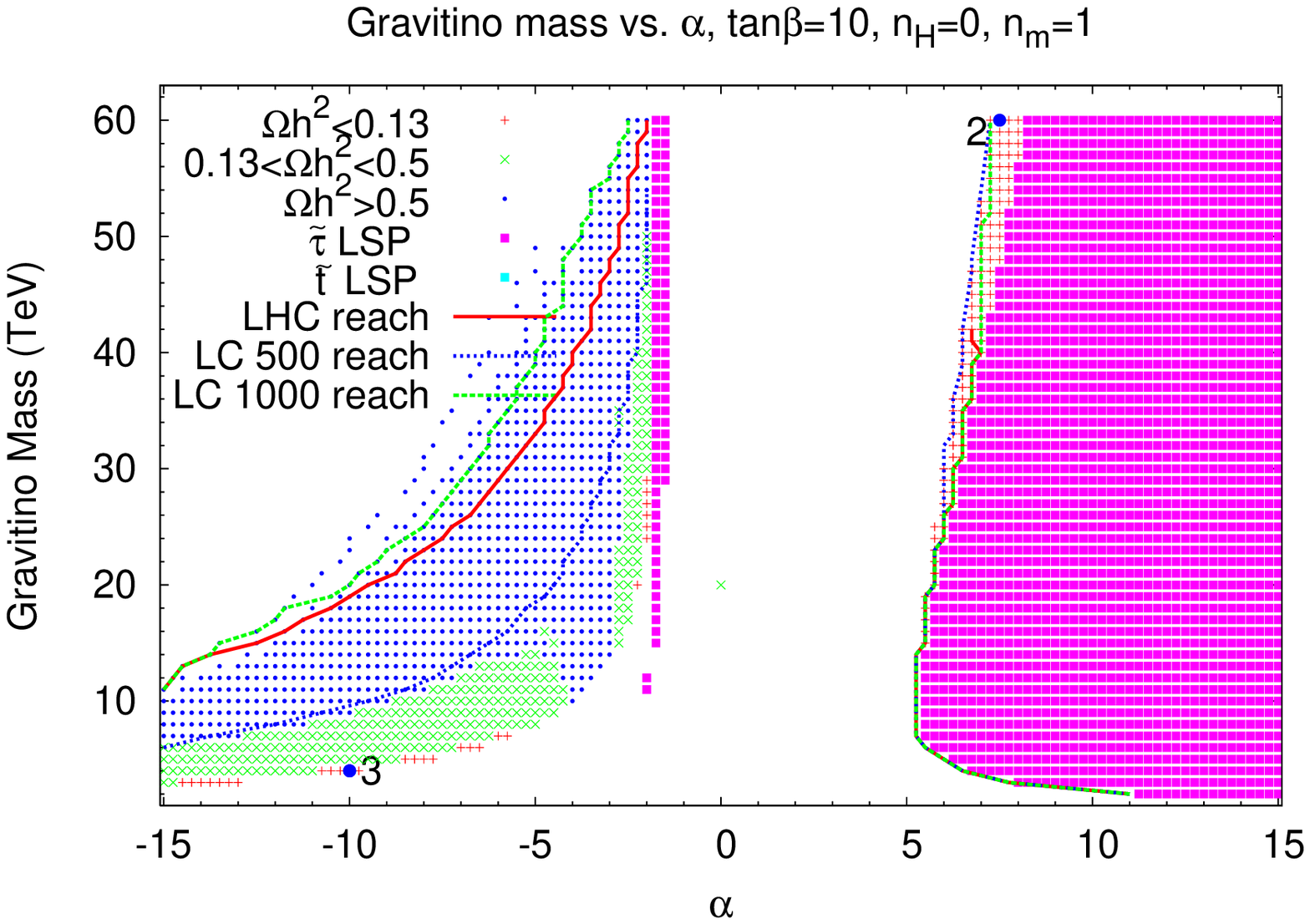,height=5.cm, angle=0}
\epsfig{file=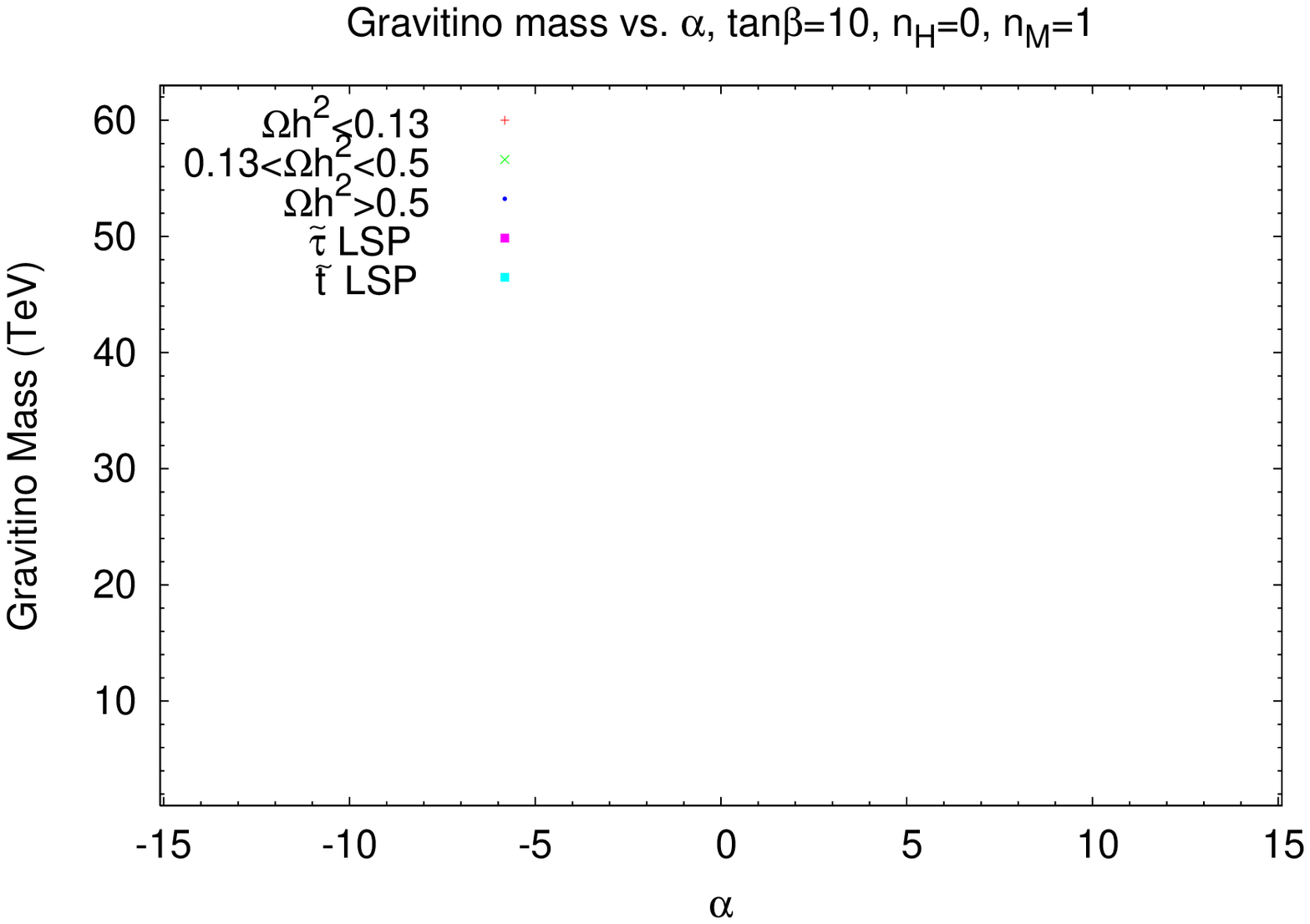,height=5.cm,angle=0}
\caption{\label{fig:psnH0}
Allowed regions of parameter space of the MM-AMSB model
with modular weight $n_H=0$ and {\it a}), {\it b}) $n_m=0$, 
{\it c}), {\it d}) $n_m=1/2$ and {\it e}), {\it f}) $n_m=1$.
Frames {\it a}), {\it c}) and {\it e}) have $\tan\beta =10$
while {\it b}), {\it d}) and {\it f}) have $\tan\beta =30$.
The plots also show the expected thermal neutralino relic density from
the Big Bang, and the reach of
the CERN LHC for 100 fb$^{-1}$ and for a $\sqrt{s}=500$ and 1000 GeV
linear $e^+e^-$ collider.
We take $\mu >0$ and $m_t=175$ GeV.}
}

In frame {\it c}), we see how the allowed parameter space changes if
instead we take matter to live on the $D3-D7$ brane intersection, for
which the modular weight $n_m=1/2$. A value of $n_m=1/2$ acts to {\it
reduce} the magnitude of the $A_i$ parameters in Eq.~(\ref{eq:A}), and to
reduce the moduli contribution to the matter scalar masses in
Eq.~(\ref{eq:m2}). As a result of the reduced value of $A_t$ at $M_{\rm
GUT}$, there is not so strong of an RG push on the top squark soft
masses, and so we are left with somewhat heavier top squarks in the mass
spectrum, and no top squark coannihilation region appears.  Instead, new
allowed regions appear: the vertical band near $\alpha\sim 6$
corresponds to an $A$-funnel annihilation region\cite{Afunnel}, where
$2m_{\tz_1}\simeq m_A$.  The upper part of the $A$-funnel is somewhat
beyond the 100 fb$^{-1}$ LHC reach.  In addition, the upper part of the
vertical band near $\alpha\sim 4$ occurs where the value of $|\mu |$
drops, and the $\tz_1$ becomes mixed higgsino-bino dark matter (MHDM).
This is due in part to the relatively low value of $|M_3|$ at $M_{\rm
GUT}$, which feeds through the RGEs to yield a low $\mu$ value (see 
Ref. \cite{lm3dm1,lm3dm2} for a discussion of the
low $|M_3|$ dark matter model).  The MHDM
has a large annihilation cross section to $W^+W^-$ and $ZZ$, which acts
to reduce the relic density, as in the focus point region of the mSUGRA
model.  The broad allowed region at low $m_{3/2}$ is where $\mu \sim
M_1\sim M_2$, and so here we actually find mixed bino-wino-higgsino dark
matter\cite{winodm}.  In this region, LHC collider events should be rich
in $b$ and $\tau$ jets, since $\tst_1$ is produced in many cascade
decays and $\tst_1\to b\tw_1$, while $\tz_2$ also gets produced,
followed by $\tz_2\to\ttau_1\bar{\tau}+\overline{\ttau}_1\tau$. 
A sample point with a mixed
bino-wino-higgsino LSP is shown as Point 1 in Table 1.  This point would
likely be excluded due to $BF(b\to s\gamma )$ constraints (see Sec. 3)
and possibly by direct dark matter search constraints (see Sec. 5).  The
BWCA region at $\alpha\sim -2$ is also present in this frame.

%
\begin{table}
\begin{tabular}{lcccc}
\hline
parameter & Point~1 & Point~2 & Point~3 & Point~4 \\
\hline
$(n_H,n_m)$        & $(0,{1\over 2})$ & $(0,1)$ & $(0,1)$ & $(0,1)$  \\
$\alpha$        & 4 & 7.5 & -10 & -10  \\
$m_{3/2} (TeV)$ & 10.0 & 60.0 & 4.0 & 4.0 \\
$\tan\beta$     & 10 & 10 & 10 & 10 \\
$\mu$       & 161.9 & 490.0 & 416.4 & -416.3 \\
$m_{\tg}$   & 393.6 & 4699.1 & 701.7 & 701.0 \\
$m_{\tu_L}$ & 354.2 & 3979.7 & 652.9 & 652.5 \\
$m_{\tst_1}$& 127.7 & 2631.5 & 436.8 & 420.8 \\
$m_{\tb_1}$ & 312.7 & 3531.2 & 582.4 & 584.0 \\
$m_{\te_L}$ & 210.7 & 1493.0 & 198.1 & 198.0 \\
$m_{\te_R}$ & 191.0 & 837.5 & 108.9 & 108.8 \\
$m_{\ttau_1}$ & 183.0 & 800.7 & 97.6 & 92.3 \\
$m_{\tw_1}$ & 131.6 & 513.9 & 185.8 & 178.1 \\
$m_{\tz_2}$ & 169.5 & 502.8 & 185.5 & 178.1 \\ 
$m_{\tz_1}$ & 118.5 & 499.3 & 66.8 & 64.9  \\ 
$m_A$       & 305.5 & 3236.7 & 510.1 & 515.5 \\
$m_h$       & 110.7 & 124.2 & 111.4 & 113.0 \\
$\Omega_{\tz_1}h^2$& 0.001 & 0.04 & 0.11 & 0.11 \\
$BF(b\to s\gamma)$ & $2.9\times 10^{-5}$ & $3.3\times 10^{-4}$ & 
$5.4\times 10^{-4}$ & $2.9\times 10^{-4}$ \\
$\Delta a_\mu    $ & $34.0\times 10^{-10}$ & $0.5\times 10^{-10}$ & 
$-39.4 \times  10^{-10}$ & $41.6 \times  10^{-10}$ \\ 
$BF(B_s \to \mu^+\mu^-)$ & $4.4\times 10^{-9}$ & $3.8\times 10^{-9}$ & 
$3.7\times 10^{-9}$ & $4.1\times 10^{-9}$ \\
$\sigma_{sc} (\tz_1p )$ & $3.1\times 10^{-7}\ {\rm pb}$ 
& $3.9\times 10^{-10}\ {\rm pb}$ & $1.9\times 10^{-10}\ {\rm pb}$ & 
$8.9\times 10^{-10}\ {\rm pb}$  \\
\hline
\end{tabular}
\caption{Masses and parameters in~GeV units for four cases of the
MM-AMSB model.
Also shown are predictions for low energy observables, together with
the cross section for direct detection of dark matter. 
In all cases, we take $m_t=175$ GeV. 
}
\label{tab:cases}
\end{table}

In frame {\it d}), with the same modular weights as {\it c}) but with
$\tan\beta =30$, the overall picture is similar to {\it c}) for
$0<\alpha\alt 5$, but the Higgs funnel disappears into the stau LSP
forbidden region. The entire WMAP allowed region is accessible to the
LHC.  For $\alpha <0$, the BWCA region present for the $\tan\beta=10$
case  disappears, leaving no WMAP-allowed region. 


In frame {\it e}), we show results for modular weights $n_H=0$, $n_m=1$
with $\tan\beta =10$. Since $n_m=1$, in Eq.~(\ref{eq:m2}) the leading
contribution proportional to $\alpha^2$ for matter scalars is absent, and so
squark and slepton masses are suppressed, and usually we find tachyonic or
charged LSPs in parameter space. A narrow region around $\alpha\sim 5-8$
survives where $m_{\tz_1}\sim m_{\ttau_1}$, while the $\tz_1$ is a nearly pure
higgsino state. This region of parameter space has 
tachyonic slepton masses at the GUT scale. 
The upper portion of this region is in accord with WMAP/LEP2 constraints, 
but is beyond the reach of both the LHC and ILC, but may be accessible
to ton-size dark matter detectors. A sample point
for $\alpha =7.5$ and $m_{3/2}=60$ TeV is shown as Point 2 in Table 1.
There are also some faint regions that survive for negative $\alpha$ values;
the WMAP/LEP2-allowed portion of these has sleptons of mass $\sim 100$ GeV, 
which leads to so-called bulk annihilation of neutralinos through light
slepton exchange. 
A sample bulk annihilation point is listed as Point 3 in Table~1. 
While this point is WMAP/LEP2 allowed, it has a large {\it negative}
contribution to $(g-2)_\mu$ (see Sec. 3), and is likely excluded.
By flipping the sign of $\mu$, the $(g-2)_\mu$ contribution flips sign,
and gives a large positive contribution: see Point 4 of Table 1.
In frame {\it f}), no parameter space points lead to a viable SUSY spectrum.
This is due to a combination of tachyonic GUT scale soft masses plus 
the downward push of a large $\tau$ Yukawa coupling, such that 
weak scale tachyonic masses are produced all over parameter space.

\subsection{${\bf n_H=1/2}$ cases}

In Fig. \ref{fig:psnHh}, we present an overview of the MM-AMSB parameter
space for $n_H=1/2$ and various $n_m$ possibilities, for $\tan\beta =10$
and 30, with $\mu >0$. Frame {\it a}) shows the $\alpha\ vs.\ m_{3/2}$
plane for $n_m=0$ and $\tan\beta =10$. The shading and labelling is as
in Fig. \ref{fig:psnH0}. We see once again that a region has appeared
where the $\tst_1$ is the LSP, although this region is smaller than in
Fig. \ref{fig:psnH0}{\it a}). The reason again is that the $A_t$
parameter is quite large (reduced only by the $n_H=1/2$ factor in
Eq.~(\ref{eq:A})) at the GUT scale, and again this serves to reduce the
top squark soft masses via RG evolution. The red-shaded region to the
right of the turquoise-forbidden region again occurs due to top-squark
co-annihilation effects. Along the region of high $m_{3/2}$ and $\alpha
\sim 4$, $\tg\tg$ will typically be followed by $\tg\to t\tst_1$, with
$\tst_1\to c\tz_1$, so that gluino pair events at the LHC will be
characterized by the presence of a pair of top quarks plus two soft
charm jets and additional $\eslt$.  The region at $\alpha\sim 2.5$ is
again a region with a $\tz_1$ that is a mixed bino-wino-higgsino state.
There also appears an $A$-annihilation funnel at $\alpha\sim 5$, and a
small BWCA region at $\alpha\sim -2$. The picture for $\tan\beta =30$ is
illustrated in frame {\it b}), where the tau slepton becomes lighter due
to the effect of the large tau-lepton Yukawa coupling. In this case, the
base of the $A$-funnel gets wider due to the addition of stau-neutralino
co-annihilation. In addition, a region of bulk annihilation through
light staus opens up at negative $\alpha$.
\FIGURE[htb]{
\epsfig{file=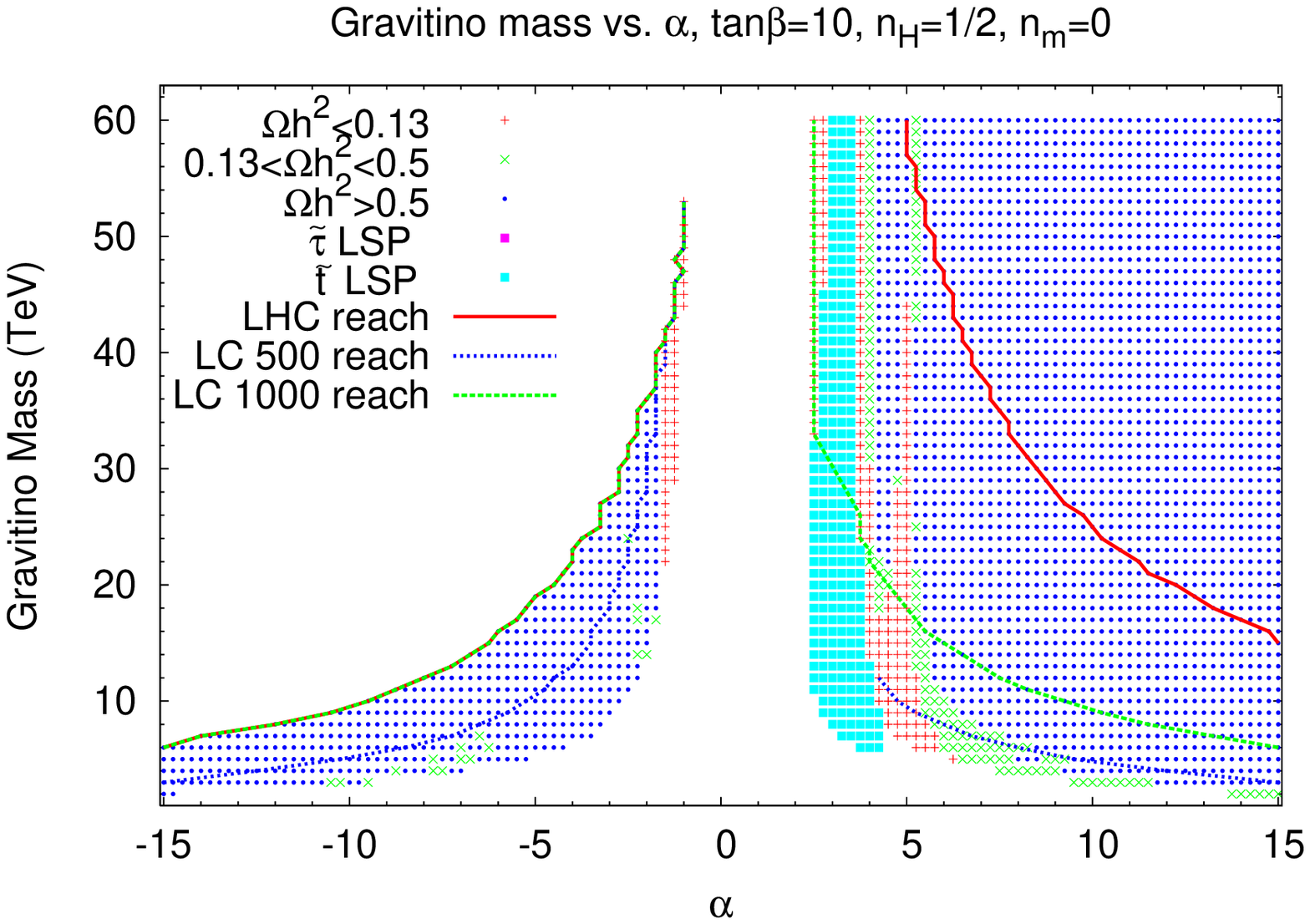,height=5.cm, angle=0}
\epsfig{file=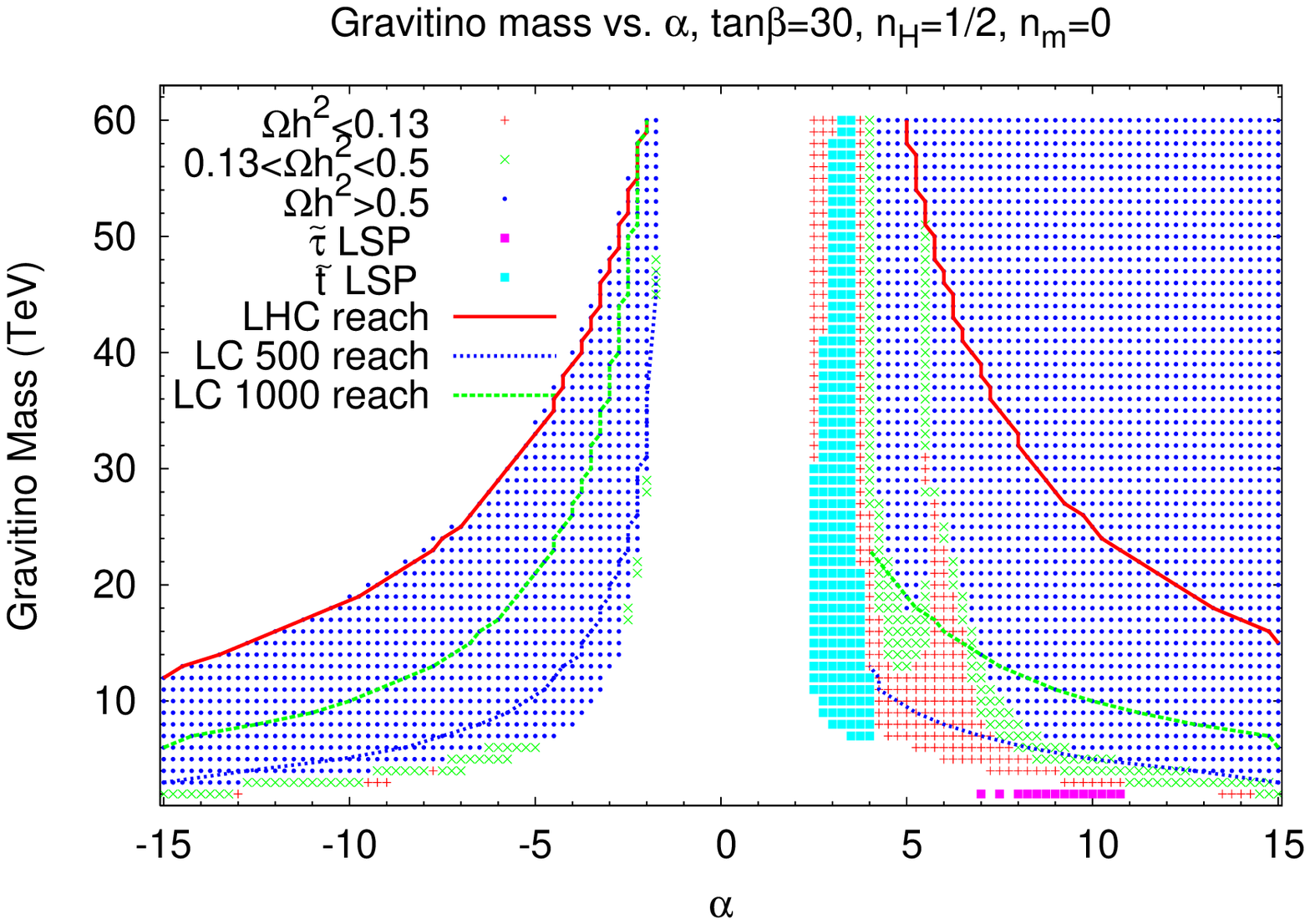,height=5.cm,angle=0} \\
\epsfig{file=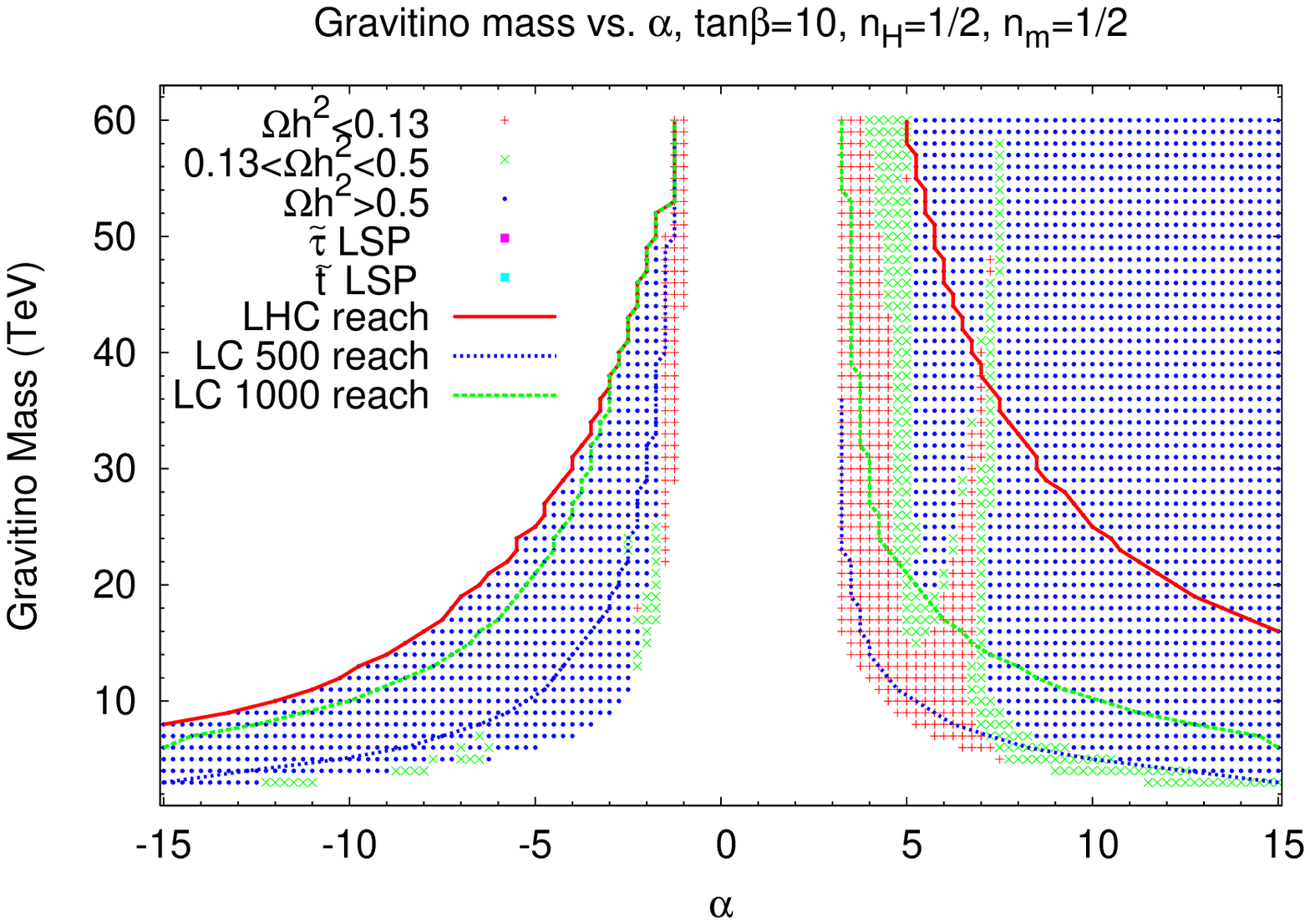,height=5.cm, angle=0}
\epsfig{file=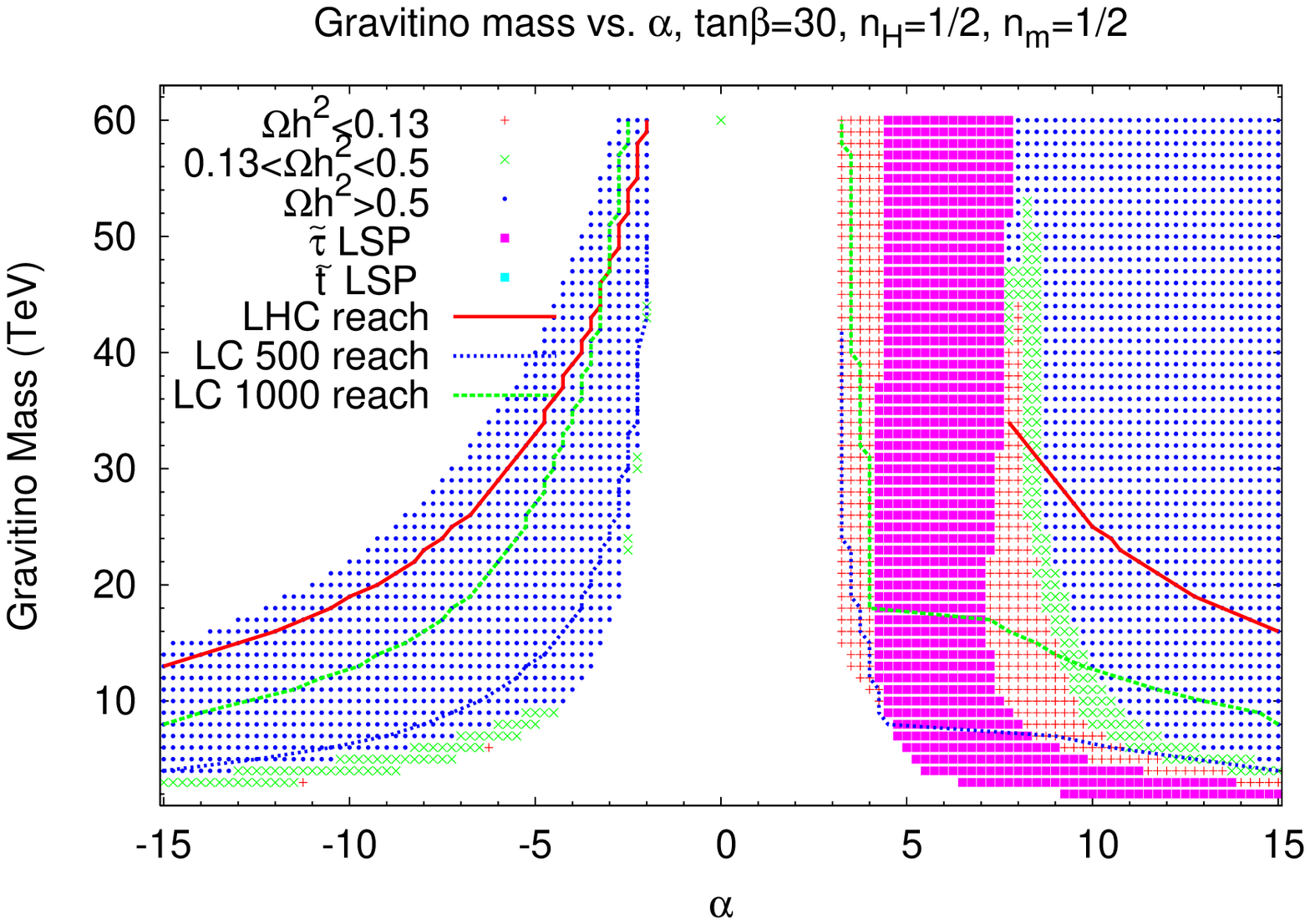,height=5.cm,angle=0} \\
\epsfig{file=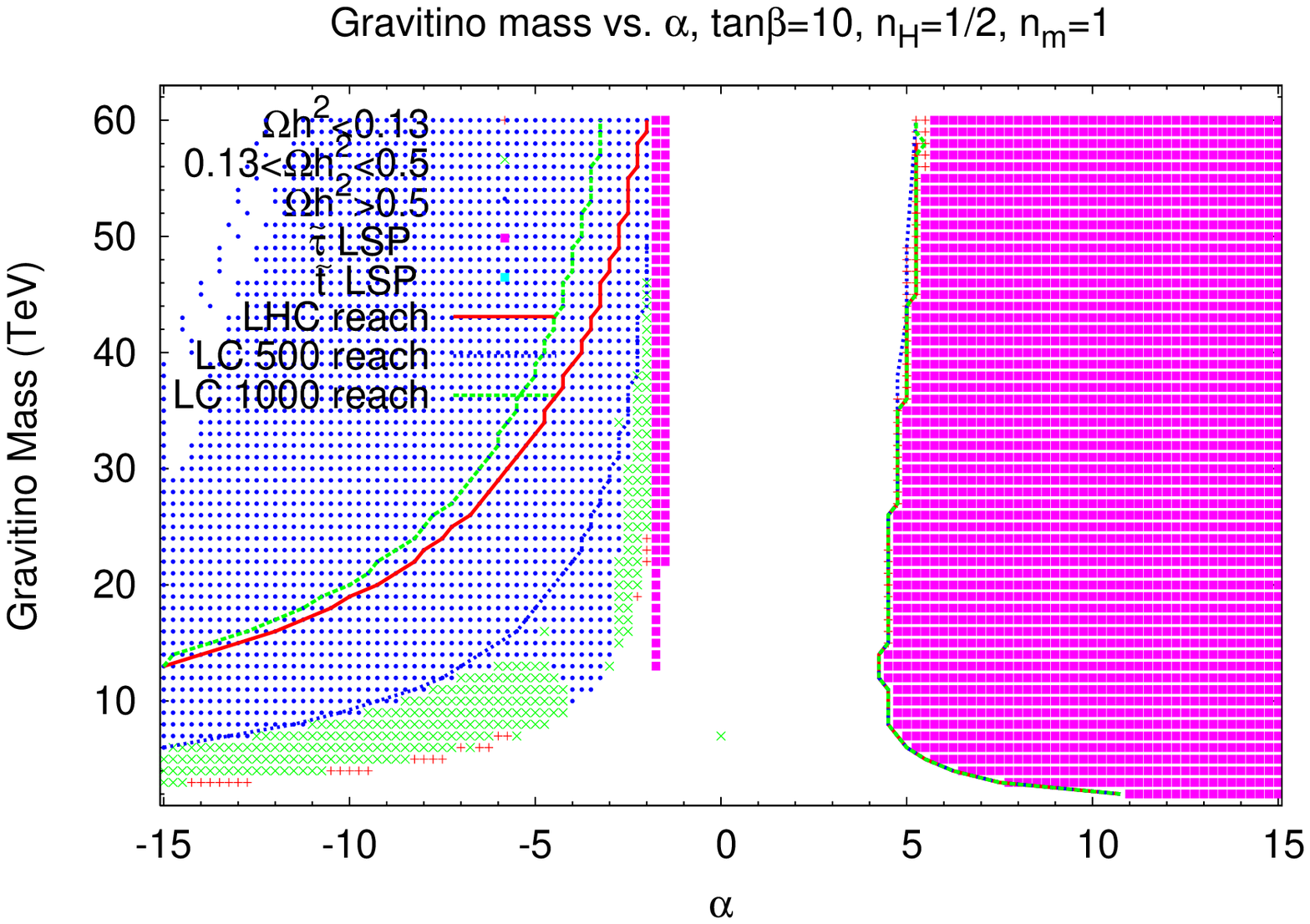,height=5.cm, angle=0}
\epsfig{file=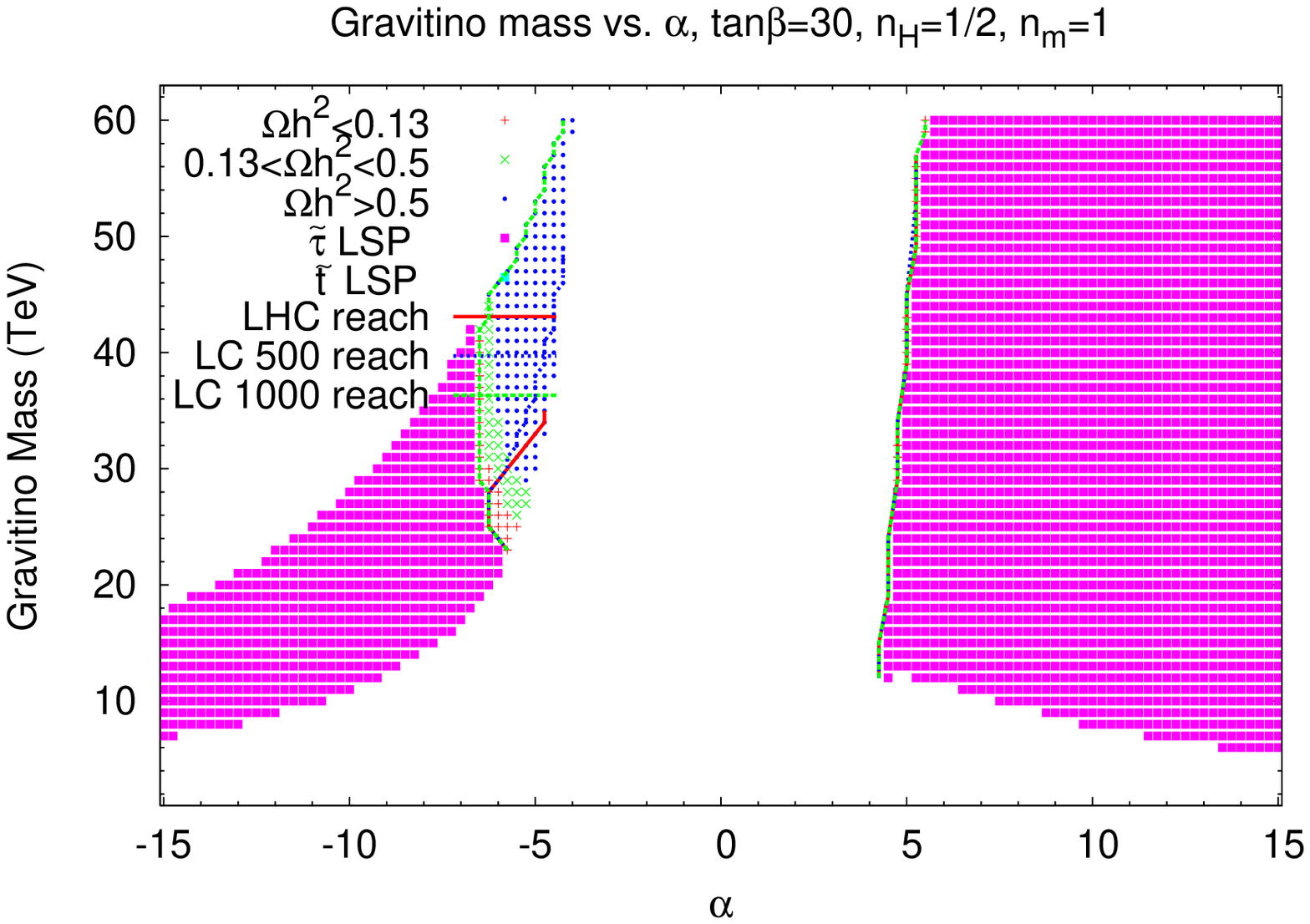,height=5.cm,angle=0}
\caption{\label{fig:psnHh}
 Allowed regions of parameter space of the MM-AMSB model
with modular weight $n_H=1/2$ and {\it a}), {\it b}) $n_m=0$, 
{\it c}), {\it d}) $n_m=1/2$ and {\it e}), {\it f}) $n_m=1$.
Frames {\it a}), {\it c}) and {\it e}) have $\tan\beta =10$
while {\it b}), {\it d}) and {\it f}) have $\tan\beta =30$.
The plots also show the expected thermal neutralino relic density from
the Big Bang, and the reach of
the CERN LHC for 100 fb$^{-1}$ and for a $\sqrt{s}=500$ and 1000 GeV
linear $e^+e^-$ collider.
We take $\mu >0$ and $m_t=175$ GeV.}}

In frame {\it c}), we see results for $n_m=1/2$ and $\tan\beta =10$. The
stop-LSP forbidden region has disappeared owing to the increased matter
modular weights.  Now there exists a broad region around $\alpha\sim
3-4$ with a mixed higgsino-bino $\tz_1$, while the low $m_{3/2}$ part
includes an LSP with a significant wino component. The $A$-annihilation
funnel is present at $\alpha\sim 6-7$, and a BWCA region occurs at
$\alpha\sim -2$.  If we shift to large $\tan\beta =30$ in frame {\it
d}), then a large excluded stau-LSP region occurs. To the right of this
region occurs an area of mixed $A$-funnel/stau co-annihilation, while
the region to the left is characterized by a higgsino-LSP with some
degree of stau co-annihilation.  We note here that some part of the
$A$-annihilation funnel extends out beyond the reach of LHC with 100
fb$^{-1}$; in this region, $m_{\tg}\sim m_{\tq}\sim 3.5$ TeV.  Any
WMAP-allowed regions at negative $\alpha$ have disappeared.

For $n_m=1$ and $\tan\beta =10$, portrayed in frame {\it e}), almost all 
allowed regions have disappeared. 
There are a few allowed points to the left of the excluded region 
at positive $\alpha$, plus a small region of
bulk annihilation via light sleptons with mass $\sim 100$ GeV 
at $\alpha <-5$. If we increase 
$\tan\beta$ to 30 as in frame {\it f}), then only a few points persist 
at positive $\alpha$, while a small band around $\alpha\sim -6$ exists
for negative $\alpha$ which is characterized by a bino-like LSP and
stau co-annihilation.

\subsection{${\bf n_H=1}$ cases}

In Fig. \ref{fig:psnH1}, we show the parameter space regions for
Higgs modular weight $n_H=1$, and various matter modular weights
with $\tan\beta =10$ and 30. In frame {\it a}), with $n_m=0$ and 
$\tan\beta =10$, it is noteworthy that the stop-LSP region which is present in
Figs. \ref{fig:psnH0}{\it a}) and \ref{fig:psnHh}{\it a}) has disappeared,
even though the matter modular weight remains the same. In this case,
just as in the $n_m=1/2, \ n_H=0$ case in Fig.~\ref{fig:psnH0}{\it c}, the 
large Higgs modular weight reduces the value of $A_t(M_{GUT})$ enough so that
the top squark soft masses are not driven to such low values. 
The large Higgs modular weight also reduces the GUT scale Higgs mass, which
also reduces the weak scale $A$ mass\cite{nuhm1}. 
Thus, the $A$ funnel region moves out to
somewhat higher values of $\alpha$ than found in Fig. \ref{fig:psnHh}{\it a}).
There also remains a large region of higgsino and mixed higgsino-wino-bino
dark matter around $\alpha\sim 3$. The BWCA region maintains a presence 
at small, negative $\alpha$ values. The CERN LHC can cover all the 
parameter space shown. Moving to $\tan\beta =30$ in frame {\it b}), 
we find the base of the $A$-funnel widened somewhat compared to frame {\it a})
due to the presence of stau co-annihilation. In addition, a region has
appeared at low $m_{3/2}$ and negative $\alpha$ where neutralino bulk 
annihilation through light staus can occur. 
\FIGURE[htb]{
\epsfig{file=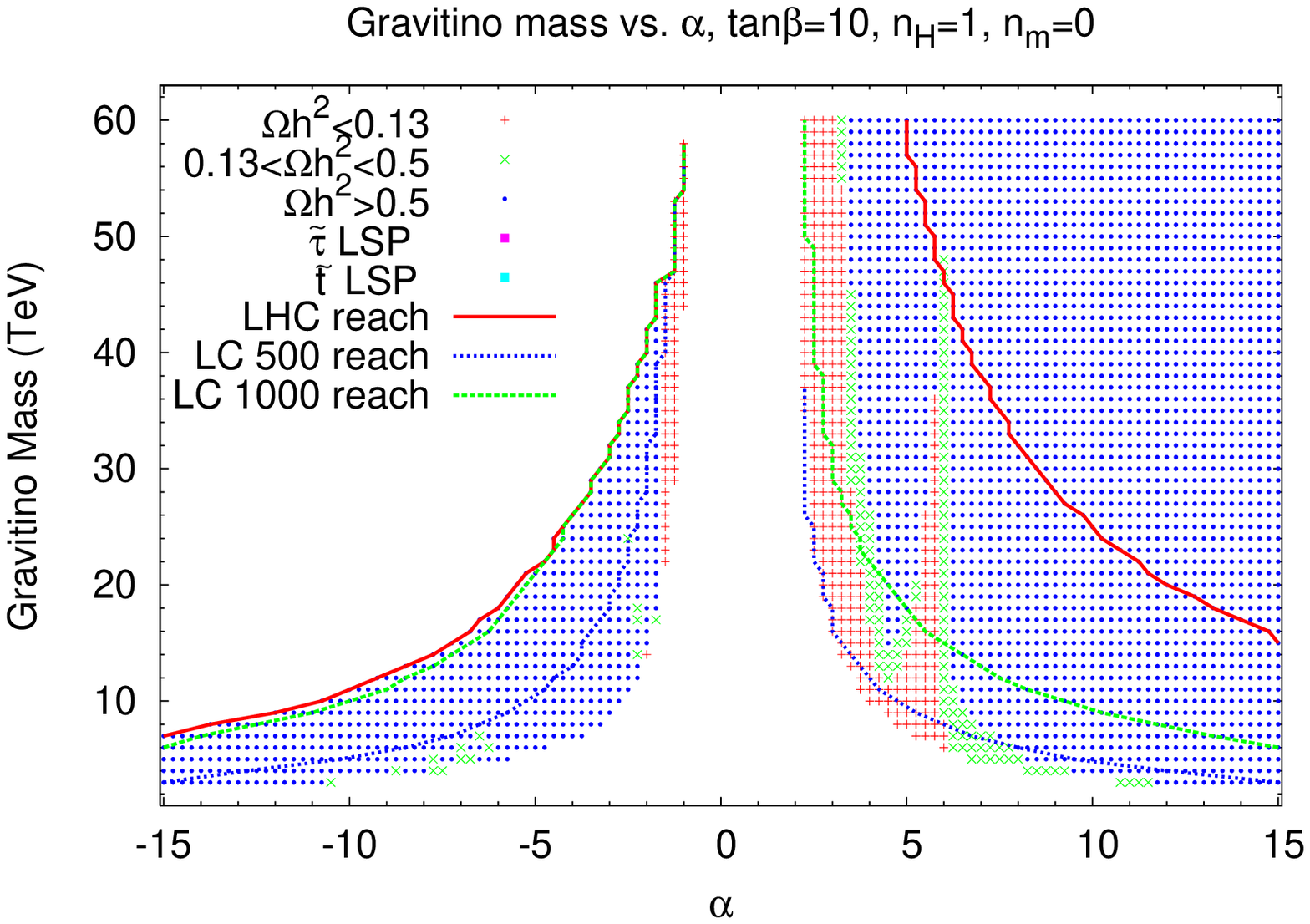,height=5.cm, angle=0}
\epsfig{file=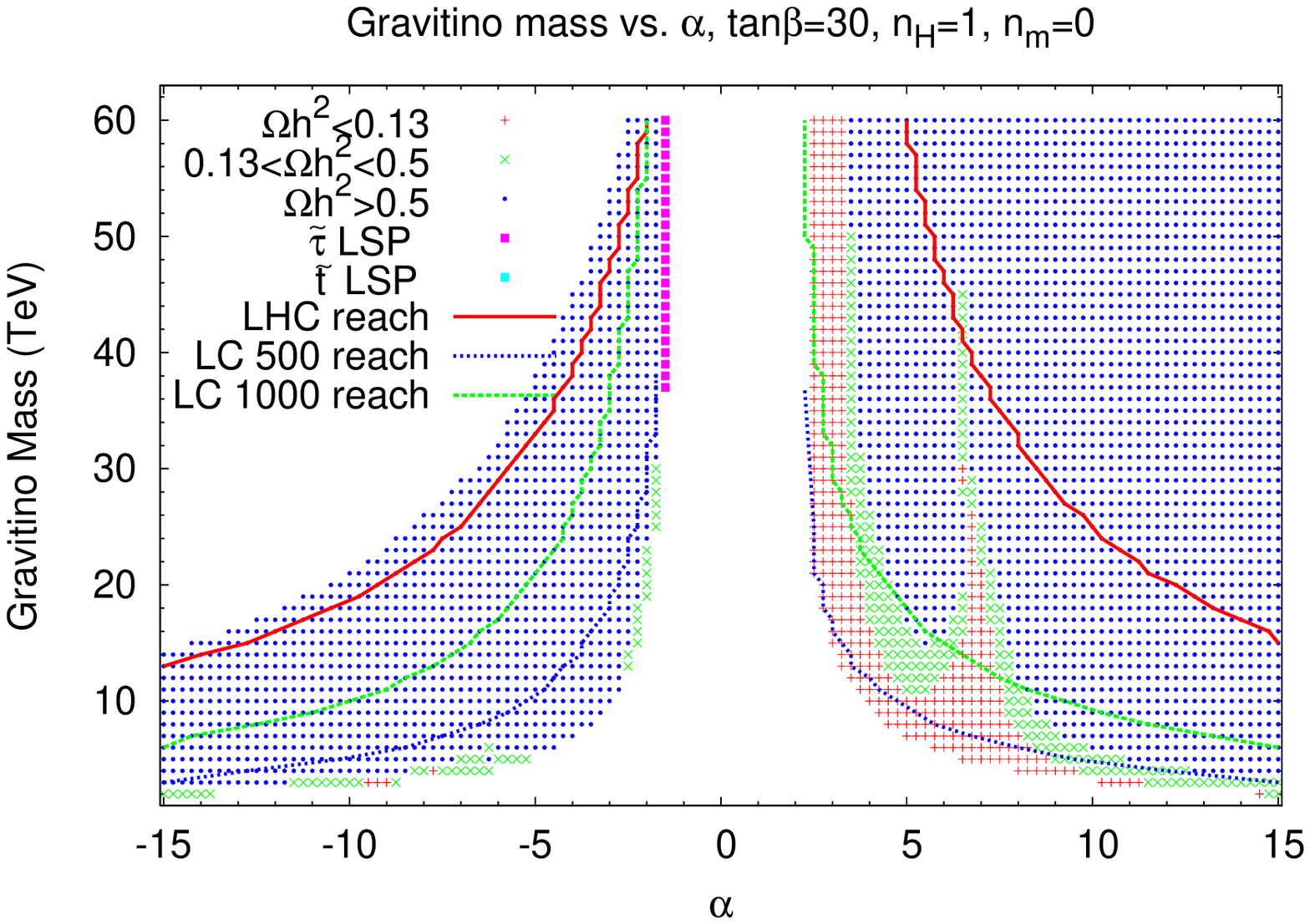,height=5.cm,angle=0} \\
\epsfig{file=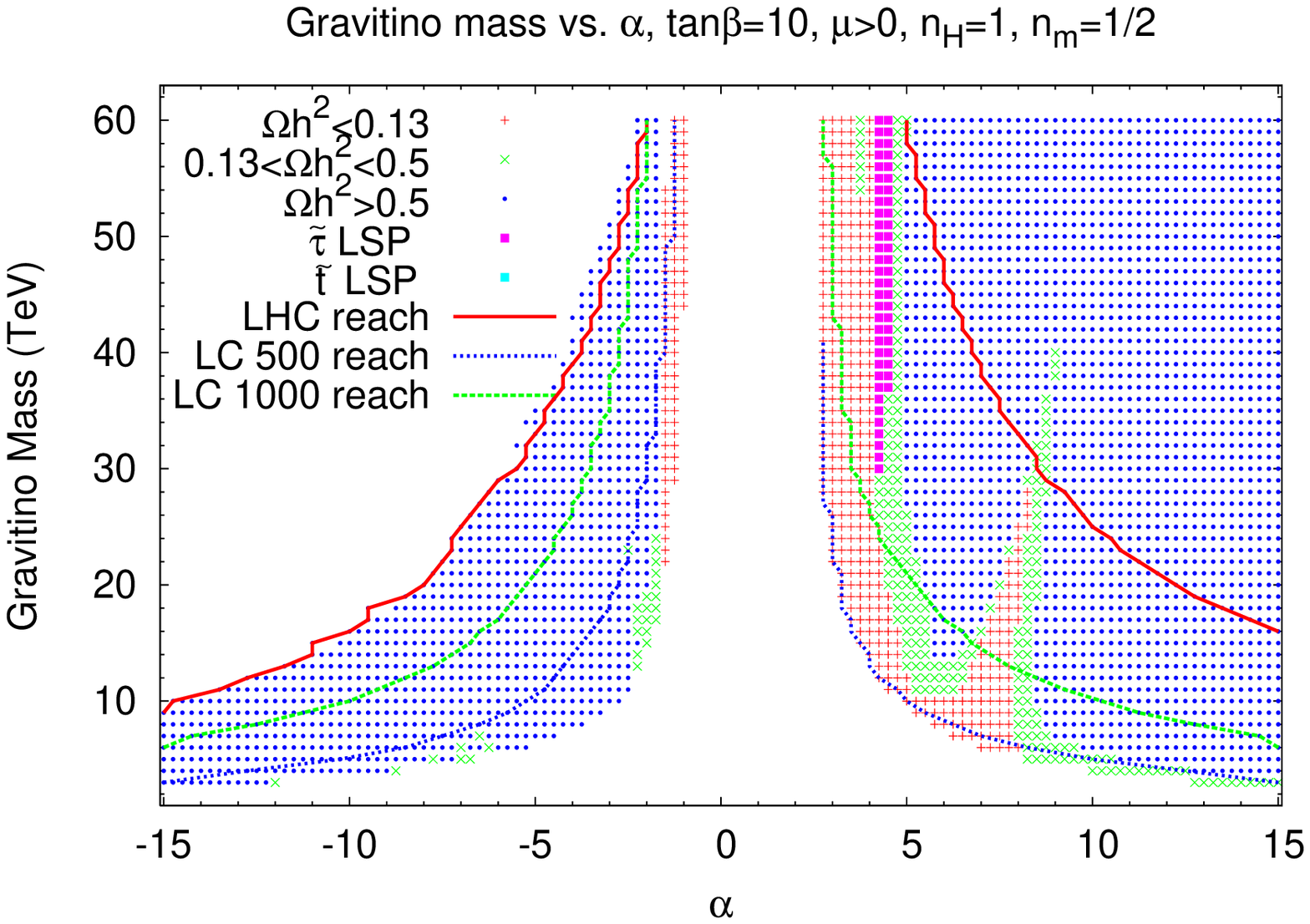,height=5.cm, angle=0}
\epsfig{file=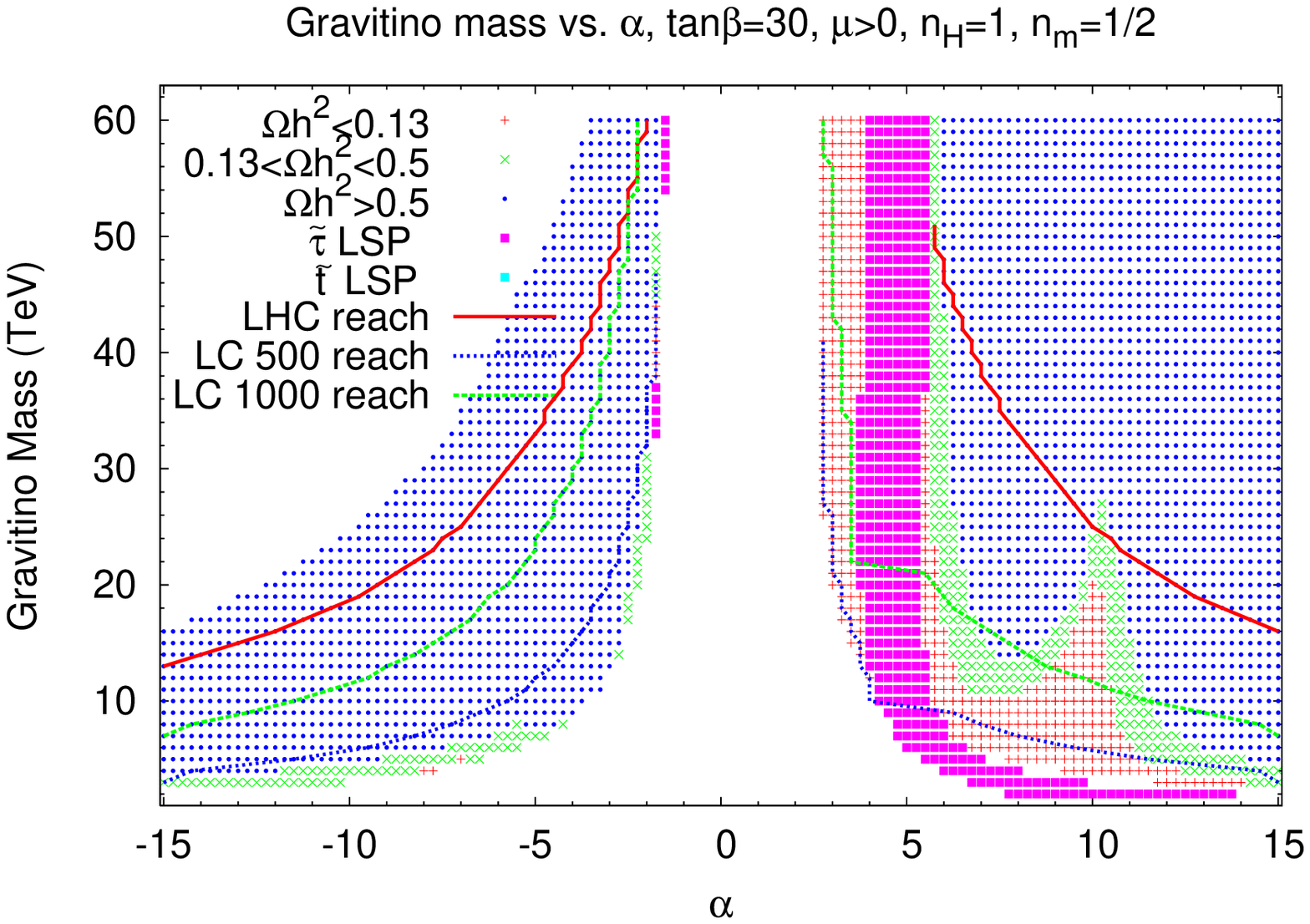,height=5.cm,angle=0} \\
\epsfig{file=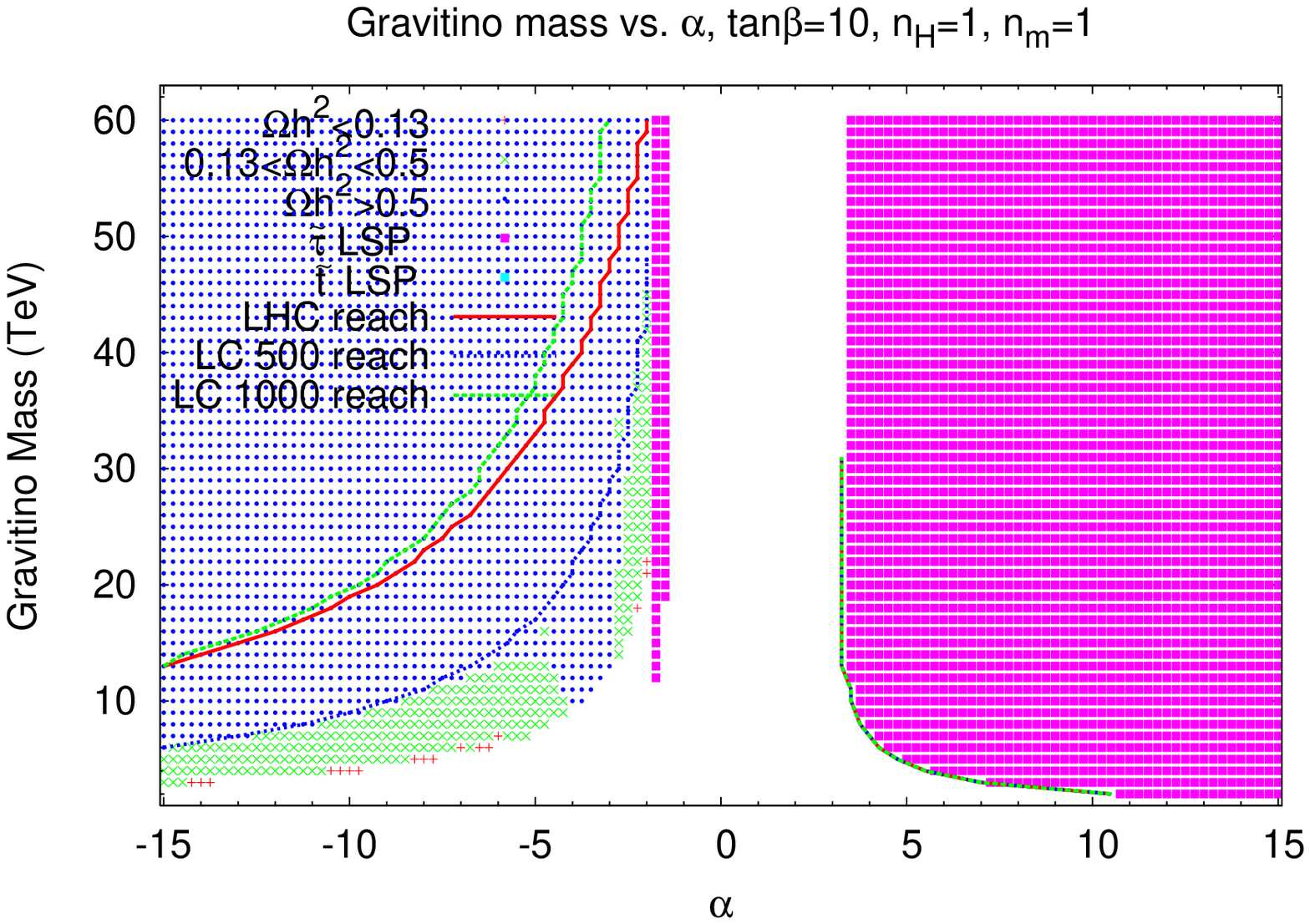,height=5.cm, angle=0}
\epsfig{file=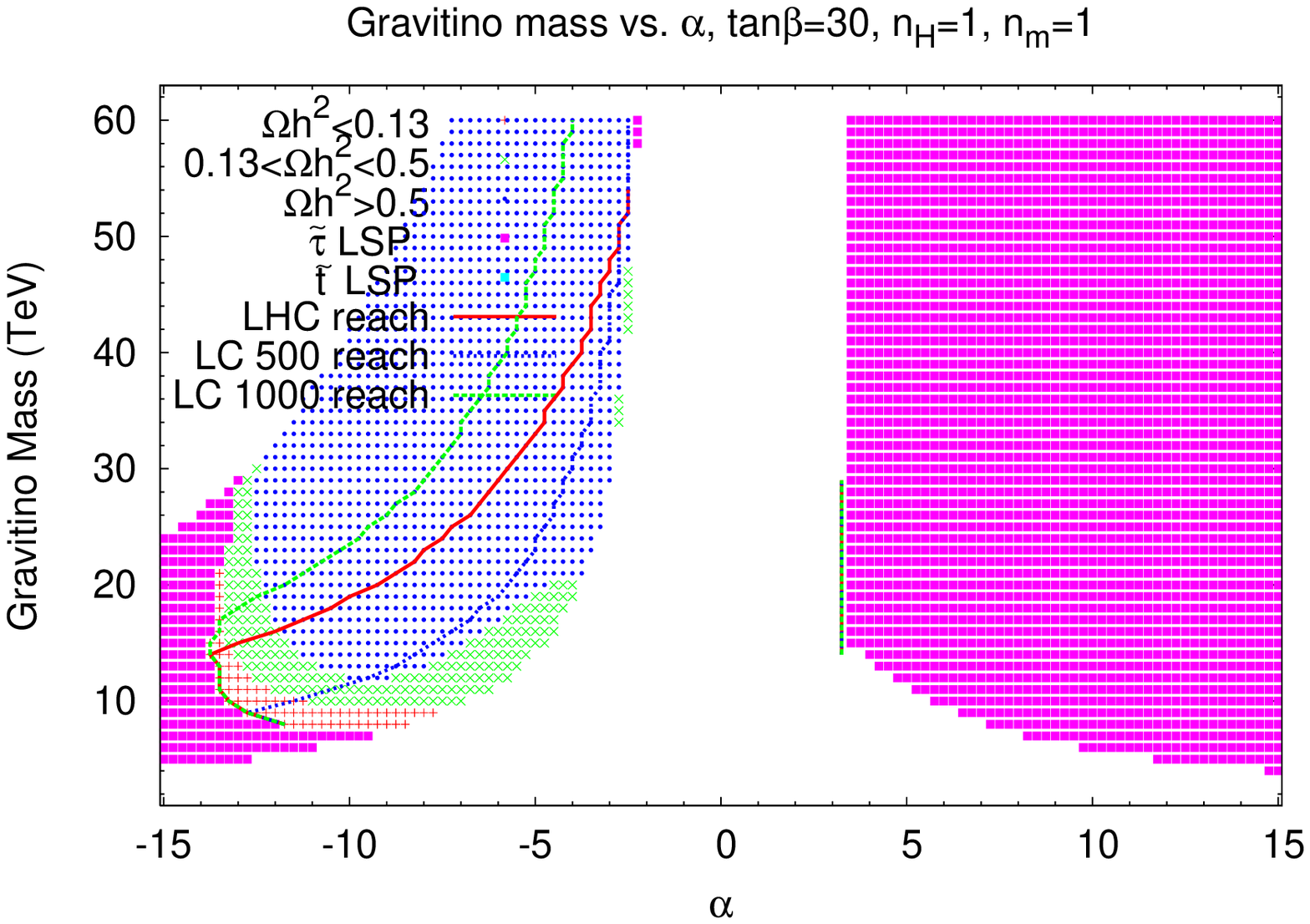,height=5.cm,angle=0}
\caption{\label{fig:psnH1}
Allowed regions of parameter space of the MM-AMSB model
with modular weight $n_H=1$ and {\it a}), {\it b}) $n_m=0$, 
{\it c}), {\it d}) $n_m=1/2$ and {\it e}), {\it f}) $n_m=1$.
Frames {\it a}), {\it c}) and {\it e}) have $\tan\beta =10$
while {\it b}), {\it d}) and {\it f}) have $\tan\beta =30$.
The plots also indicate the neutralino relic density, and the reach of
the CERN LHC for 100 fb$^{-1}$ and a $\sqrt{s}=500$ and 1000 GeV
linear $e^+e^-$ collider.
We also take $\mu >0$ and $m_t=175$ GeV.}}

In frame {\it c}), we plot parameter space for $n_H=1$, $n_m=1/2$, the same
choices adopted in the non-zero modular weight analysis of Ref. \cite{kklt1}.
Here, we have an $A$-annihilation region at large $\alpha$, a higgsino
region at small positive $\alpha$, and a BWCA region at small negative 
$\alpha$. In this case, the LHC covers the entire $A$-funnel, unlike
Figs. \ref{fig:psnH0}{\it c}) and \ref{fig:psnHh}{\it c}). 
In frame {\it d}) for $\tan\beta =30$, a large stau-LSP excluded region
has appeared. In addition, the $A$-funnel has moved to very large
$\alpha$ values, and is very broad, owing to an overlap with the stau
co-annihilation region.

Moving to $n_H=1$ and $n_m=1$ portrayed in frame {\it e}) for $\tan\beta =10$,
we see that all the $\alpha >0$ region is now excluded due to a stau LSP.
In this case, the moduli contribution to scalar masses only comes from 
the mixing term in Eq. \ref{eq:m2}, and so the scalar 
spectrum is similar to AMSB, where sleptons have tachyonic masses.
For negative $\alpha$ values, the AMSB-moduli interference term in
Eq. \ref{eq:m2} becomes positive, so some spectra are allowed. The only
WMAP-allowed points occur at very low $m_{3/2}$ values, and are typified
by bulk neutralino annihilation through light sleptons.
In the case of large $\tan\beta =30$ shown in frame {\it f}), 
again, no allowed regions appear for $\alpha >0$, while some
points at large negative $\alpha$ are allowed where
neutralino annihlation occurs through light slepton exchange in the 
$t$-channel. This region, we will see, turns out to be largely excluded 
because it yields a 
large, {\it negative} value of $(g-2)_\mu$, in contradiction with experimental
measurements whose discussion we now turn to.

\section{${\bf (g-2)_\mu}$ and ${\bf BF(b\to s\gamma})$ in the MM-AMSB model}
\label{sec:bsg}

\subsection{${\bf (g-2)_\mu}$}

Current measurements of the muon anomalous magnetic moment show an apparent
deviation from SM predictions. Combining QED, electroweak, hadronic
(using $e^+e^-\to {\rm hadrons}$ to evaluate hadronic loop contributions)
and light-by-light contributions, and comparing against measurements
from E821 \cite{g-2expt} at BNL, a {\it positive} deviation in 
$a_\mu\equiv \frac{(g-2)_\mu}{2}$ of
\be
\Delta a_\mu =a_\mu^{exp} -a_\mu^{SM} =22(10)\times 10^{-10} 
\ee
is reported in the Particle Data Book\cite{pdb}, {\it i.e.} a
$2.2\sigma$ effect. 
Within this framework, corrections from Standard Model expectations arise
from one loop diagrams with either
$\tw_i-\tnu_\mu$ or $\tz_i-\tmu_{1,2}$ in the loop, possibly
accounting for the (rather weak) discrepancy. 

In Fig. \ref{fig:amu1} we show $a_\mu^{\rm SUSY}$\cite{gm2}, 
the SUSY contribution
versus $m_{\tmu_L}$ arising from the various modular weight choices in
the MM-AMSB model.  Only points with $\Omega_{\tz_1}h^2<0.13$ are
plotted in this figure.  The various colors/shapes of the plotting
symbols correspond to different choices of modular weights, as listed on
the figure.  We see from Fig.~\ref{fig:amu1}{\it a}) and
Fig.~\ref{fig:amu1}{\it b}) that, as expected, the SUSY contributions
from models with large $m_{\tmu_L}$ (and concomitantly large
$m_{\tnu_{\mu}}$ and $m_{\tmu_R}$) give almost no contributions, since
the contributions are suppressed by the large sparticle masses. However,
depending on the value of $\tan\beta$, models with $m_{\tmu_L}$ ranging
from about 200~GeV to several hundred GeV can give sizable
contributions, and even accommodate the theory-experiment deviation. For
the larger $\tan\beta$ case, very light sleptons in fact give too large
a SUSY contribution to the muon magnetic moment. In
addition, several sets of models give large {\it negative} contributions
to $a_\mu^{\rm SUSY}$. These are the models with $n_m=1$ and any choice of
$n_H$ but with moderate to large values of negative $\alpha$, wherein matter
scalar masses are suppressed, and the points are WMAP-allowed by bulk
annihilation of neutralinos via light $t$-channel slepton exchange. Even
though we take $\mu >0$ in these models, we know that
\be
a_\mu^{\rm SUSY}\sim \frac{m_\mu^2 \mu M_i\tan\beta}{M_{\rm SUSY}^4}
\ee
and since the weak scale gaugino masses $M_{1,2}$ are 
usually {\it negative} in these cases, a negative contribution
to $a_\mu^{\rm SUSY}$ arises. Thus (even allowing for considerable
theoretical uncertainties) these models with $a_\mu^{\rm SUSY}\sim -40$ 
are likely ruled out. 

The situation is more clearly illustrated in Fig. \ref{fig:amu2}, where
we plot $a_\mu^{\rm SUSY}$ versus $\alpha$ for {\it a})~$\tan\beta=10$
and {\it b})~$\tan\beta=30$. We see that models with $\alpha<-5$ give
large negative contributions to $a_\mu^{\rm SUSY}$.  Thus, when $\alpha
\alt -5$, models with $\mu <0$ should give $\Delta a_\mu$ allowed
regions. Models where agreement with the measured CDM relic density is
obtained via BWCA (these do not occur for the $\tan\beta=30$ cases)
yield a smaller value of $a_{\mu}^{\rm SUSY}$, in part because the
sleptons are heavier, and in part because of the relative sign between
$M_1$ and $M_2$ that leads to a negative interference between the
chargino and neutralino loop diagrams.
\FIGURE[htb]{
\epsfig{file=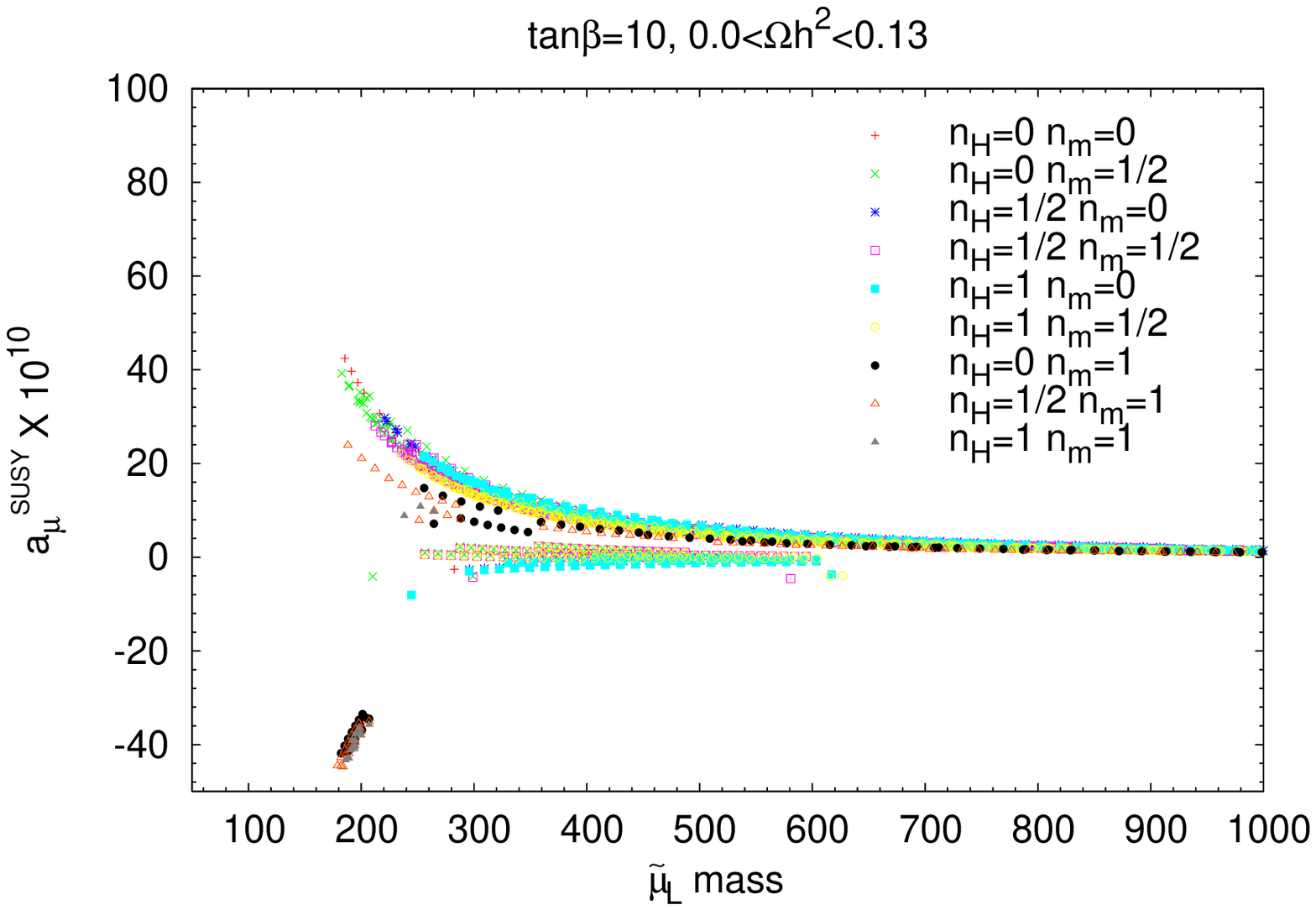,height=5.cm, angle=0}
\epsfig{file=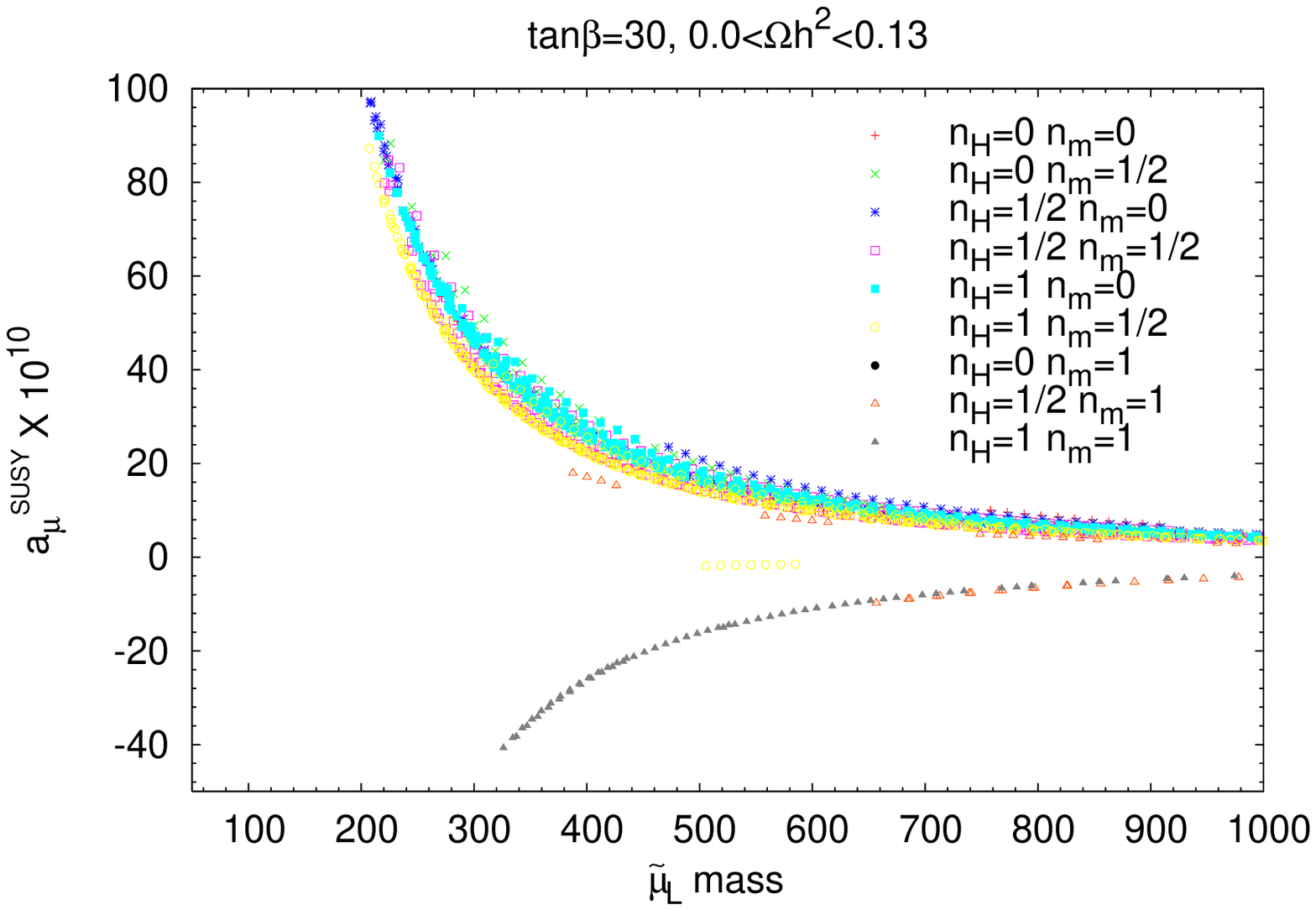,height=5.cm,angle=0} \\
\caption{\label{fig:amu1}
Plot of $a_\mu^{\rm SUSY}\ vs.\ m_{\tmu_L}$ in the MM-AMSB model for various
modular weight choices with $\mu >0$ 
and $m_t=175$ GeV. Only models with $\Omega_{\tz_1}h^2<0.13$ are
retained. In frame {\it a}), we show results for $\tan\beta =10$, while in 
frame {\it b}) we take $\tan\beta =30$. }}
\FIGURE[htb]{
\epsfig{file=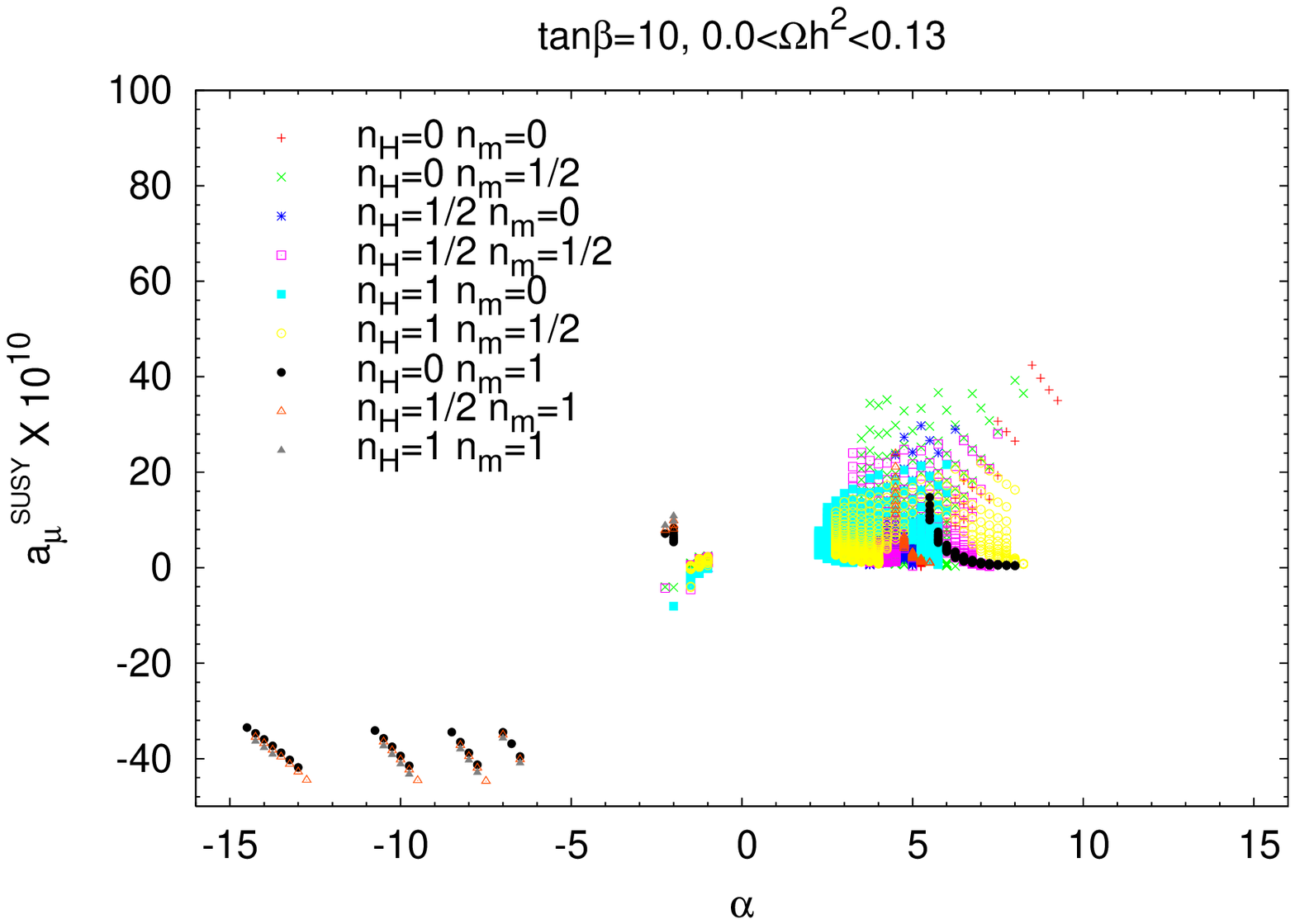,height=5.cm, angle=0}
\epsfig{file=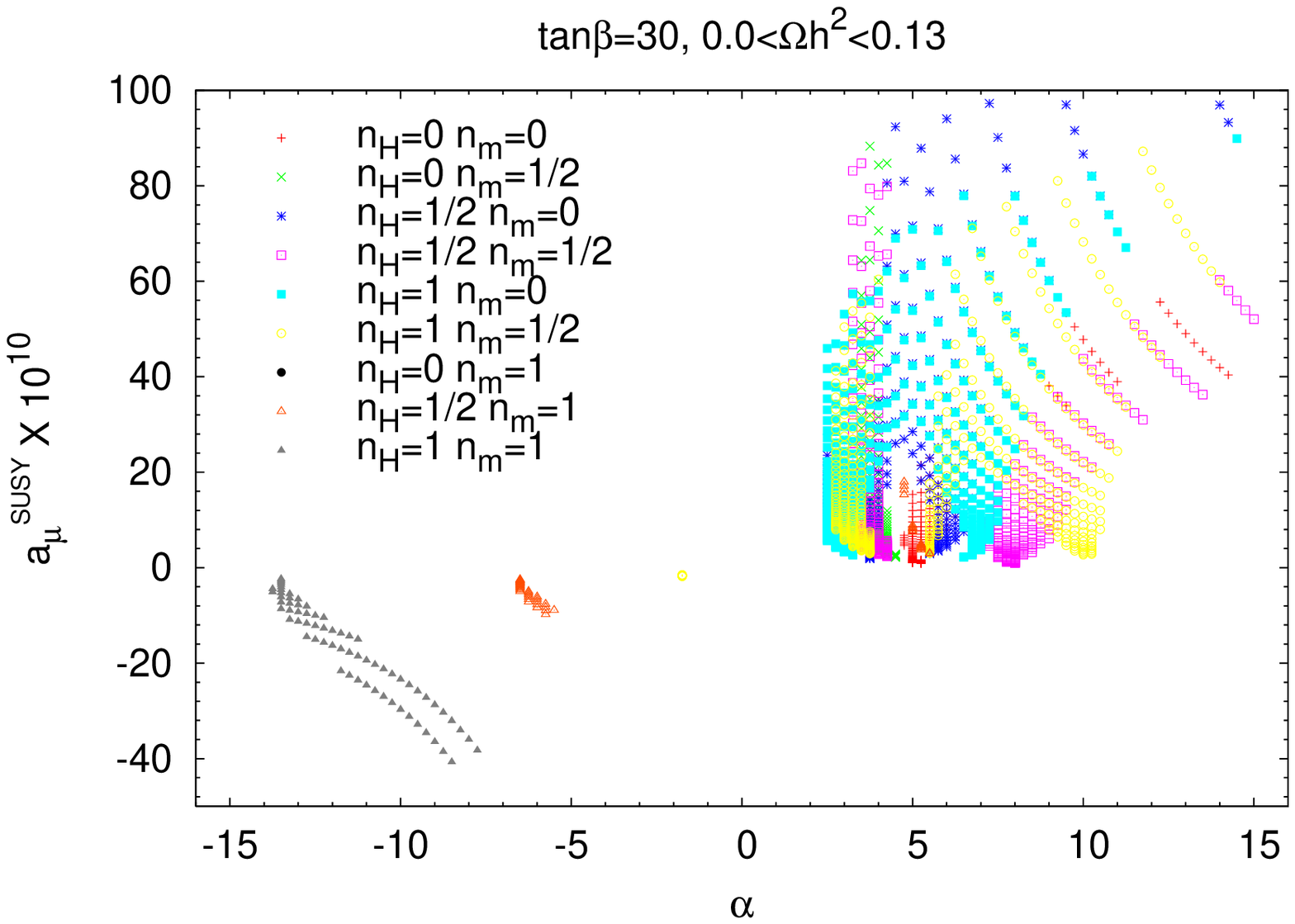,height=5.cm,angle=0} \\
\caption{\label{fig:amu2}
Plot of $a_\mu^{\rm SUSY}\ vs.\ \alpha$ in the MM-AMSB model for various
modular weight choices with $\mu >0$ 
and $m_t=175$ GeV. Only models with $\Omega_{\tz_1}h^2<0.13$ are
retained. In frame {\it a}), we show results for $\tan\beta =10$, while in 
frame {\it b}) we take $\tan\beta =30$. }}
%


\subsection{${\bf BF(b\to s\gamma )}$}

The branching fraction $BF(b\to s\gamma )$ serves as a strong constraint
on SUSY models
in part because, if $m_{\rm SUSY}\simeq M_W$,
supersymmetric contributions mediated by 
$\tw_i\tst_j$ and $bH^+$ loops are expected to occur at similar rates\cite{bsg}
to the SM contribution, mediated by a $tW$ loop. 
The measured branching fraction-- from a combination of CLEO, Belle and
BABAR experiments\cite{bsg_ex}--
is $BF(b\to s\gamma )=(3.55\pm 0.26)\times 10^{-4}$, while the
latest SM calculations find\cite{bsg_th} $BF(b\to s\gamma )=
(3.29\pm 0.33)\times 10^{-4}$. Thus, any SUSY contribution to
$BF(b\to s\gamma )$ seems highly suppressed. 

The results from the MM-AMSB models for various modular weight choices
and with $\Omega_{\tz_1}h^2<0.13$ are shown in Fig. \ref{fig:bsg1},
where we plot $BF(b\to s\gamma )\ vs.\ m_{\tg}$ for {\it a}) $\tan\beta
=10$ and {\it b}) $\tan\beta =30$. Many of the MM-AMSB models with
$m_{\tg}\alt 1000$ GeV predict $BF(b\to s\gamma )<2\times 10^{-4}$, and
thus are likely ruled out (unless small flavor-changing contributions to
squark mass matrices are invoked). Since $m_{\tg}$ is strongly
correlated with $m_{\tq}$ (see Fig.~\ref{fig:sq_gl} below), these are
models with relatively light squarks (and sleptons).  Models with
$\alpha < 0$ that are in agreement with the measured CDM relic density
due to bulk annihilation via very light sleptons (and correspondingly
light squarks) typically give {\it too high} a prediction for $BF(b\to
s\gamma )$, and are again likely ruled out.  In frame {\it b}) with
$\tan\beta =30$, the contributions to $BF(b\to s\gamma )$ can be even
more anomalous, and a much larger fraction of models with $m_{\tg}\alt
1000$ GeV are likely ruled out.

\FIGURE[htb]{
\epsfig{file=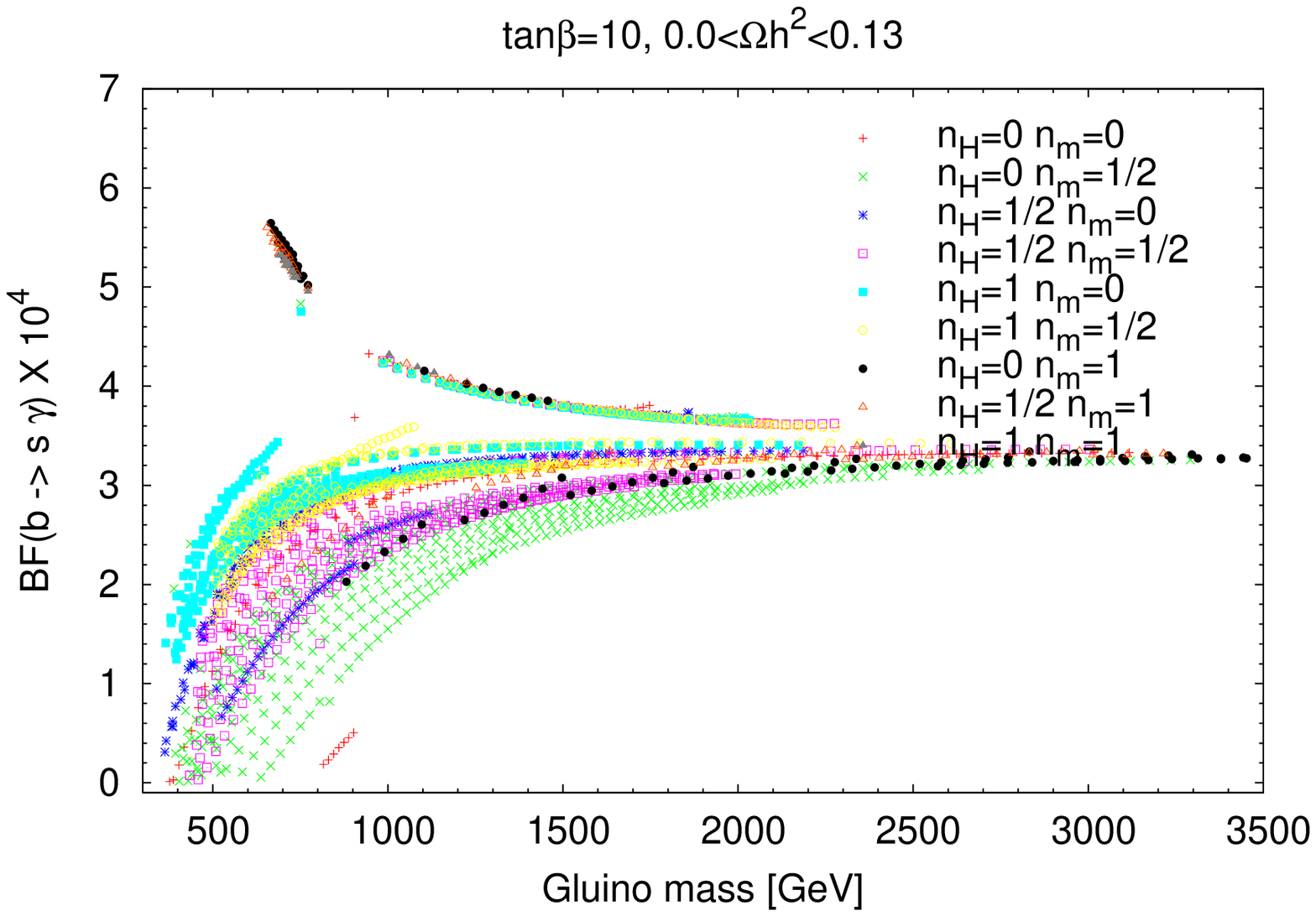,height=5.cm, angle=0}
\epsfig{file=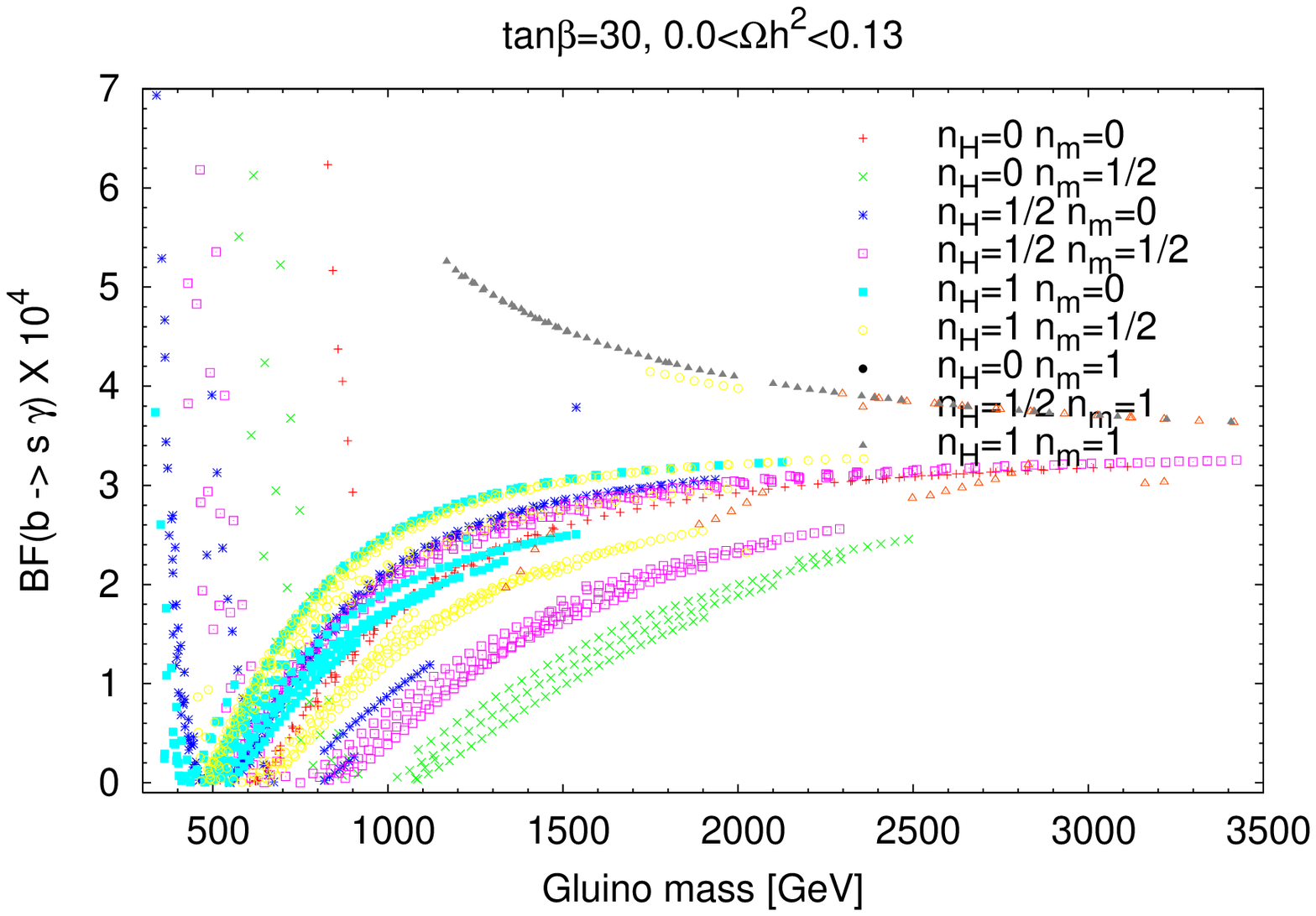,height=5.cm,angle=0} \\
\caption{\label{fig:bsg1} Plot of $BF(b\to s\gamma )\ vs.\ m_{\tg}$ in
the MM-AMSB model for the different modular weight choices, for {\it
a})~$\tan\beta =10$, and {\it b})~$\tan\beta =30$.  Only models with
$\Omega_{\tz_1}h^2<0.13$ are included.  We take $\mu >0$ and $m_t=175$
GeV.}}

\section{Collider searches in the MM-AMSB Model}
\label{sec:col}

\subsection{Fermilab Tevatron $p\bar{p}$ collider}

It is possible to search for gluino and squark pair production at
the Fermilab Tevatron $p\bar{p}$ collider by looking for
multi-jet$+\eslt$ signals. While many Tevatron $\eslt +jets$ searches for
models with gaugino mass universality have been
pre-empted by LEP2 searches for chargino pair production, in models with 
non-universal gaugino masses, gluino and squark searches 
may still be of interest. Indeed, in the so-called ``low $M_3$ dark matter 
model'' (LM3DM)\cite{lm3dm1}, 
the Tevatron with 5 fb$^{-1}$ of integrated luminosity was
found to be sensitive to $m_{\tg}\sim 200-350$ GeV in models
beyond the reach of LEP2\cite{lm3dm2}. The MM-AMSB models share with 
LM3DM models the characteristic that $M_3$ is reduced relative to $M_1$ and
$M_2$ at the GUT scale, since the AMSB contribution to $M_3$
subtracts from the moduli-mediated contribution. 

In Fig. \ref{fig:sq_gl}, we plot $m_{\tu_R}\ vs.\ m_{\tg}$ for MM-AMSB
models with {\it a}) $\tan\beta =10$ and {\it b}) $\tan\beta =30$, and
all modular weight choices, where we require $\Omega_{\tz_1}h^2<0.13$
and $m_{\tw_1}>103$ GeV.  The diagonal dashed line denotes where
$m_{\tu_R}=m_{\tg}$.  One lesson from this plot is that generic first
and second generation squark masses are {\it comparable} in mass to
$m_{\tg}$ for {\it all} modular weight choices; the possibility that
$m_{\tq}\gg m_{\tg}$ -- which occurs in the large $m_0$ region of the
mSUGRA model-- does not occur here.  We also find that in some cases
(such as $n_H=1$, $n_m=0$) the value of $m_{\tg}\sim m_{\tq}$
can be as low as $\sim 300$ GeV, and hence may be amenable to Tevatron
gluino and squark searches, where the current limit for $m_{\tq}\simeq
m_{\tg}$ is $\sim 325$~GeV within the mSUGRA framework \cite{tevgl}.
\FIGURE[htb]{
\epsfig{file=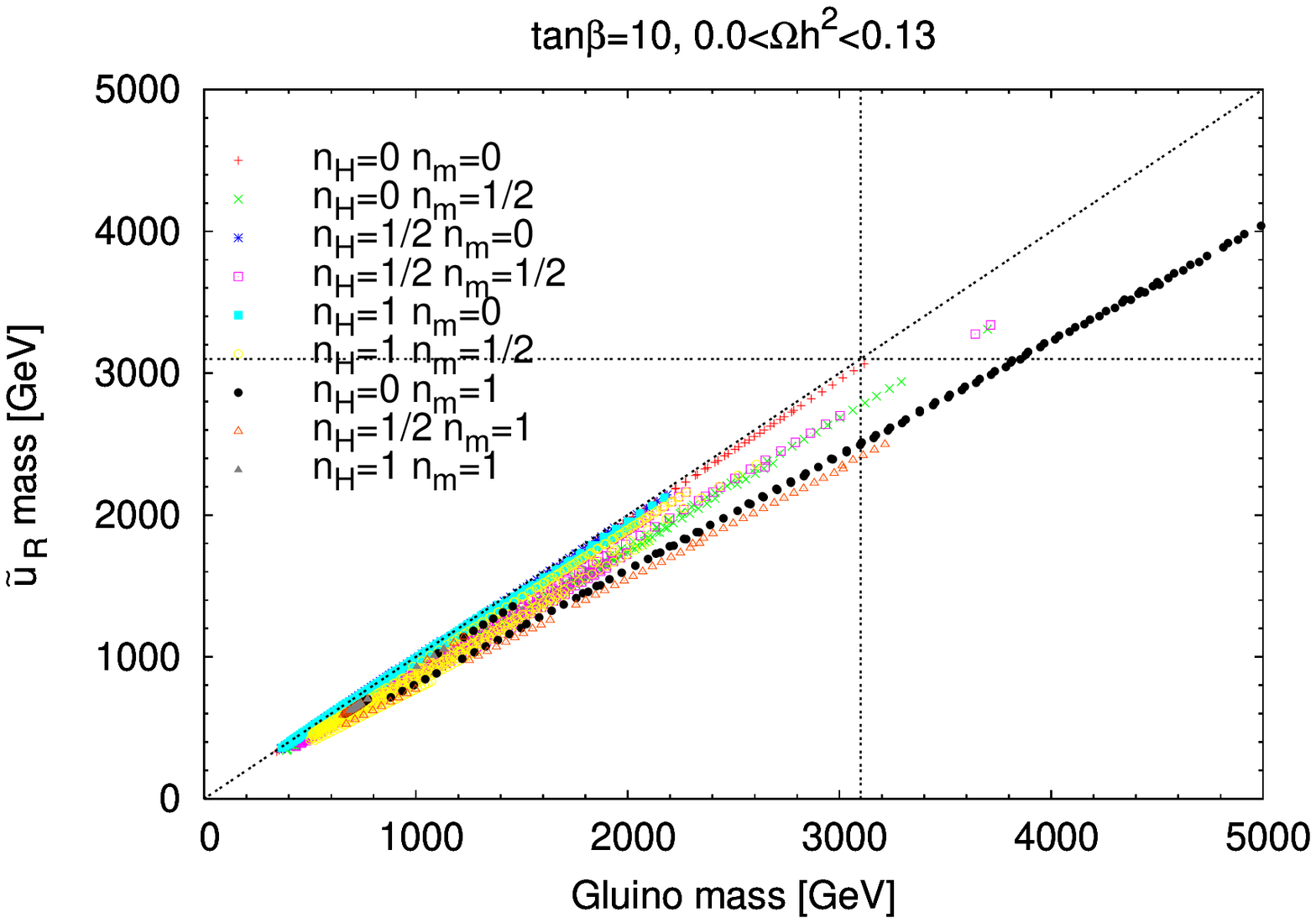,height=5.cm, angle=0}
\epsfig{file=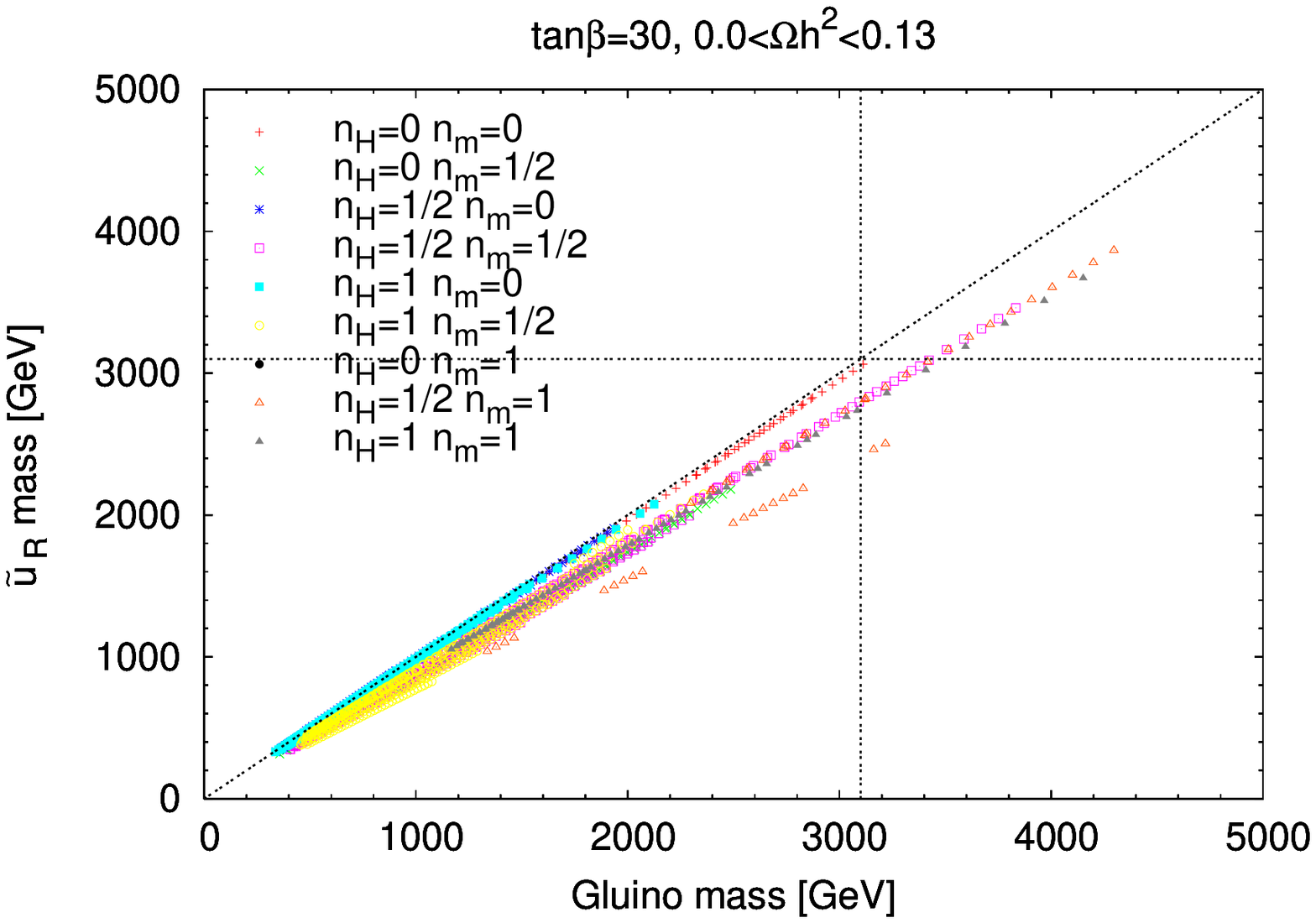,height=5.cm,angle=0} \\
\caption{\label{fig:sq_gl}
Plot of $m_{\tu_R}\ vs.\ m_{\tg}$ in the MM-AMSB model for various
modular weight choices. Only models with $\Omega_{\tz_1}h^2<0.13$ 
and LEP2 constraints satisfied are
included. In frame {\it a}), we plot models for $\tan\beta =10$, while in 
frame {\it b}) we plot models for $\tan\beta =30$.}}

It is also possible for Tevatron experiments to search for SUSY via
$p\bar{p}\to\tw_1^\pm\tz_2 X\to 3\ell+\eslt +X$ events\cite{trilep}.
The clean trilepton search is viable when the ``spoiler'' decay modes
$\tz_2\to\tz_1 h$ and $\tz_2\to\tz_1 Z$ are closed\cite{nanop}, {\it
i.e.} when $m_{\tz_2}-m_{\tz_1}<M_Z$ and $m_h$. We plot in
Fig. \ref{fig:z2z1_w1} the $m_{\tz_2}-m_{\tz_1}$ mass difference versus
$m_{\tw_1}$ for all modular weight choices and WMAP/LEP2-allowed models,
for {\it a}) $\tan\beta =10$ and {\it b}) $\tan\beta =30$.  The region
above the horizontal dashed line is where the spoiler decay $\tz_2\to
\tz_1 Z$ turns on.  We see a large fraction of MM-AMSB models of all
modular weights are below this line, and thus have closed spoiler modes,
so that $\tz_2\to\tz_1\ell\bar{\ell}$ is likely to have a significant
branching fraction (an exception occur if interference effects suppress
the leptonic branching decay or if the decay $\tz_2 \to \tau\ttau_1$ 
becomes accessible). 
Depending
on $BF(\tz_2\to\tz_1\ell\bar{\ell})$, a portion of the lower
range of $m_{\tw_1}\sim 100-200$ GeV and $m_{\tz_2}-m_{\tz_1}<M_Z$ may
thus be accessible to Tevatron clean trilepton searches.
\FIGURE[htb]{
\epsfig{file=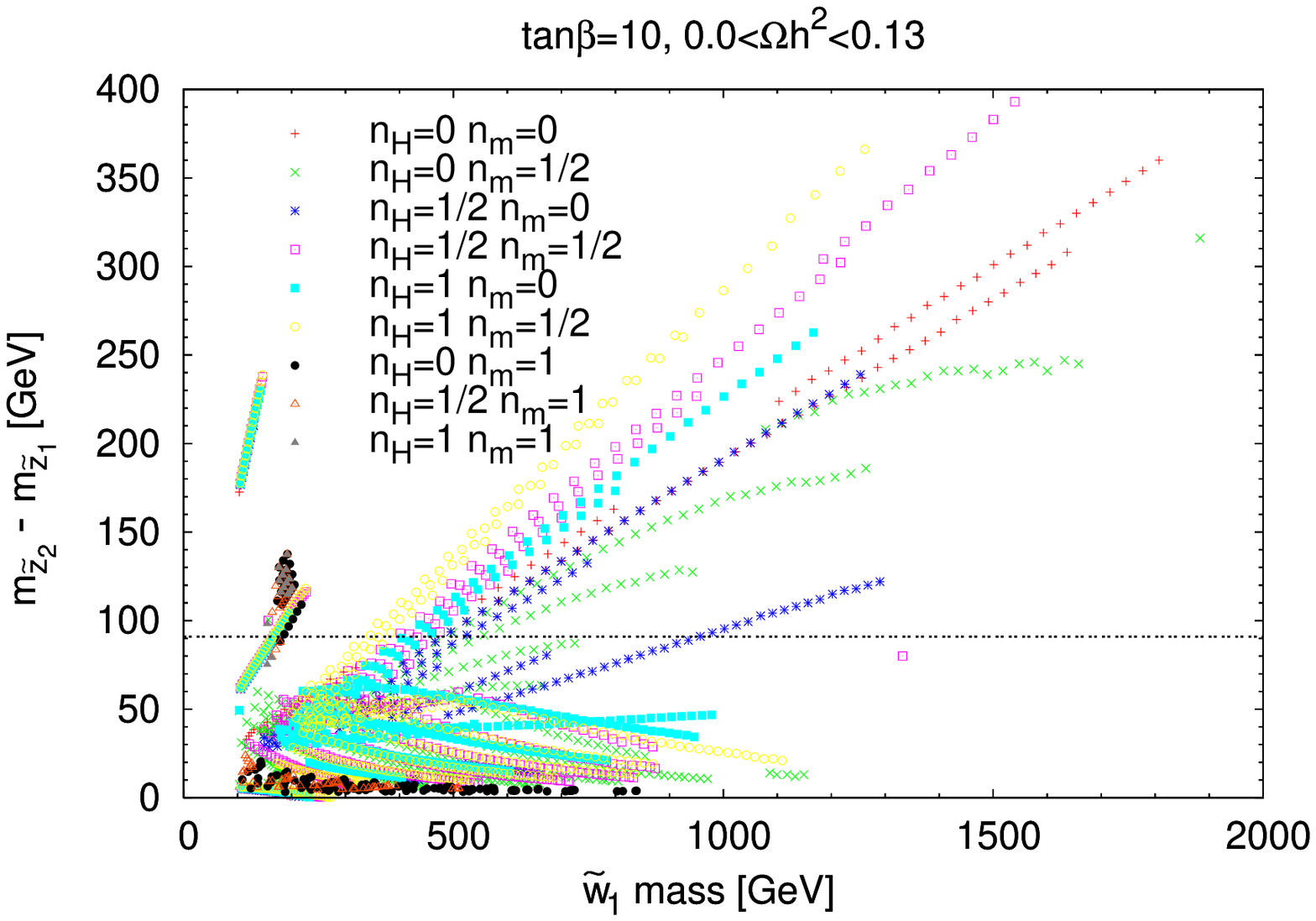,height=5.cm,angle=0}
\epsfig{file=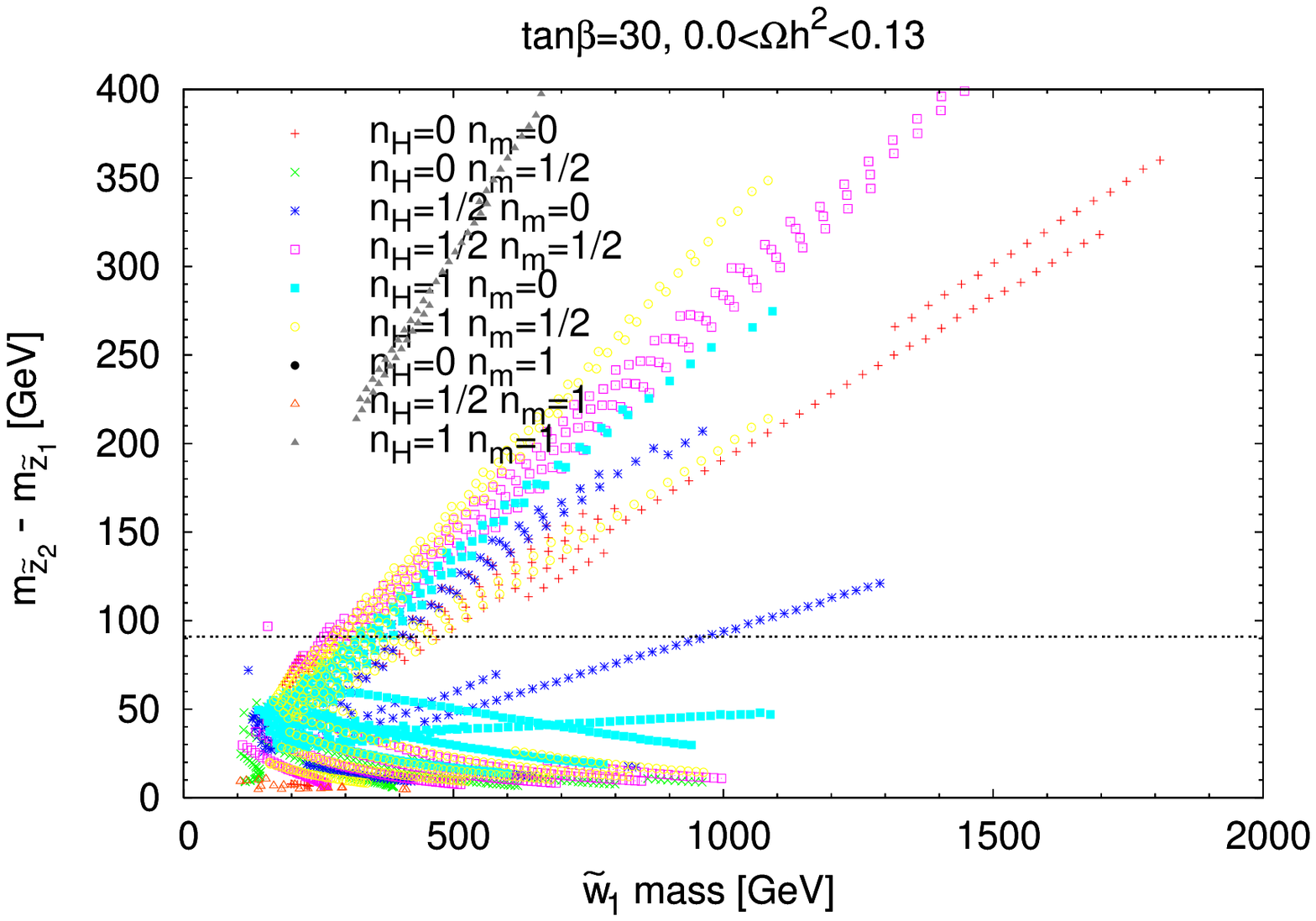,height=5.cm,angle=0} \\
\caption{\label{fig:z2z1_w1}
Plot of $m_{\tz_2}-m_{\tz_1}\ vs.\ m_{\tw_1}$ in the MM-AMSB model for various
modular weight choices. Only models with $\Omega_{\tz_1}h^2<0.13$ 
and LEP2 constraints satisfied  are
included. In frame {\it a}), we plot models for $\tan\beta =10$, while in 
frame {\it b}) we plot models for $\tan\beta =30$.}}

Finally, in Fig. \ref{fig:t1_higgs} we plot the values of $m_h\ vs.\
m_{\tst_1}$ for WMAP/LEP2-allowed MM-AMSB models and all modular weight
choices. Here, we see that $m_h\alt 125-130$ GeV, so that Tevatron Higgs
searches may see evidence of $Wh$, $Zh$ production depending on the
integrated luminosity achieved. In addition, the
$\tst_1$ mass can range down to $\sim 100$ GeV in MM-AMSB models and may
be accessible to Tevatron searches\cite{bst}.  However, in these cases,
the $m_{\tst_1}-m_{\tz_1}$ mass gap is usually quite small, in which
case the $\tst_1$ decay products will be soft, making hadron collider
searches more difficult.
\FIGURE[htb]{
\epsfig{file=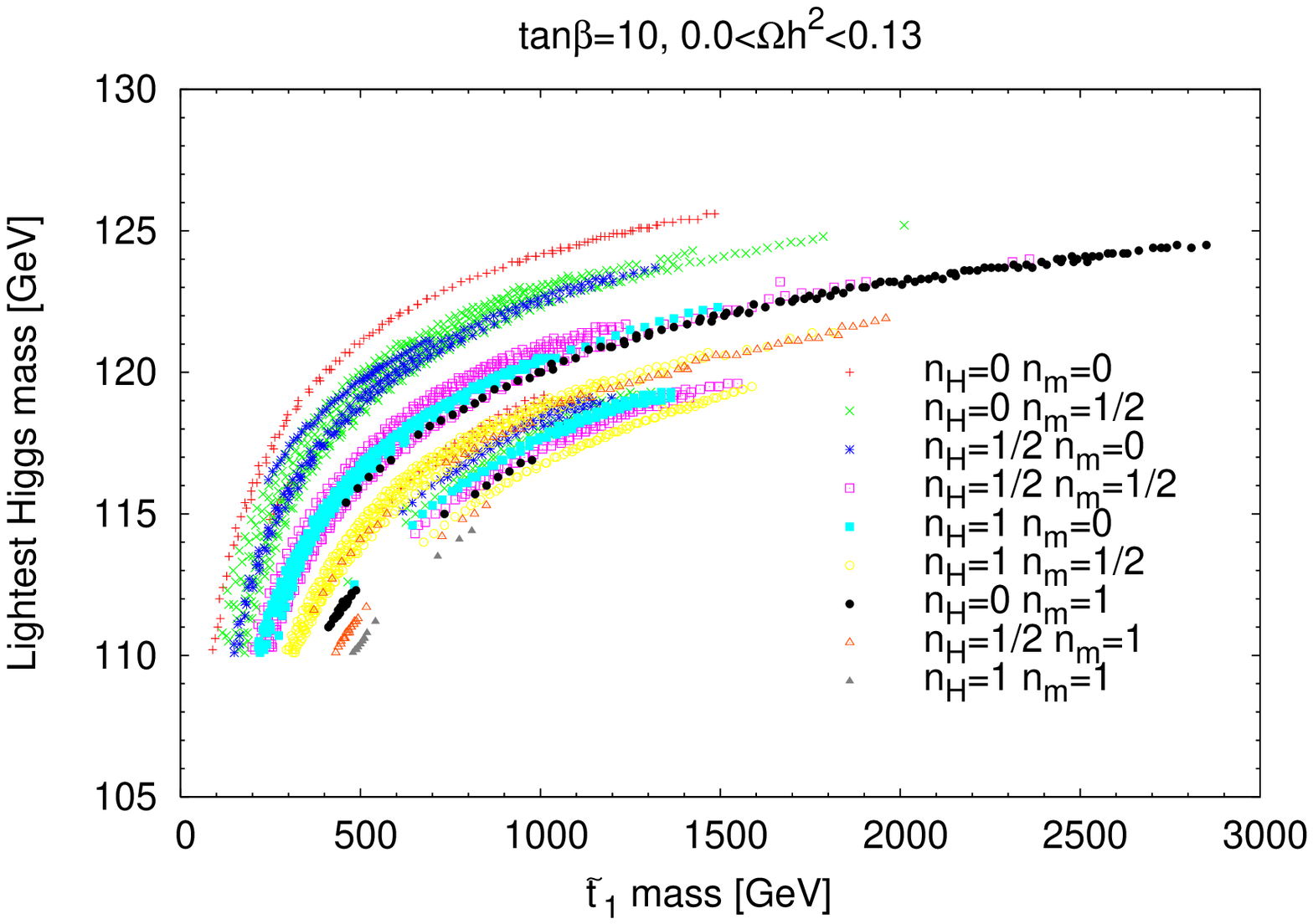,height=5.cm,angle=0}
\epsfig{file=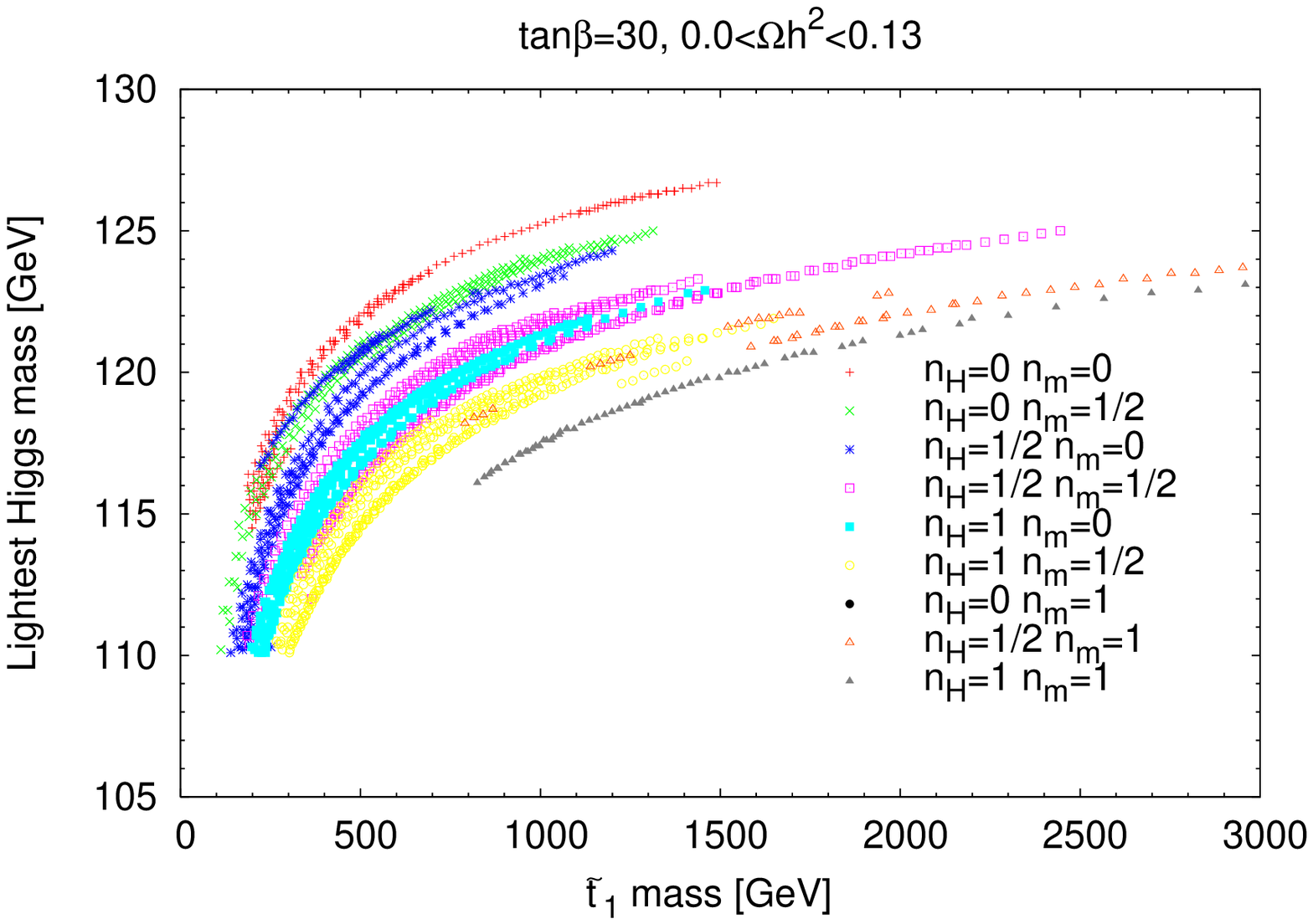,height=5.cm,angle=0} \\
\caption{\label{fig:t1_higgs}
Plot of $m_h\ vs.\ m_{\tst_1}$ in the MM-AMSB model for various
modular weight choices. Only models with $\Omega_{\tz_1}h^2<0.13$ 
and consistent with LEP2 constraints satisfied are
included. In frame {\it a}), we plot models for $\tan\beta =10$, while in 
frame {\it b}) we plot models for $\tan\beta =30$.}}

We note here also that in the MM-AMSB models with $m_{\tst_1}<m_t$ and $m_h$ is
light, some of the conditions for successful electroweak baryogenesis are
satisfied. However, in Ref. \cite{carena}, it is noted that the soft term
$m_{\tst_R}^2<0$ condition also ought to be satisfied. We find that in MM-AMSB
models $m_{\tst_R}^2>0$ and the light $\tst_1$ arises in part due to the
large $A_t$ term; this apparently contradicts the requirements for
successful EW baryogenesis in these models. 
Of course, this would be moot unless
there is also a significant $CP$ violating phase ($\agt 0.05$) 
either via the $b$-term, or via corrections to other SSB parameters.

\subsection{CERN LHC $pp$ collider}

The CERN LHC $pp$ collider will begin data taking in earnest in 2008
running at $\sqrt{s}=14$ TeV. The search for supersymmetry by Atlas
and CMS will initially focus on gluino and squark production followed
by cascade decays. It has been shown in Ref. \cite{lhcreach} that
in the case where $m_{\tq}\sim m_{\tg}$ as in MM-AMSB models, 
LHC experiments should have a reach to $m_{\tg}\alt 3$ TeV for
an integrated luminosity of 100 fb$^{-1}$. 
Except for possibly cases where the sparticle masses are all roughly degenerate
(where $\mu_{\rm mir}\sim 1$ TeV) 
so that the visible decay products are significantly
softer than the expectation in the mSUGRA model\cite{nojiri}, we would
expect a similar reach within the MM-AMSB framework.
We can see from
Fig. \ref{fig:sq_gl} that {\it most} MM-AMSB models should be detectable
at CERN LHC. Exceptions occur in the cases of $n_H=0$, $n_m=1$
models for $\tan\beta =10$, and $n_H=1/2$, $n_m=1$ models for
$\tan\beta =30$. In these instances, the thin allowed region to the left of
the excluded region for $\alpha\sim 5-7$ gives rise to cases with
$m_{\tg}$ ranging up to $4-5$ TeV, which is well-past the LHC reach
for any projected integrated luminosity value.

Fig.~\ref{fig:z2z1_w1} shows also that a large fraction of models contain
a mass difference $m_{\tz_2}-m_{\tz_1}<M_Z$. Thus, decays of $\tz_2$s
produced mostly via cascade decays of gluinos and squarks, should lead
to a measurable dilepton mass edge with $m(\ell\bar{\ell})$ bounded by
$m_{\tz_2}-m_{\tz_1}$ unless the branching fraction
for the decay is strongly suppressed. This mass edge can serve as a
starting point for gluino and squark cascade decay
reconstruction\cite{mll_lhc}.

\subsection{Linear $e^+e^-$ collider}

Experiments at a linear $e^+e^-$ collider will most easily
be able to discover charginos, and sleptons (including
sneutrinos, if these decay visibly) via their pair production.
Moreover, unless the mass gaps
$m_{\tw_1}-m_{\tz_1}$ or $m_{\tell}-m_{\tz_1}$
are very small, 
signals from their production should be readily visible over Standard
Model backgrounds essentially all the way to the kinematic limit for
their production. Specialized analyses allow the signal to be extracted
for smaller values of the sparticle-LSP mass gap\cite{bbkt}.
We show in Fig. \ref{fig:sl_ch} the $\ttau_1$ mass versus
$\tw_1$ mass for all MM-AMSB models consistent with WMAP/LEP2.
A $\sqrt{s}=500$ GeV machine should have a reach to 
$m_{\tw_1}$ or $m_{\ttau_1}\sim 250$ GeV, while a $\sqrt{s}=1000$ GeV
machine will have a reach to $m_{\tw_1}$ or $m_{\ttau_1}\sim 500$ GeV,
as indicated by the dashed reach lines.
Thus, while the ILC would certainly see the light Higgs boson $h$ in
all MM-AMSB cases, it would only be able to access sparticles
for a fraction of the models plotted. 
%
\FIGURE[htb]{
\epsfig{file=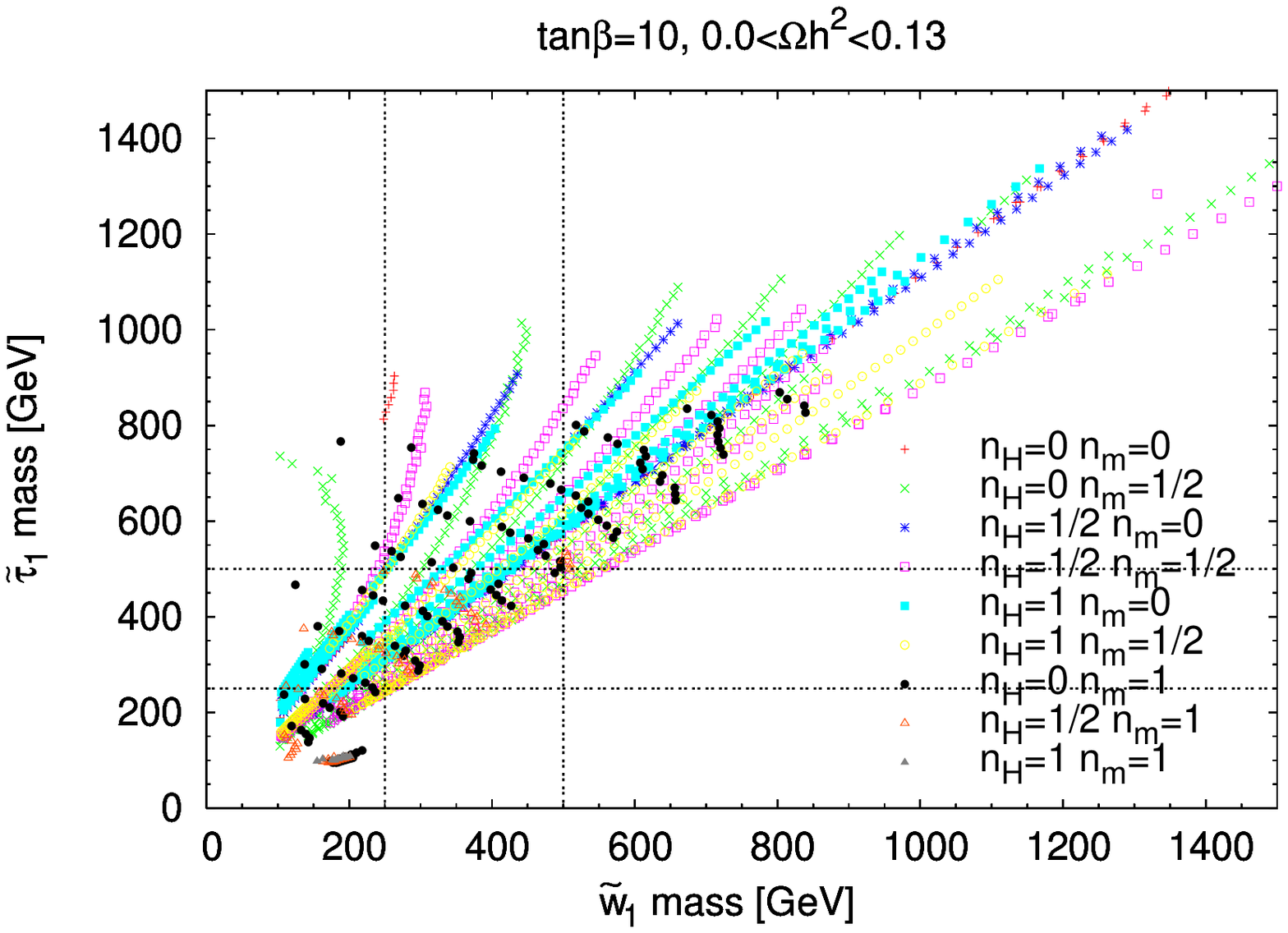,height=5.cm,angle=0}
\epsfig{file=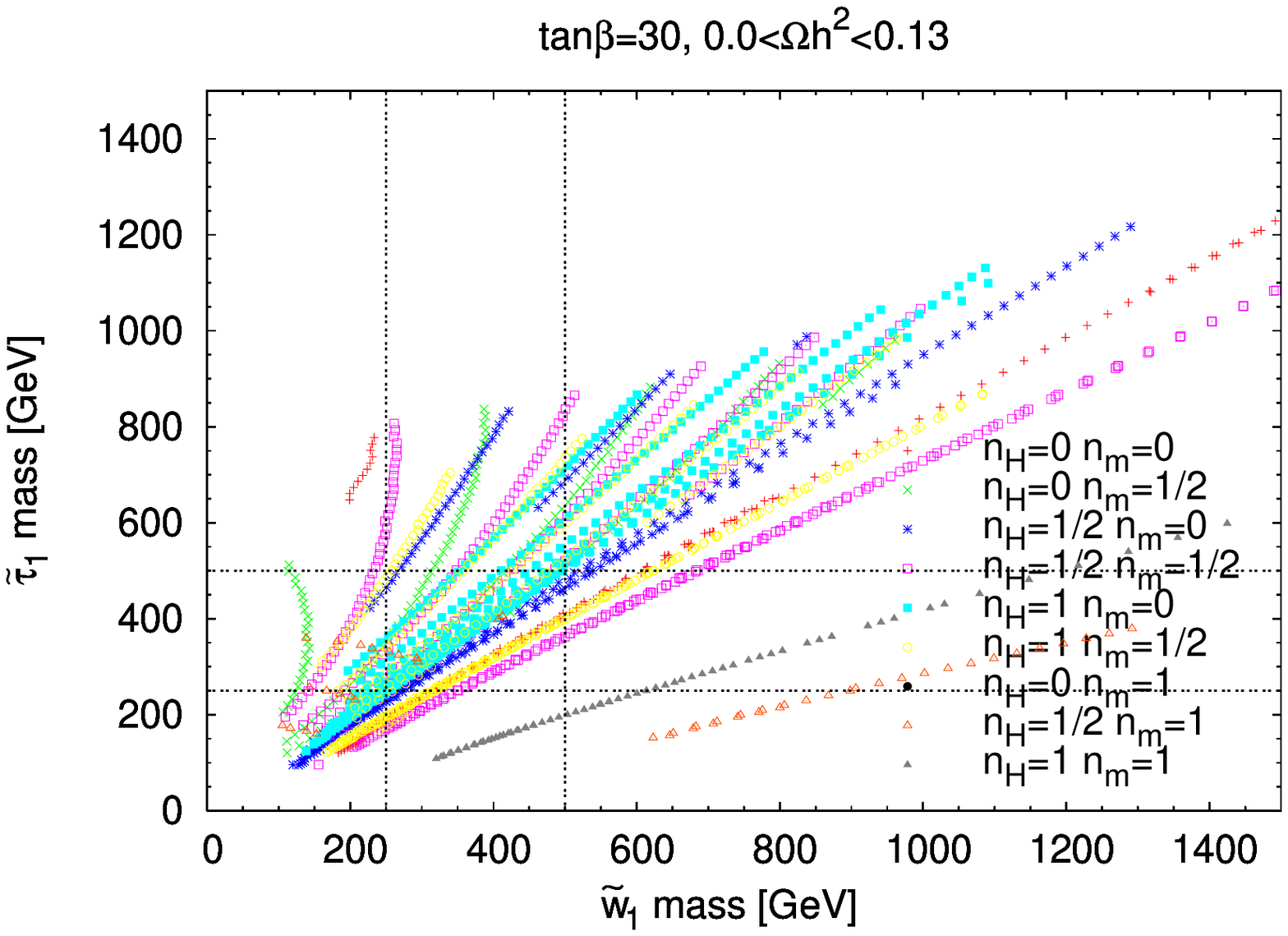,height=5.cm,angle=0} \\
\caption{\label{fig:sl_ch}
Plot of $m_{\ttau_1}\ vs.\ m_{\tw_1}$ in the MM-AMSB model for various
modular weight choices. Only models with $\Omega_{\tz_1}h^2<0.13$ 
and consistent with LEP2 constraints satisfied are
plotted. In frame {\it a}), we plot models for $\tan\beta =10$, while in 
frame {\it b}) we plot models for $\tan\beta =30$.}}

\section{Dark matter searches}
\label{sec:dm}

In addition to collider searches for sparticles and Higgs bosons,
searches for dark matter relics from the Big Bang, either via direct
detection experiments at underground dark matter detectors, or
indirectly via searches for neutralino annihilation to high energy
neutrinos in the core of the sun, or neutralino annihilations to high
energy gamma rays or antimatter in the galactic halo, may provide an
independent signal for new physics.  The results of these DM searches are
sensitive to the composition of the LSP, and combining these with
information from collider experiments may help us to zero in on the
underlying SUSY model.

\subsection{Direct dark matter searches}

In the case of particle physics models with $R$-parity conserving
supersymmetry and a neutralino LSP, dark matter detectors look for rare
neutralino-nucleus collisions in experiments located deep underground,
where the neutralino is expected to deposit of order  ten  keV of
energy in an elastic scattering event. The neutralino-nucleus scattering
cross section naturally breaks up into spin-independent (SI) and
spin-dependent (SD) pieces, where at present the greatest experimental
sensitivity is to SI cross sections, since in this case the neutralino
couples coherently to the entire nucleus, so that the cross section
increases as $A^2$, where $A$ is the mass number of the nucleus. To
facilitate comparision between different detector materials, the
spin-independent scattering cross section off a single proton is usually
used as a figure of merit.
We use the IsaRes code (part of the Isajet/Isatools
package\cite{bbb,bbbo}) to evaluate this associated direct dark matter
detection cross section in MM-AMSB models.

In Fig.~\ref{fig:dd}, we plot the expectation for $\sigma_{SI}(\tz_1 p)$
as a function of $m_{\tz_1}$ for all nine sets of modular weights,
retaining only those points consistent with constraints from both WMAP
and LEP2 data for {\it a})~$\tan\beta =10$, and {\it b})~$\tan\beta
=30$.  Presently, the most stringent limit on this cross section comes
from the CDMS collaboration\cite{cdms}, and is shown by the top-most
contours labelled CDMS. Since the neutralino can transfer the maximum
energy to the nucleus if its mass is $\sim$  the mass of the nuclear
target $m_T$, the sensitivity is the greatest when $m_{\tz_1}\sim m_T$.
The maximum energy that can be transferred reduces rapidly if $m_{\tz_1}
\ll m_T$, and saturates if the neutralino is very heavy,
accounting for the general shape of this contour (as well as of the
projected reach contours discussed below).
We see that, especially for the larger value of $\tan\beta$, the CDMS
experiment has already excluded neutralinos as heavy as 300-400~GeV for
some choices of modular weights.  CDMS II is now operating, and is
expected to reach cross sections at the $10^{-8}$~pb level which will
begin probing MM-AMSB parameter space with neutralinos as heavy as
500-800~GeV depending on $\tan\beta$.  CDMS plans an upgrade to 7
supertowers located in the Sudbury underground site, where a sensitivity
to cross sections at the $10^{-9}$~pb level is anticipated
\cite{supercdms}. To go further, ton-scale noble liquid detectors will
have to be used. As an example, we plot the reach of the proposed Warm
Argon Project (WARP) 1400 kg detector \cite{warp}, which will endeavour
to probe cross sections as low as $10^{-10}$~pb. We see that if proposed
ton-sized noble gas detectors do achieve the projected sensitivity, they
should be able to probe the bulk of MM-AMSB parameter space unless
$\tz_1$ is rather heavy.  Exceptions where light neutralinos ($m_{\tz_1}
\alt 500$~GeV ) would escape detection even at these facilities occur
when the neutralino is essentially bino-like, and the WMAP relic density
is attained via squark or slepton co-annihilation, or via BWCA.

%
\FIGURE[htb]{
\epsfig{file=det-10.eps,height=5.cm,angle=0}
\epsfig{file=det-30.eps,height=5.cm,angle=0} \\
\caption{\label{fig:dd}
Plot of $\sigma_{SI}(\tz_1 p)\ vs.\ m_{\tz_1}$ in the MM-AMSB model for various
modular weight choices for {\it a})~$\tan\beta =10$, and 
{\it b})~$\tan\beta =30$.
Only models with $\Omega_{\tz_1}h^2<0.13$ 
and LEP2 constraints satisfied are
included. 
The region above the contour labelled CDMS is excluded by the 
non-observation of a signal
in the CDMS experiment. The other contours show
the projected reach of CDMS (labelled CDMS II) along with the
projections for its proposed upgrade (superCDMS) and for 
the Warm Argon Project (WARP1400),
taken here as a representative of proposed ton-sized noble
gas detectors.}}

\subsection{Indirect dark matter searches}

In addition to direct detection experiments discussed above, there are
also indirect searches for dark matter that rely on the detection of a
flux of high energy particles, or of anti-particles, produced via the
annihilation of neutralinos in our galactic halo\cite{bo}.

The IceCube\cite{icecube} and ANTARES\cite{antares} experiments
are designed to detect high energy neutrinos from the annihilation of
neutralinos gravitationally trapped in the core of the
Sun\cite{neut_tel}.  In Fig.~\ref{fig:icecube}, we present rates for
detection of $\nu_\mu \to \mu$ conversions in the Antarctic ice, with
the muons being detected in the IceCube experiment.  We show results for
{\it a})~$\tan\beta=10$, and {\it b})~$\tan\beta=30$. The reference
experimental sensitivity that we show is for a flux of 40 muons per 
${\rm km}^2$ per year with $E_\mu >50$ GeV. 
We see that while many sets of modular weights lead to a
potentially detectable signal for neutralinos as heavy as 1~TeV ({\it
e.g.} $n_H=1, n_m=0$), there are other models where the signal is {\it
always} below the observable level ({\it e.g.} $n_H=0$, $n_m=0$). This may
be understood if we recognize that the signal at IceCube will be largely
governed by the higgsino content of the neutralino since this controls
the neutralino-nucleon scattering cross section (essential for trapping
the neutralinos in the first place) and, along with the wino content,
also affects the neutralino annihilation rate. Thus, when we can only
have a bino LSP (as, for instance, for $n_H=n_m=0$ case) the signal will
be small. In contrast, regions with mixed higgsino or mixed
wino-higgsino-bino DM will typically lead to the largest signals at
IceCube.

\FIGURE[htb]{
\epsfig{file=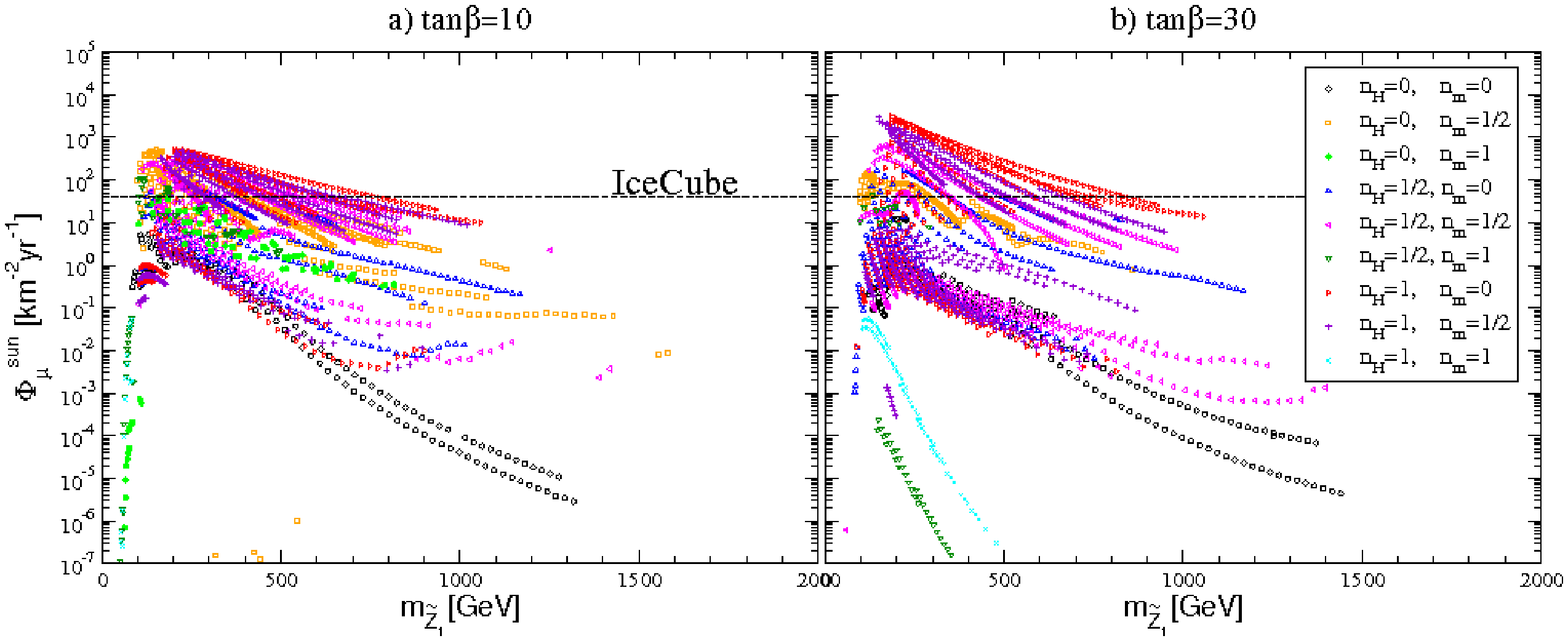,height=16.cm,angle=-90}
\caption{\label{fig:icecube} The integrated 
muon flux and the projection for the ultimate reach of the
IceCube experiment for the various choices of Higgs and matter modular
weights as labelled on the figure for {\it a})~$\tan\beta=10$, and {\it
b})~$\tan\beta=30$ with $\mu>0$ and $m_t=175$. The points above the
horizontal dashed line are projected to be within the reach of IceCube,
assuming the sensitivity is given by 40 events/km$^2$/yr for
$E_{\mu}>50$~GeV. Only points consistent with constraints from WMAP and
LEP2 are included in the figure.}}

Neutralinos in our galactic halo may also annihilate into anti-particles
or anti-nuclei, thus providing new ways for their indirect detection,
though signals from neutrinos not pointing from the sun are below their
background. Positron detection experiments include HEAT\cite{heat},
Pamela\cite{pamela} and AMS-02\cite{ams}; antiprotons may be detected by
BESS\cite{bess}, Pamela, AMS-02; anti-deuterons
by BESS\cite{bessdbar}, AMS-02 and GAPS\cite{gaps}. For positrons and
anti-protons we evaluate the averaged differential antiparticle flux in
a projected energy bin centered at a kinetic energy of 20~GeV, where we
expect an optimal statistics and signal-to-background ratio at
space-borne antiparticle detectors\cite{statistical}. We take
the experimental sensitivity to be that of the Pamela experiment after
three years of data-taking as our benchmark.  Finally, the average
differential anti-deuteron flux has been computed in the $0.1<T_{\bar
D}<0.25$ GeV range, where $T_{\bar D}$ stands for the antideuteron
kinetic energy per nucleon, and compared to the estimated GAPS
sensitivity for an ultra-long duration balloon-borne experiment
\cite{gaps}.

Our projection for the sensitivity of Pamela ($\bar{p}$ and $e^+$) and
GAPS experiments ($\overline{D}$) is shown in Fig.~\ref{fig:antip}. We
show these for $\tan\beta=10$ (left column) and $\tan\beta=30$ (right
column). The expected fluxes depend on the (unknown) details of the
neutralino distribution in our galactic halo, and are shown for the
Burkert profile \cite{burkert} obtained if it is assumed that the
central cusp of the DM halo seen in numerical simulations is smoothed
out by heating of colder particles. This less clumpy halo distribution
yields lower fluxes for the anti-particles so that our projections for
the reach of these indirect search experiments may be regarded as
conservative.

\FIGURE[htb]{
\epsfig{file=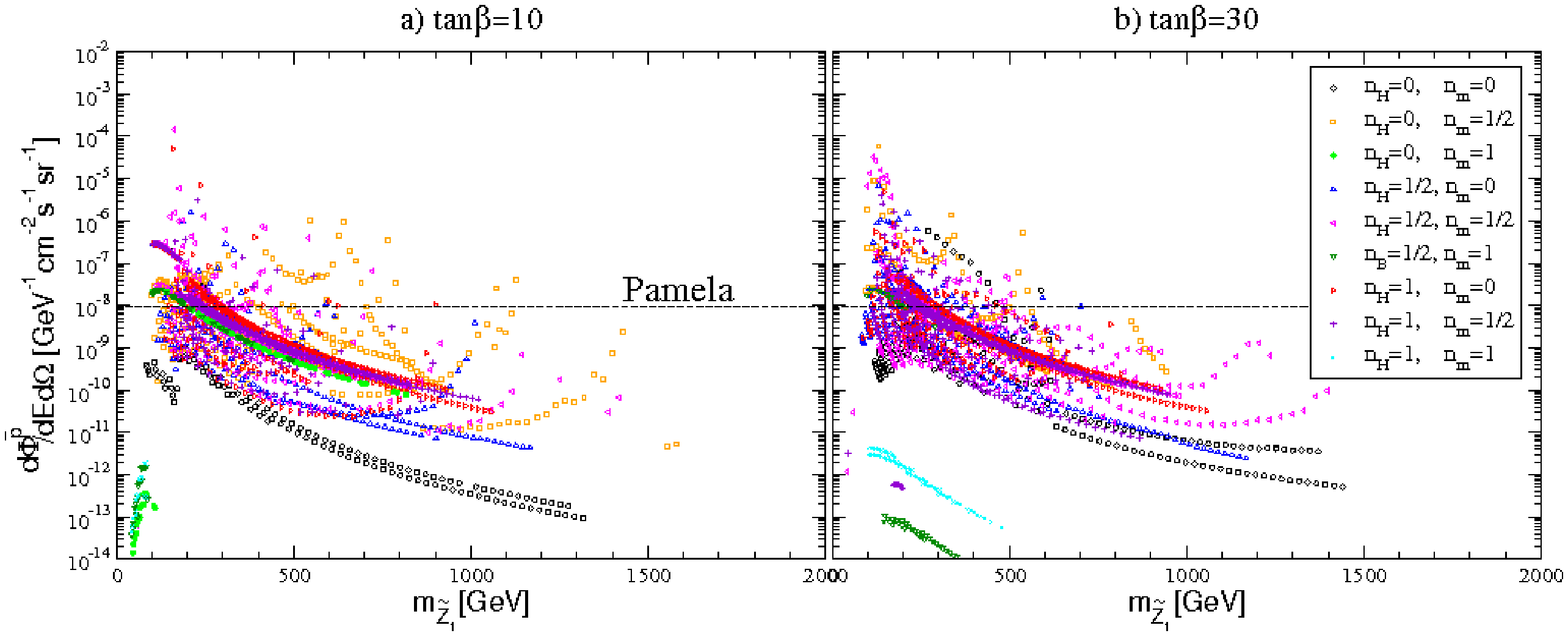,height=12cm,angle=-90}\\
\epsfig{file=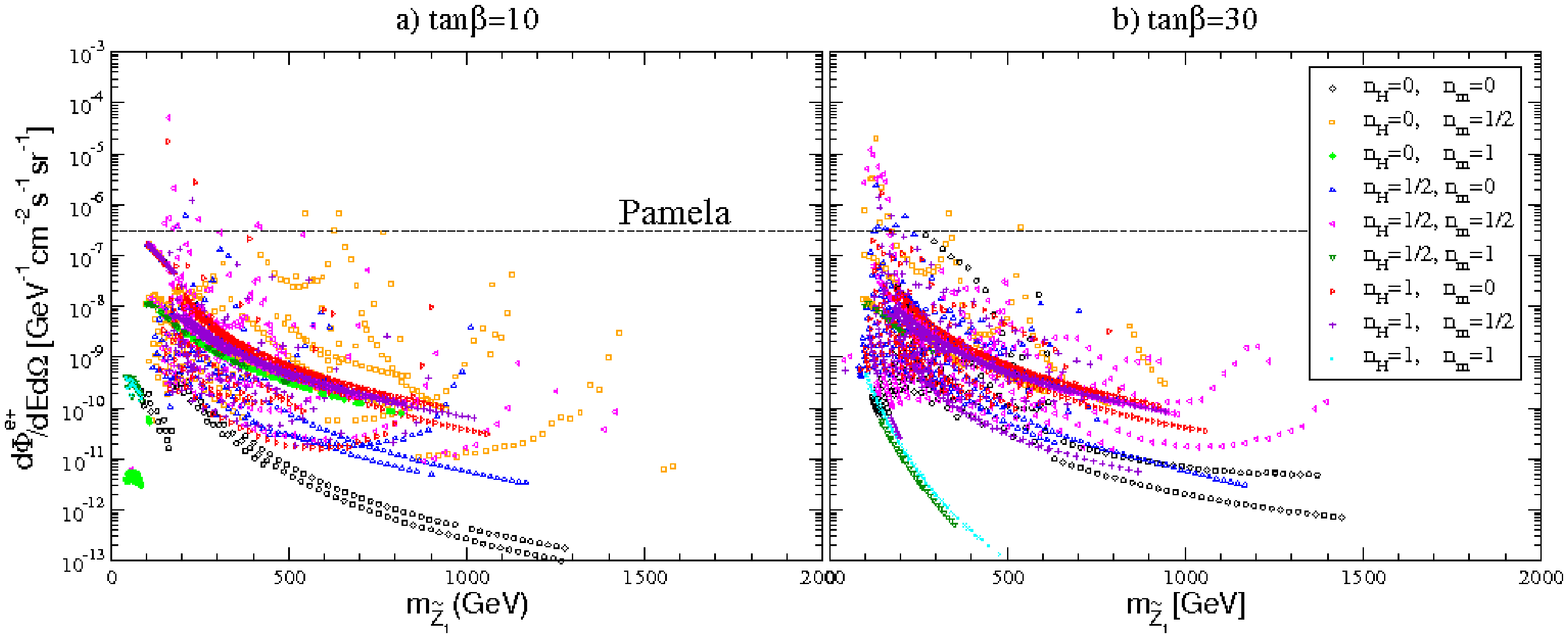,height=12cm,angle=-90}\\
\epsfig{file=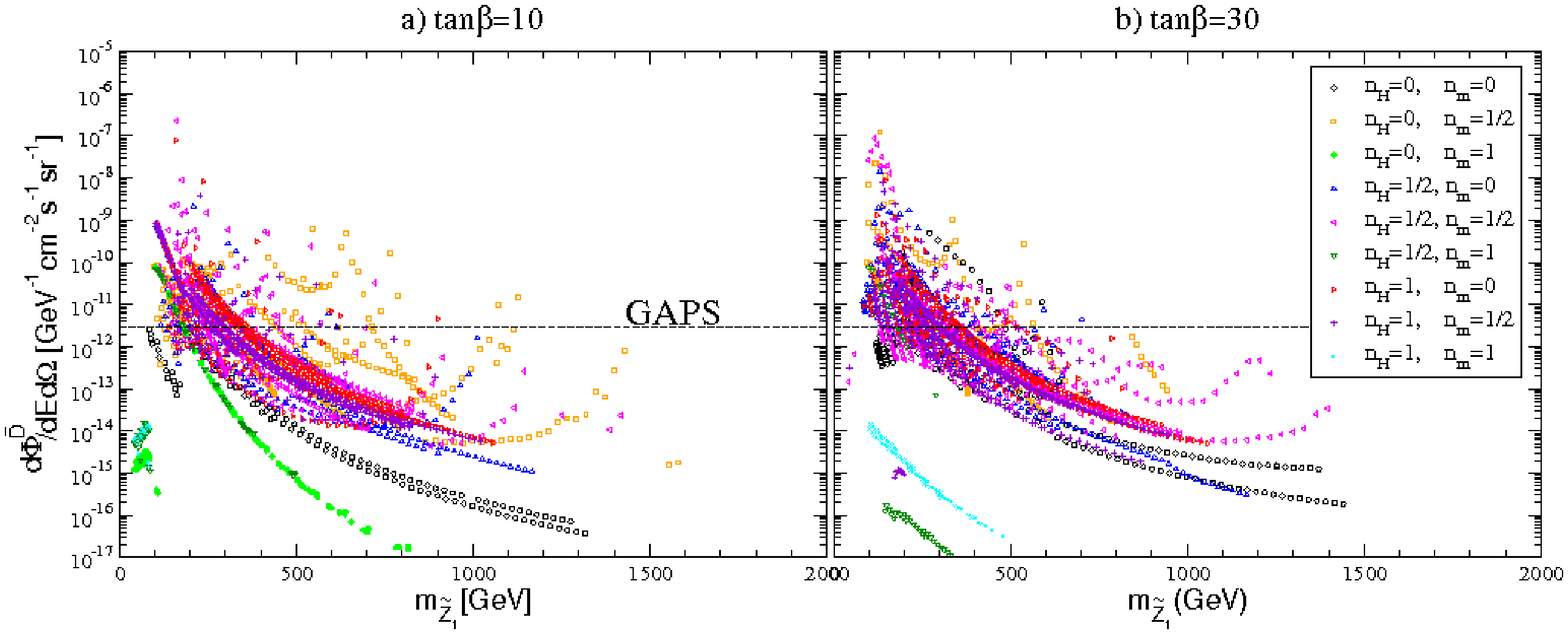,height=12cm,angle=-90}\\
\caption{\label{fig:antip} Expected fluxes
and projected sensitivities of the Pamela
experiment for the detection of anti-protons or positrons, and of the
GAPS experiment for the detection of anti-deuterons, from
neutralino annihilation in our galactic halo, assuming the Burkert
profile for the halo DM distribution for the various choices of Higgs
and matter modular weights as labelled on the figure. We take $\mu>0$
and $m_t=175$ and show results for $\tan\beta=10$ (left column), and
$\tan\beta=30$ (right column). These experiments should be sensitive to
points above the horizontal dashed lines with criteria described in the
text.  Only points consistent with constraints from WMAP and LEP2 are
included in the figure.  }}

We see that typically only a small fraction of the models will be within
the reach of Pamela, with a somewhat larger fraction within the reach of
GAPS. Perhaps, more pertinently, most models with $m_{\tz_1} \agt
300-400$~GeV are projected to be beyond the reach of these experiments.
We should mention, however, that {\it this conclusion is sensitive} to
our assumption of the Burkert halo profile. If instead we assume a
scenario where baryon infall causes progressive deepening of the
gravitational potential well, a considerably clumpier halo distribution
is obtained. For the resulting Adiabatically Contracted $N03$ Halo
Model\cite{achm}, we have checked that we get larger fluxes by factors
of $\sim$15-20 (anti-protons), $\sim$5-7.5 (positrons) and $\sim$10-15
(anti-deuterons) so that the reach is correspondingly increased.

Indirect detection of neutralinos is also possible via the detection of
high energy gamma rays \cite{egret} from produced by neutralino
annihilation in the centre of our Galaxy \cite{gammas}. These will also
be searched for by the GLAST collaboration \cite{glast}.  We have evaluated
expectations for the integrated continuum $\gamma$ ray flux above a
$E_\gamma=1$ GeV threshold in the MM-AMSB model. Since these are
extremely sensitive to the assumed neutralino halo distribution, we show
in Fig.~\ref{fig:glast} results for both the Burkert profile (upper
frames) and the Adiabatically Contracted $N03$ Halo Model (lower
frames), again for $\tan\beta=10$ (left frames) and $\tan\beta=30$
(right frames). The horizontal line marks the expected sensitivity of
GLAST, 1.0$\times10^{-10}\ {\rm cm}^{-2}{\rm s}^{-1}$. We see that while
it almost appears that the entire MM-AMSB parameter space will be
accessible to GLAST for the optimistic projection using the $N03$
profile, the inference using the Burkert profile is completely
different. We need to better understand the halo profile before any
definitive statement can be made, but at least for some models, the
reach of GLAST even very conservatively extends out to
$m_{\tz_1}=500$~GeV.

\FIGURE[htb]{
\epsfig{file=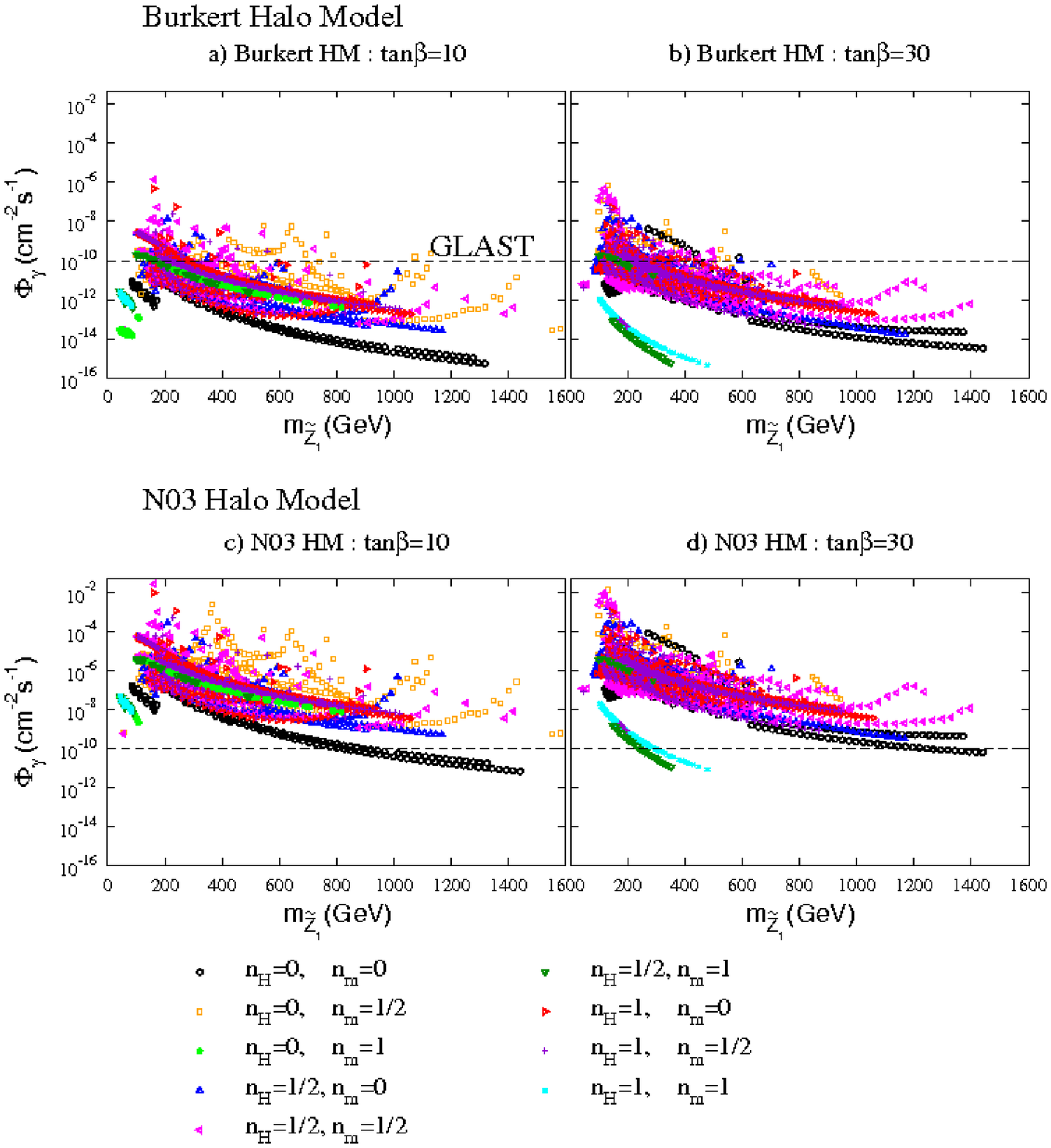,height=16cm, angle=0}
\caption{\label{fig:glast}
The integrated flux and 
projected sensitivity of the GLAST experiment for the detection of gamma
rays with $E_{\gamma}>1$~GeV
from neutralino annihilation in the centre of our Galaxy, for the
the Burkert profile (upper frames) and for the Adiabatically
Contracted $N03$ halo profile for the galactic DM distribution, 
for the various choices of Higgs and matter modular
weights as labelled on the figure. We take $\mu>0$ and $m_t=175$ and
show results for $\tan\beta=10$ (left column), and $\tan\beta=30$ (right
column). GLAST should be sensitive to points above the
horizontal dashed lines with criteria described in the text.
Only points consistent with constraints from WMAP and
LEP2 are included in the figure.}}

\section{Testing mirage unification for ${\bf \alpha < 0}$ and very
  large ${\bf \mu_{\rm mir}}$ }

The distinguishing characteristic of MM-AMSB models is that the SSB
gaugino mass parameters, when extrapolated to high scales using {\it one
loop RGEs} are expected to exhibit unification at the scale $Q=\mu_{\rm
mir}$ given by Eq.~(\ref{eq:mumir}). We emphasize that there is no
physical threshold at this scale and the phenomenon is therefore
referred to as mirage unification. 
For positive values of
$\alpha$, $M_{\rm weak} \le \mu_{\rm mir} \le M_{\rm GUT}$, and it is
straightforward to test the mirage unification
of gaugino mass parameters since the weak scale
SSB gaugino masses can be extracted from
measurements at the proposed electron-positron linear collider. By
evolving the corresponding weak scale sfermion SSB parameters from the
weak scale,
the expected concomitant mirage unification of first generation scalar
masses-- truly a smoking gun confirmation of the scenario-- may also be
revealed\cite{kklt2}. 
An example of the evolution of gaugino
mass parameters evolved from $M_Z$ to $M_{\rm GUT}$, clearly
exhibiting mirage unification, is illustrated in 
Fig.\ref{fig:Mievol}{\it a}), for the MM-AMSB model with $n_H=1$,
$n_m=1/2$ with $\alpha=6$,
for which  Eq.~(\ref{eq:mumir}) gives $\mu_{\rm mir}\sim 4\times
10^{10}$~GeV.
As discussed in 
Ref. \cite{kklt2}, the unified value of gaugino masses at $\mu_{\rm mir}$
is given by $M_s$, while the values of the first/second generation 
scalar mass parameters
extrapolated to $\mu_{\rm mir}$ are given by 
$m_i^2(\mu_{\rm mir})=(1-n_i)M_s^2$, so that the {\it ratio} of 
scalar-to-gaugino masses at $\mu_{\rm mir}$ yields
\be
\left.\frac{m_i^2}{M_a^2}\right|_{\mu_{\rm mir}}=1-n_i,
\ee
which directly measures that matter field modular weights, and hence their
location in the extra-dimensional geometry.
\FIGURE[htb]{\vspace{0.8cm}
\epsfig{file=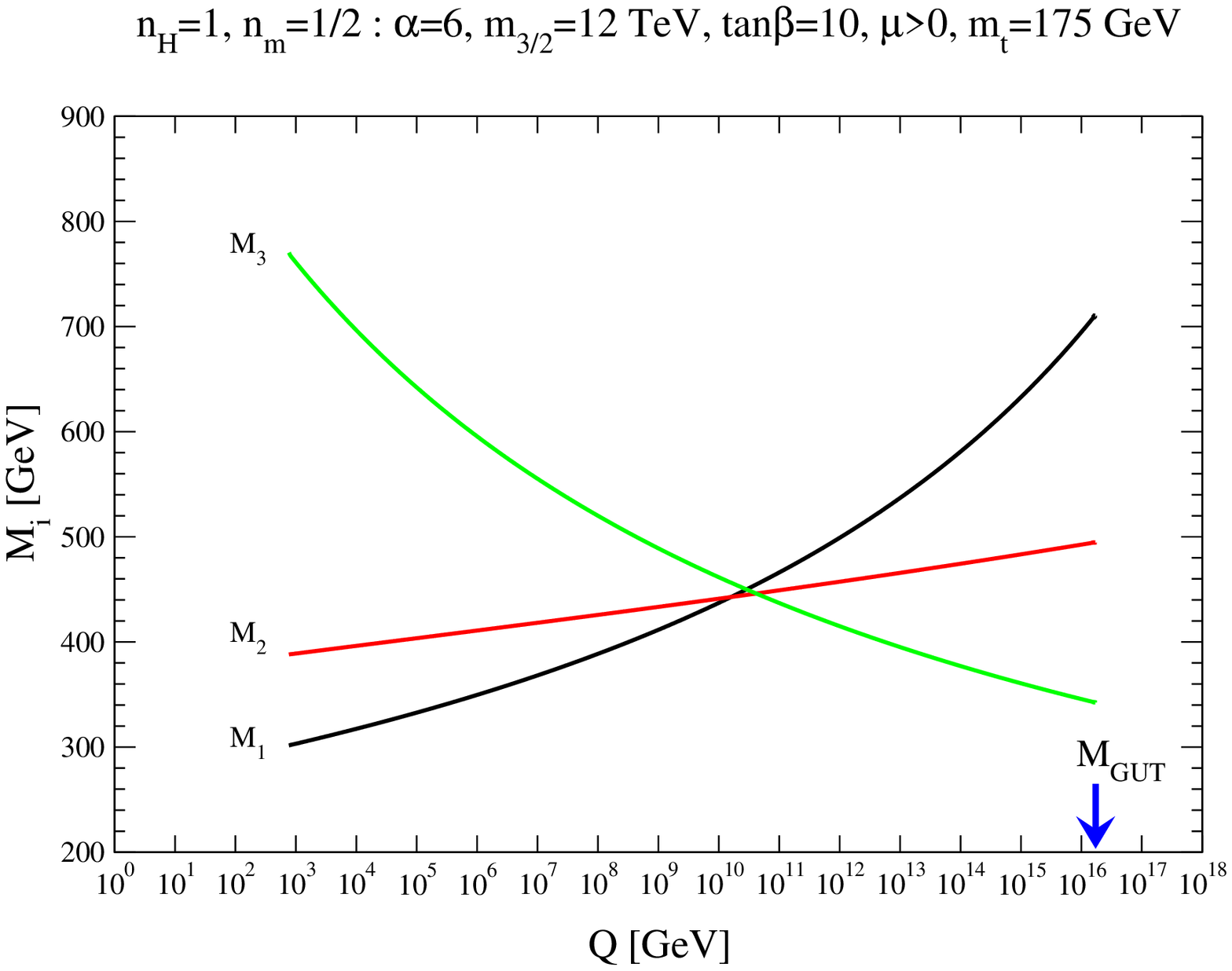,height=8cm,angle=-90}\\
\epsfig{file=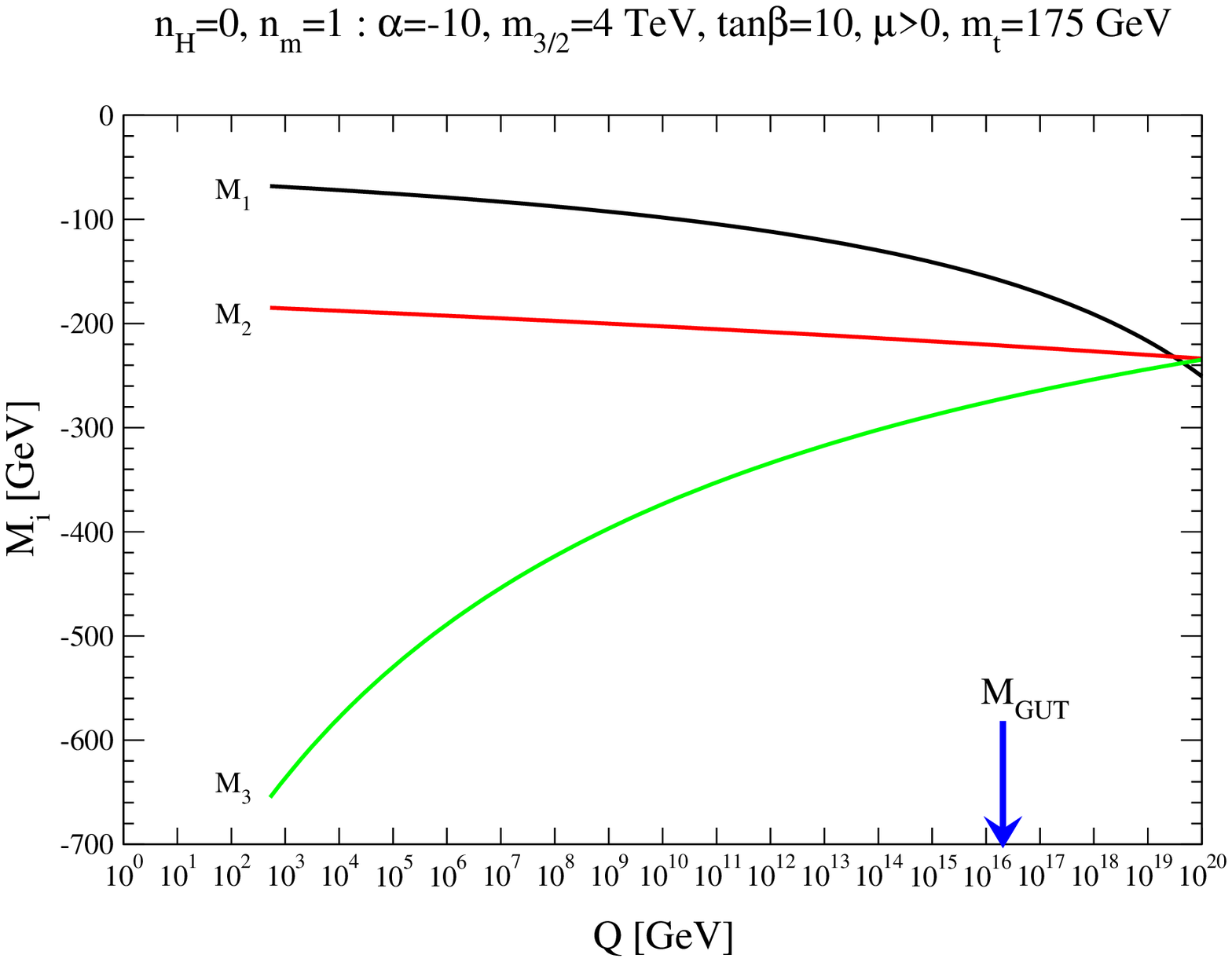,height=8cm,angle=-90}\\
\epsfig{file=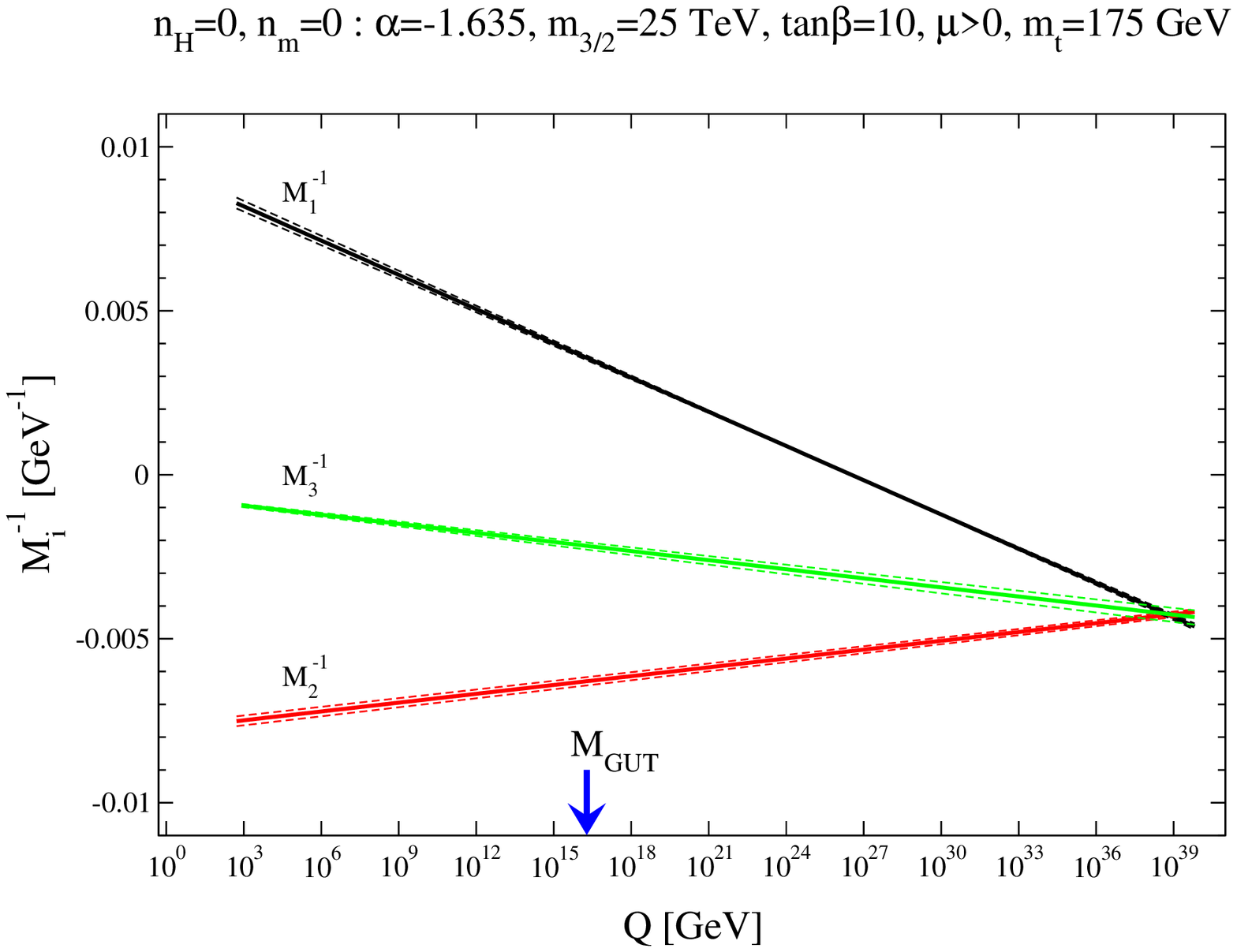,height=8cm,angle=-90}\\
\caption{\label{fig:Mievol} Plot of evolution of gaugino masses versus
energy scale $Q$ from $Q=M_Z$ to $Q=M_{\rm GUT}$ and beyond for three
cases of MM-AMSB model: {\it a}) $(n_H,n_m)=(1,{1\over 2})$ with $\alpha
=6$ and $m_{3/2}=12$ TeV, {\it b}) $(n_H,n_m)=(0,1)$ with $\alpha =-10$,
$m_{3/2}=4$ TeV and {\it c}) $(n_H,n_m)=(0,0)$ with $\alpha =-1.635$ and
$m_{3/2}=25$ TeV.  All figures take $\tan\beta =10$ and $\mu >0$.  We
adopt error projections of $\pm 2\%$ on $M_1$ and $M_2$, and $\pm 5\%$
on $M_3$ for frame {\it c}).Note that we continue MSSM evolution even
beyond the scale $Q=M_{\rm GUT}$
for reasons that we discuss in the text.  }}
We should mention though that even with ideal
experimental measurements, we would not expect mirage unification to be 
perfect because,
\begin{enumerate}

\item mirage unification is a 1-loop phenomenon, and even if the
  RGEs are evolved at 1-loop, the measured physical masses, and hence
  the extracted SSB parameters at the weak scale include ``higher loop
  effects'', and 

\item gauge couplings do not unify exactly because of weak and GUT scale
  threshold corrections. 
\end{enumerate}

If $\alpha < 0$ then $\mu_{\rm mir} > M_{\rm GUT}$ and gaugino and scalar mass
parameters would have to be evolved beyond $M_{\rm GUT}$ using RGEs
corresponding to $SU(3)\times SU(2) \times U(1)$ MSSM evolution to
reveal mirage unification. For large negative values of
$\alpha$, this is again straightforward, as illustrated
in Fig.~\ref{fig:Mievol}{\it b}) for Point~3 in Table~1. We
see that the gaugino mass parameters exhibit mirage unification {\it
beyond} $Q=M_{\rm GUT}$ at $\mu_{\rm mir}\sim 10^{20}$ GeV. 
Matter scalar mass parameters can be similarly
evolved to $\mu_{\rm mir}$ in order to extract matter modular weights,
within errors expected from the extrapolation. 

However, if $\alpha$ is  small in magnitude,
but negative, we face a new problem because $\mu_{\rm mir}$ becomes
so large that
one (or more) of the one-loop-evolved gauge couplings
diverges for $Q< \mu_{\rm mir}$, making the numerical integration of the RGEs
from the weak scale to $\mu_{\rm mir}$ difficult. We can circumvent this 
difficulty because it is possible to analytically integrate the one loop
RGEs for gaugino masses and gauge couplings to obtain\cite{wss}
\be
\frac{M_i(Q)}{g_i^2(Q)}=K_i, \ \ \ {\rm or \ equivalently}, \frac{1}{M_i(Q)}=
K_i^{-1} \times \frac{1}{g_i^2(Q)},
\label{eq:Mevol}
\ee
where $K_i$ is a constant independent of the energy scale $Q$, and
$i$ labels the gauge group. Indeed, if gaugino masses and
gauge couplings both
unify at $M_{\rm GUT}$, this leads to the well-known gaugino mass unification
relation. 
From the experimentally determined weak scale values of $M_i$
and $g_i$ we can obtain $K_i$; these can then be used together with
$1/g_i^2(Q)$ obtained using
\be
{1 \over {g_i^2(Q)}} -{1 \over {g_i^2(Q_0)}} = -{{b_i}\over {8\pi^2}}
\ln\left({Q\over {Q_0}}\right)\;
\label{eq:gevolve}
\ee 
to get the values of $1/M_i(Q)$ at any higher scale.\footnote{Notice
that ${1 \over {g_i^2(Q)}}$ vanishes ({\it i.e.} $g_i^2(Q)$ blows up) when
${1 \over {g_i^2(Q_0)}} -{{b_i}\over {8\pi^2}} \ln\left({Q\over
{Q_0}}\right) =0$, and becomes negative for yet larger values of $Q$. It
is these (unphysical) negative values of $g_i^2(Q)$, obtained from the
solution of the one-loop RGEs, that must be used in evaluating
$1/M_i(Q)$. The ratio of two different gaugino mass parameters can
flip sign precisely because one of the corresponding $g_i^2(Q)$s becomes
negative.}
In Fig.\ref{fig:Mievol}{\it c}) we show the evolution of $1/M_i(Q)^2$
using the procedure just described
for a WMAP/LEP2-allowed case with 
$(n_H,n_m)=(0,0)$ with $\alpha =-1.635$, $m_{3/2}=25$ TeV, $\tan\beta =10$
and $\mu >0$ (Point~4 from Table~1 of Ref. \cite{kklt}), which
should have a value $\mu_{\rm mir}\sim 2\times 10^{37}$ GeV, to be
compared with $Q\sim 10^{39}$~GeV in the figure.  Notice also 
that by working with $M_i(Q)^{-1}$ and $1/g_i^2(Q)$, we circumvent the
divergence that would have been present at $Q\sim
10^{27}$~GeV. The bands correspond to a hypothetical uncertainty of $\pm
2$\% on the extraction of the weak scale values of $M_1$ and $M_2$, and
of $\pm 5$\% on the extraction of $M_3$. 

Turning to the possibility of checking mirage unification for
first/second generation scalar masses (for which Yukawa interactions can
be neglected) when $\alpha < 0$, we observe that
\be
m_i^2(Q) = Z_i - 2\sum_a \frac{C_2^a(f_i)}{b_a} M_a^2(Q)
\label{eq:modwt}
\ee
 is a solution to the 1-loop RGEs for scalar mass squared, where $Z_i$
are scale-independent constants, and $C_2^a(f_i)$ is the quadratic
Casimir for the a$^{th}$ gauge group that we have encountered below
Eq.~(\ref{eq:m2}). Since $Z_i$ are scale-independent, by taking the
difference of $m_i^2(\mu)$ at the two scales, we obtain, 
\be
m_i^2(\mu_{\rm mir}) =  m_i^2(\mu_{\rm weak}) + 2\sum_a
\frac{C_2^a(f_i)}{b_a}\left[ M_a^2(\mu_{\rm weak}) - M_a^2(\mu_{\rm
mir})\right]\;,
\label{eq:negscal}
\ee where $m_i^2(\mu_{\rm weak})$ is the weak scale SSB parameter for
the $i^{th}$ sfermion, which we can obtain from its measured mass by
removing the $D$-term.\footnote{In practice we may not know the value of
$\tan\beta$ very well. Fortunately, the change in the $D$-term is
negligibly small as long as $\tan\beta \agt 5$, and often even for
smaller values of $\tan\beta$. In our case study described below, we use
the $D$-term with $\cos 2\beta= -1$. } Eq.~(\ref{eq:modwt}), or
equivalently, Eq.~(\ref{eq:negscal}) can then be used to obtain
$m_i^2(\mu_{\rm mir})$ using the values of gaugino mass parameters at
the weak and mirage unification scales that we have already obtained,
and the corresponding modular weights can be extracted. Specifically, we
first use the weak scale SSB mass parameters to obtain $Z_i$, and then
use this value of $Z_i$ along with the gaugino SSB parameters at
$Q=\mu_{\rm mir}$ determined above, to obtain $m_i^2(\mu_{\rm mir})$.

%
\begin{table}\begin{center}
\begin{tabular}{lccc}
\hline & $n_m=0$ & $n_m=\frac{1}{2}$ & $n_m=1$ \\ 
\hline $\tQ$ &
$465^{+203}_{-576}$ & $429^{+211}_{-631}$ & $388^{+222}_{-654}$\\
$n_{\tQ}^{\rm exp}$ &[$-7.5$, 1.2] & [$-6.9$, 1.7]& [$-6.2$, 2.4]\\ \hline
$\te_L$ & $280.1^{+8.6}_{-8.9}$ & $212.8^{+9.6}_{-10.0}$ &
$110.4^{+14.6}_{-17.1}$ 
\\ $n_{\te_L}^{\rm exp}$ &[$-0.58$, $-0.29$]
&[0.05, 0.27] & [0.70, 0.84]\\ \hline
$\te_R$ & $257.6^{+3.4}_{-3.5}$ &
$181.5^{+3.0}_{-3.1}$ & $-32.3^{+7.6}_{-6.1}$ \\ 
$n_{\te_R}^{\rm exp}$
& [$-0.26$, $-0.13$] &[0.34, 0.43] &[1.01, 1.03]\\ \hline
\end{tabular}
\caption{Values of sfermion mass parameters at $Q=\mu_{\rm mir}$
  obtained from the intersection of $M_1(\mu_{\rm mir})=M_2(\mu_{\rm
  mir})$ for the case studies with $\alpha=-1.635$, $m_{3/2}=25$~TeV,
  $\tan\beta=10$, $\mu>0$, $n_H=0$, and $n_m=0$, $\frac{1}{2}$, and 1. 
  For each sparticle, we show the value of
  $sign[m_i^2(\mu_{\rm mir})] \sqrt{|m_i^2(\mu_{\rm mir})|}$. The
  range corresponds to the projected experimental uncertainties
  discussed in the text. For each sparticle, the second line gives the
  range of the modular weight consistent with the range of the
  corresponding $m_i^2(\mu_{\rm mir})$. The negative sign in the value
  of $m(\te_R)$ for $n_m=1$ really refers to the negative sign of
  $m^2(\te_R)$.}
\label{tab:mod}\end{center}
\end{table}

We have performed a case study to illustrate this procedure, adopting
the point chosen in Fig~\ref{fig:Mievol}{\it c}), except that we perform
the analysis for $n_m=0$, $\frac{1}{2}$ and 1 to study how well the
matter modular weights can be extracted in this difficult case.  Very
optimistically, we assume that it will eventually be possible to obtain
colored sparticle masses to $\pm 5$\%, that selectron masses will be
measured with a precision of $\pm 1$\%, and finally, that we will be
able to extract weak scale values of $M_1$ and $M_2$ to $\pm 2$\%. From
these sfermion masses we obtain the corresponding weak scale SSB
parameters upon removing the $D$-terms using $\cos 2\beta=-1$, and then
use Eq.~(\ref{eq:modwt}) to get the SSB parameters at the mirage
unification scale, $\mu_{\rm mir}$ obtained from the intersection of
$M_1$ and $M_2$ (as {\it e.g.} in Fig.~\ref{fig:Mievol}{\it c}). The
results of our computation are shown in Table~\ref{tab:mod}. The quoted
range on the entries corresponds to the variation we obtain by varying the weak
scale SSB parameters within their assumed errors. For each sparticle, on
the second line, we show the range of modular weights compatible
with the extracted values of the SSB mass squared parameters at the
mirage unification scale. The following seem to be worthy of note.
\begin{itemize}
\item Even with the optimistic error of only $\pm 5$\% on the extraction
of colored sparticle masses, we see that squark SSB parameters are very
poorly restricted at the mirage scale. The reason is clear from
Eq.~(\ref{eq:negscal}): squarks and gluinos each have masses $\sim
1000$~GeV, so that a very large cancellation between the terms on the
right hand side is needed to get the left hand side to be $\sim (100-250
\ {\rm GeV})^2$, the value of the matter mass parameters at $\mu_{\rm
mir}$. The effect of higher loop terms, not-exact-unification of gauge
couplings and various threshold corrections spoil the delicate
cancellations, leading to a very large error on the extraction of squark
modular weights in this case.

\item We see that while slepton masses do appear to unify within $\alt
10$\% at least for $n_m=0$ and $n_m=1/2$ cases in Table~\ref{tab:mod},
the ranges of the modular weights that we extract using our procedure are
shifted systematically to lower values. We have traced this shift to
the fact that the value of $M_a(\mu_{\rm mir})\simeq 232.5$~GeV that we
obtain from Fig.~\ref{fig:Mievol}{\it c}) is too small by about 10\%,
causing a systematic shift in the extracted modular weights. We had
already remarked that $\mu_{\rm mir}$ obtained from this figure was
about 50 times larger than its theoretically expected value. For
$n_m=1$, it appears that $m^2(\te_R)$ and $m^2(\te_L)$ unify very
poorly. Note, however, that $m_i^2(\mu_{\rm mir})$ is expected to be
very small in this case, so that we must again have a large cancellation
between the two terms on the right-hand-side of Eq.~(\ref{eq:negscal})
for this to occur. As with the squarks, systematic effects then play an
important part. It is, however, interesting to see that $m_i^2(\mu_{\rm
  mir}) \ll M(\mu_{\rm mir})$, so that the extracted matter modular
weights are close to unity.

\end{itemize}
We conclude that while our simple procedure provides a good indication
of mirage unification of gaugino mass parameters, and a qualitative
indication for the mirage unification of scalar mass parameters,
quantitative extraction of the model parameters have significant
systematic uncertainties in this difficult case of very large $\mu_{\rm
mir}$. We presume that this is because of higher loop and finite
correction effects that are present in the weak scale SSB parameters. In
principle, a determination of {\it all} weak scale SSB parameters may
make it possible to extract ``one-loop'' weak scale gaugino and
first/second generation SSB parameters, but until this is possible,
quantitative extraction of modular weights (particularly with a
precision to enable a distinction between toroidal and more complicated
compactifications) may prove difficult if
$\mu_{\rm mir} \gg M_{\rm GUT}$.

\section{Summary and Conclusions}
\label{sec:conclude}

Mixed modulus-anomaly mediation of SUSY breaking, also referred to as
mirage-mediation, provides a new, theoretically well-motivated, and
phenomenologically viable framework for analyses of the implications of
SUSY. The MM-AMSB model is completely specified by just three parameters
along with the discrete set of modular weights for chiral super-fields
that are determined by their location in the extra spatial
dimensions. 
The naturally heavy moduli and gravitinos can be consistent with
cosmology, assuming that there is a second late period of inflation that
dilutes
their density. The observed dark matter may then comprise principally
of neutralinos that are thermally produced upon reheating after this
epoch.
The
framework also provides a plausible solution to the SUSY flavor
problem, and also ameliorates the SUSY $CP$ problem. 

We have examined the phenomenology of these models under the assumption
(motivated by the absence of large flavor changing neutral currents)
that matter superfields all reside on either a $D7$ brane, a $D3$ brane,
or on their intersection: using examples with toroidal compactifications, the
matter modular weights are  fixed to be 0, 1 or ${1\over 2}$,
respectively. Higgs superfields may lie at a different location, giving
us 9 discrete choices for set ($n_H$, $n_m$) of the matter and Higgs
modular weights. More complicated configurations that lead to other
choices for the modular weights ({\it e.g.} 2/3) may be possible, but we
expect that the phenomenology will then ``lie in between'' the cases that we
examine. 

The phenomenological implications of the model differ from that of other
models because, (1)~the combination of modulus and anomaly-mediated
contributions to SUSY breaking lead to novel patterns of sparticle
masses, and (2)~the composition of the lightest neutralino differs
because the weak scale gaugino masses are {\it not} in the ratio $\sim
1:2:6$ predicted by the gaugino mass unification condition, as is the
case in many models. These features of the spectrum make it possible to
obtain a neutralino thermal relic density consistent with WMAP
measurements in a variety of interesting ways that have been proposed in
previous {\it bottom-up} studies, not all of which are possible in the
much studied mSUGRA model. These include co-annihilation with staus or
stops, resonance annihilation via $A$ and $H$, bino-wino
co-annihilation, mixed higgsino and mixed bino-wino-higgsino DM and low
$|M_3|$ (mixed-higgsino) DM, but not all mechanisms are realized for
every choice of matter and Higgs modular weights. Interestingly, mixed
wino DM (without also a significant higgsino content), or of mixed
higgsino DM via non-universal Higgs mass parameters, does not seem to be
realized within this framework.

Within this framework, it may be possible for Tevatron experiments to
discover supersymmetry in portions of the parameter space not accessible
to LEP2 searches. Possibilities include searches via the
$E_T^{\rm miss}$ channel for say ($n_H$, $n_m$) = (1, 0), where gluinos
and squarks as low as about 300~GeV may have escaped detection up to
now, searches for light stops and, depending the branching ratio for
$\tz_2\to\ell\ell\tz_1$ decay, trilepton events from $\tw_1\tz_2$
production. As can be seen from
Fig.~\ref{fig:psnH0}-Fig.~\ref{fig:psnH1}, assuming 100~fb$^{-1}$ of
integrated luminosity, LHC experiments should be able to detect
SUSY signals over {\it most} of the parameter space of the model
consistent the relic density measurement, with some exceptions for
$n_H=0$ or $1/2$, and $n_m=1$. Experiments at a 1~TeV linear collider
will be able to make precision measurements over part of the region
probed by the LHC. Interestingly, these experiments
will be able to explore charginos and neutralinos if
the measured relic density arises due to BWCA: this is important because
the small gap, $m_{\tw_1} - m_{\tz_1}$, may make exploration of charginos
difficult at the LHC. 

We have also examined prospects for direct and indirect detection of DM
for the nine cases of modular weights. Our results for direct detection
are shown in Fig.~\ref{fig:dd}, where we see that the CDMS experiment
has already begun to exclude a portion of the parameter space for
$m_{\tz_1}$ as heavy as 300~GeV, for the larger value of $\tan\beta$.
It is also striking that proposed ton size noble gas detectors (we use
the proposed Warm Argon Project with 1,400~kg of argon as a benchmark)
will be able to explore most of the parameter space of this model, even if
$m_{\tz_1}$ is as heavy as $\sim 1000$~GeV. Prospects for indirect
detection are somewhat less certain. Models with significant higgsino
components yield the largest signals, while models with a bino
LSP will generally not give an observable signal. There may be
detectable signals from high energy muons from the sun at IceCube, as
well as from anti-particles at Pamela and Gaps, or from gamma rays in
GLAST. There is, however, considerable theoretical uncertainty (factor
10-20) in the prediction of the anti-particle signals from the unknown
clumping of neutralinos in our galactic halo, and up to four orders of
magnitude uncertainty for signals at GLAST, so that projections for
these experiments should be viewed with care. Since this clumping
provides a {\it common uncertainty} for these experiments, observing a
signal in one of these will greatly facilitate a much more certain
prediction for the other experiments, assuming that the observed signal
is from neutralinos annihilating in our galactic halo. 

Finally, we have explored the possibility of testing mirage unification
of SSB parameters, which is the smoking gun for this framework. The case
for small or intermediate mirage unification scale had already been
explored and it had been shown that mirage unification of both gaugino
and scalar mass parameters would be readily testable, and allow extraction
of the modular weights. Here, we have
explored the much more difficult case that occurs if $\alpha$ is small
in magnitude, but negative, so that $\mu_{\rm mir} \gg M_{\rm GUT}$. In
this case, we find that mirage unification of gaugino masses should
still be testable, although extraction of underlying values of $\mu_{\rm
mir}$ and $M_s$ has significant systematic uncertainty due to the large
``distance'' between weak and mirage scales. Combining measurements from
the LHC with measurements that will be possible with initial runs of a
linear collider may allow a qualitative test of mirage unification of
scalar masses, but unlike the case of intermediate values of $\mu_{\rm
mir}$, extraction of matter modular weights is fraught with systematic
uncertainty.

In summary, we have made an exhaustive exploration of the phenomenology
of the MM-AMSB model, a promising new framework for SUSY phenomenology.
The novel patterns of sparticle masses that are possible allow various
mechanisms that lead to a DM relic density in agreement with
observations to be incorporated into a top-down framework, something not
previously possible. This can qualitatively change the correlations
between various experimental signals from expectations in the much
studied mSUGRA framework. In view of its attractive and economic
theoretical structure as well as its attractive phenomenological
attributes, we believe it is at least as attractive (and theoretically,
perhaps, more compelling) as mSUGRA as a framework for analyses of SUSY.

\section*{Acknowledgments}

We thank Kiwoon Choi for clarifying discussions and correspondence. We
are also grateful to J.~Conlon, M.~Endo and F.~Takahashi for drawing
our attention to their work, and for helpful correspondence about the same.
This research was supported in part by the U.S. Department of Energy
grants DE-FG02-97ER41022 and DE-FG02-04ER41291.

%

\end{document}